# Mid- and Far-Infrared Photometry of Galactic Planetary Nebulae with the AKARI All-Sky Survey

J.P. Phillips, R.A. Marquez-Lugo

Instituto de Astronomía y Meteorología, Av. Vallarta No. 2602, Col. Arcos Vallarta, C.P. 44130 Guadalajara, Jalisco, México   e-mails : jpp@astro.iam.udg.mx, alejmar.astro@gmail.com,

**Abstract**

We provide mid- and far-infrared photometry of 857 Galactic planetary nebulae (PNe) using data derived from the AKARI All-Sky Survey. These include fluxes at 9 and 18 $\mu$m obtained with the Infrared Camera (IRC), and at 65, 90, 140 and 160 $\mu$m using the far-Infrared Surveyor (FIS). It is noted that the IR luminosities of the youngest PNe are comparable to the total luminosities of the central stars, and subsequently decline to ~ $510^2$ $L_\odot$ where D > 0.08 pc. This is consistent with an evolution of PNe dust opacities, and appreciable absorption in young and proto-PNe. We also note that there is little evidence for the evolution in IR/radio flux ratios suggested by previous authors. The fall-off in dust temperatures is similar to that determined in previous studies, whilst levels of Ly$\alpha$ heating are < 0.5 of the total energy budget of the grains. There appears to be an evolution in the infrared excess (IRE) as nebulae expand, with the largest values occurring in the most compact PNe.

**Key W**ords:   (ISM:) dust, extinction --- (ISM:) planetary nebulae: general --- ISM: jets and outflows --- infrared: ISM

# 1. Introduction

Several large-scale surveys at infrared wavelengths have been undertaken in the last thirty or so years, permitting a clearer understanding of the properties of dust in planetary nebulae (PNe). Thus for instance, Pottasch et al. (1984) investigated the thermal and physical characteristics of grains using photometry taken with the Infrared Astronomical Satellite (IRAS; Neugebauer et al. 1984). They found that dust temperatures decline with increasing nebular radius R, and that dust/gas mass ratios decrease with expansion of the shells. A further analysis by Lenzuni et al. (1989) found that grain radii also decreased with radius, whilst the volume density of particles increased as $\propto R^{4.19}$; tendencies which were attributed to grain-grain collisions, sputtering by stellar winds, or the selective destruction of larger grains. A further analysis by Stasinska & Szczerba (1999), however, found no change in either of the latter two parameters.

Subsequent surveys of PNe have been undertaken in the near infrared (NIR) using the 2MASS all-sky survey (Ramos-Larios & Phillips 2005), whilst a more restrictive survey, taken with the Spitzer Space Telescope (Spitzer; Werner et al. 2004), has enabled mid-infrared (MIR) observations of PNe close to the Galactic plane (see e.g. Cohen et al. 2005; Phillips & Ramos-Larios 2008; Ramos-Larios & Phillips 2008; Zhang & Kwok 2009; Phillips & Marquez-Lugo 2010). These latter results have shown the importance of polycyclic aromatic hydrocarbons (PAHs) in enhancing MIR emission at 5.8 and 8.0 μm, and revealed the presence of extensive emission outside of the ionized regimes, likely arising from photo-dissociation regimes (PDRs).

A more recent AKARI infrared survey has now completed the major phase of its observational program, and the point source catalogue (PSC) was made available in March 2010. This represents an updated version of the IRAS project, containing as it does more photometric channels, higher levels of spatial resolution, and higher sensitivity detectors. We present photometry for 857 Galactic PNe taken with the infrared camera (IRC) and far-infrared surveyor (FIS), and use this to analyse the evolution in grain temperatures and FIR luminosities.

## 2. Observations

The AKARI space mission (Murakami et al. 2007) was launched in 2006, and contains the IRC (Onaka et al. 2007) for observations in the range 2-26 μm, and the FIS (Kawada et al. 2007) for the 50-200 μm regime. The Ritchey-Chretian type telescope has a 0.685 m primary mirror, cooled to 6 K by liquid helium and mechanical coolers. One of the primary objectives of this mission was an all sky IRC/FIS survey, undertaken between 2006 May and 2007 August. The resulting PSCs (Kataza et al. 2010 (IRC); Yamamura et al. 2010 (FIS)) were published in 2010 March.

The IRC filter profiles extend between 6.76 and 11.6 μm (in the case of MIR-S), with an effective wavelength of 9 μm, and between 13.9 and 25.6 μm (MIR-L) with an effective wavelength of 18 μm. The camera field of view (FOV) was 10x9.6 arcmin$^2$ for MIR-S, and 10.7x10.2 arcmin$^2$ for MIR-L, whilst the pixel scales were 2.34x2.34 arcsec$^2$ (MIR-S) and 2.31x2.39 arcsec$^2$ (MIR-L). Each of four adjacent pixels were subsequently binned to create so-called "virtual" pixels, with areas 4x4 times greater than those of the original pixels. However, subsequent processing of two independent survey grids resulted in an improvement in spatial resolution, and point spread functions (PSFs) with FWHM ~ 5.5 arcsec at 9 μm, and 5.7 arcsec for the 18 μm channel.

The FIS survey, by contrast, permitted observations at effective wavelengths of 65 μm (filter N60, bandwidth $\Delta\lambda$ = 21.7 μm), 90 μm (Wide-S, $\Delta\lambda$ = 37.9 μm), 140 μm (Wide-L, $\Delta\lambda$ = 52.4 μm) and 160 μm ($\Delta\lambda$ = 34.1 μm). The 65 and 90 μm results were taken with a monolithic Ge:Ga detector having a pixel size 26.8 arcsec, whilst the longer wave (140 and 160 μm) data were acquired using a stressed Ge:Ga detector, and pixel scale of 44.2 arcsec. The FOVs were in all cases ~ 1 arcmin, whilst the FWHM of the PSFs were 37 arcsec (at 65 μm), 39 arcsec (90 μm), 58 arcsec (140 μm) and 61 arcsec (160 μm). Detector sensitivities are important in determining the rates of source detection, and quoted values range between 0.55 Jy at 90 μm to as high as 25 Jy for the 160 μm results. This question is further examined in Sect. 3.4, where we consider its importance in determining nebular emission trends.

Details of the source selection procedure are provided in the following sections, together with a description of the catalogue results, the infrared dimensions of the PNe, and a comparison of the AKARI and IRAS photometry.

**Planetary Nebula Data Base and Search Radius**

The positions of the PNe investigated in this present study are taken from the listing of coordinates by Kerber et al. (2003). This provides positions for 1086 sources based on the 2$^{nd}$ generation General Star Catalogue. The coordinates for sources which are compact (or have central stars) are quoted as being accurate to 0.35 arcsec, whilst the positions for 226 more extended PNe are accurate to ~ 5 arcsec.

By contrast, the AKARI/IRC All-Sky Survey Release Point Source Catalogue – Release Note (Rev. 1) (Kataza et al. 2010) suggests that the mean angular separation of the AKARI and 2MASS sources is of order $\cong$ 0.77 arcsec, with 95% of the sources have angular separations of < 2 arcsec. Similarly, the positional errors of the longer wave FIS sources (Yamamura et al. 2010) is determined through a comparison with the SIMBAD database, yielding flux independent uncertainties of order ~ 6 arcsec.

Given that the PNe data base is identical for the FIS and IRC searches, it follows that search radii need to be larger for FIS than for the IRC results.

That this is the case is apparent from Fig. 1, where it is seen that the rates of source detection increase with increasing search radius, eventually reaching plateau values of 696 sources for the IRC detections, and 678 sources for the FIS results. It is also apparent that whilst most of the PNe are detected within radii $\theta_S$ ~ 3 arcsec for the IRC results, a radius of order $\theta_S$ ~ 14 arcsec is required for comparable rates of FIS detection. We have adopted a search radius $\theta_S$ = 6 arcsec for the IRC results, and $\theta_S$ = 10 arcsec for the FIS photometry.

There are two reasons to suspect that background contamination is extremely modest. It is apparent for instance that rates of source detection are flat as $\theta_S$ increases from ~15 arcsec to 30 arcsec, implying that any field source contamination is likely to be < 2 %. We also note that the AKARI FIS catalogue includes a parameter NDENS, corresponding to the number of sources within 5 arcmin of the central (PNe) positions. This takes a mean value of ~2.2 for 415 estimates of the parameter in the most sensitive (90 μm) channel, suggesting again that background contamination is extremely low. A similar conclusion applies for the NDENS parameters at 9 and 18 μm, corresponding to the numbers of sources within 45 arcsec of the search positions. We determine mean values NDENS09 = 0.068 (for 396 estimates of the parameter), and NDENS18 = 0.020 (for 612 estimates of NDENS), implying that very few of our identifications are likely to be spurious.

**Catalogue Results for Galactic PNe**

Photometric results for 857 PNe detected in the IRC catalogue are presented in Table 1, where all of the fluxes have FQUAL values of 3 (i.e. all of the data is classified as being valid). An analysis by the AKARI team of trends in photometric errors shows that the most probable uncertainties are of the order of ~2-3 %, with 80% of the data having errors of <15 %. It is found that the signal to noise ratio (S/N) for the 9 μm data is of order ~19 for sources having F(9μm) > 0.5 Jy, and falls as approximately S/N ~ $32F_\nu(9\mu m) + 4$ for smaller flux levels. Similarly, it is found that the S/N for 18 μm results is ~ 15 where F(18μm) > 0.8 Jy, and decreases as ~ $17F_\nu(18\mu m) + 3$ below this limit.

We have, apart from this, specified the numbers of the PNe in the Kerber et al. list of coordinates (column 1); the commonly accepted names of the sources (column 2); and the right ascensions and declinations of the nebulae (columns 3 & 4). The angular diameter θ is based upon radio and optical results, where we have taken values from Acker et al. (2002), Zhang (1995), Cahn & Kaler (1971), Tylenda et al. (2003), Perek & Kohoutek (1967), and Siodmiak & Tylenda (2001). Where maximal and minimal diameters $\theta_{MAX}$ and $\theta_{MIN}$ are quoted, then we have employed harmonic mean values $\theta = (\theta_{MAX}\theta_{MIN})^{0.5}$. Similarly, the AKARI mean radii at 18 μm (denoted in the catalogue by MEAN_AB18) are provided in column 8, and labeled MN_18. The 1.4 GHz results (column 9) are taken from the uniform data base of Condon et al. (1998), whilst the 5 GHz measures (column 10) derive from the compilations of Acker et al. (2002), Zhang (1995) and Siodmiak & Tylenda (2001). Finally, the distances of PNe remain notoriously uncertain, and the numbers of more reliable distances (i.e. those based on trigonometric, reddening, gravitational and kinematic measures, among others) is still pitifully small. We have therefore quoted distances based on larger-scale statistical measures, whilst recognizing that these are prone to systematic and random errors. Typical uncertainties in distance are probably of the order ≈ 30 %. We have used values quoted by Phillips (2004) and Zhang (1995) in that order of priority, and these are listed in column 11 of Table 1. These estimates are also combined with the angular diameters in column 7 to determine the physical diameters in column 12.

The longer wave FIS identifications are listed in Table 2, where apart from the Kerber et al. number, and PNe names and coordinates (columns 1-4), we include photometry at 65, 90, 140 and 160 $\mu$m (columns 5-8). The data has been assigned four levels of quality FQUAL, where FQUAL = 0 corresponds to "non-detections"; FQUAL = 1 implies that the source was not confirmed; FQUAL = 2 is for sources which are confirmed, but for which fluxes are unreliable; and FQUAL = 3 is for high quality fluxes, where the source is confirmed and the flux is reliable. We list sources for which at least one of the four channels has a FQUAL value of 3, and for these cases, include all of the fluxes corresponding to FQUAL = 1-3. As an example, the detection at 90 $\mu$m may have been assigned a FQUAL value of 3, but FQUAL factors for the other three channels (65, 140 and 160 $\mu$m) may be as low as 1. However, although this photometric listing is relatively complete, and provides more and less reliable results, we have used only FQUAL = 3 fluxes for the analysis below. The FQUAL values for the 65, 90, 140 and 160 $\mu$m channels are consecutively indicated in column 9, whilst the remaining columns define the radio fluxes, distances and diameters described above.

The noise in these results again derives from absolute uncertainties in the fluxes, and relative errors deriving from a variety of instrumental causes, including glitches, noise, residuals in the responsivity corrections and other anomalies. For the brighter AKARI sources, the absolute uncertainties are of order ~15 %, whilst relative errors are ~10%, leading to total errors of ~20% in all of the longer wave bands.

Of the sources listed in Tables 1 & 2, 415 PNe are common to both of the listings. Similarly, a total number of 857 PNe has been measured in this survey, corresponding to 79 % of the sources listed in Kerber et al. (2003).

## Source Sizes in the AKARI Catalog

We have noted, in Sect. 2, that although virtual pixel sizes for the IRC results are of the order of 9-10 arcsec, the FWHM of the PSFs are ~ 5.5-5.7 arcsec. Given that many PNe have dimensions exceeding these limits, it follows that AKARI should permit us to determine the angular dimensions for many of the detected PNe.

The AKARI IRC catalogue includes the flags EXTENDED09 and EXTENDED18, which are triggered when sources have radii > 15.6 arcsec. None of the PNe in the present sample are flagged as being extended, even though a reasonable proportion (of order 13 %) have radio/optical radii which are greater than these limits. The AKARI catalogue also provides mean radii of the sources at 9 and 18 $\mu$m, indicated by the parameters MEAN_AB09 and MEAN_AB18, although here again, the values appear quite different from those in previous studies. This disparity is most clearly to be seen in Fig. 2, where we plot the AKARI parameter 2xMEAN_AB18 against the radio/optical diameters from Table 1. It is clear that there is little correlation between the two sets of parameters.

To understand what is happening more clearly, we first concentrate on the distribution of results with respect to the horizontal axis. It is apparent that most of the sources have values 2xMEAN_AB18 which fall within a narrow range of radii $0.625 < \log(\theta/\text{arcsec}) < 1.125$. The large majority of these are larger than the FWHM of the PSF. Where we now view the distribution of sources with respect to the vertical axis – that is, with respect to values of diameter taken from radio and optical observations – it becomes apparent that half of sources are smaller than the PSF, and that the distribution is very much broader than that of the AKARI results. This latter trend is also apparent in the lower panel of Fig. 2, where we show normalized histograms of the AKARI and radio/optical results. Whilst the AKARI dimensions are sharply peaked close to $\theta = 7.5$ arcsec, the radio/optical data extends over a regime which is ~60 times larger.

It is therefore clear that whilst the mean infrared sizes of the sources are similar to those in other wavelength regimes – the AKARI dimensions are ~0.86 times as large as those determined through optical and radio observations – the level of agreement for individual sources is much less impressive. It is found that the infrared dimensions are mostly very much larger, or significantly less.

It is no surprise that there is paucity of AKARI sources with dimensions less than the PSF – sources which are smaller than this limit would be unresolved. What is surprising is that so many optically compact sources have substantially larger IR diameters, and that these are close to the size of the PSF. One explanation for this may be that many smaller radio/optical sources correspond to younger PNe, in which the neutral envelopes are larger than the ionized regimes. Where a large fraction of the IR emission derives from these HI shells, then the dimensions of the sources will be correspondingly enhanced. On the other hand, it is possible that sources which are compact are prone to larger errors in their derived diameters, and that many of the MIR dimensions are significantly too large.

Similarly, it is possible that [OIV] $\lambda 25.87$ μm and [NeV] $\lambda 24.316$ μm are dominant in evolved PNe, leading to centralized emission, and smaller mean dimensions. A similar trend has been noted in MIPSGAL 24 μm mapping of PNe (Phillips & Marquez-Lugo 2010; Chu et al. 2009; Ueta 2006; Su et al. 2004, 2007), and this may explain the paucity of sources having larger angular diameters. Here again however, it is possible that weaker, more extended emission falls below the AKARI detection limits, and that this causes a significant reduction in the sizes of the PNe.

It is therefore clear that instrumental errors, and variations in the structures of the PNe are capable of explaining the disparities illustrated in Fig. 2. Further analysis is required before these differences can be understood. Given the uncertainties in the veracity of the AKARI results, however, we shall be using radio/optical estimates of diameter in our investigation of evolutionary trends.

## Comparison of the AKARI and IRAS Photometric results

We have finally undertaken a comparison between the AKARI photometry and corresponding IRAS results. The flux comparisons are provided in Fig. 3, where all of the results have quality factors of 3. For this case, the IRAS PSC was interrogated using a search radius of 18 arcsec, comparable to the positional uncertainty of sources in this catalogue. This resulted in the detection of 573 nebulae with flux quality FQUAL = 3 in at least one of the photometric channels.

The IRAS photometry was undertaken at 12, 25, 60 and 100 $\mu$m; wavelengths which are slightly different from those quoted for the present results. It is not therefore possible to undertake a direct one-to-one comparison between the fluxes, although we are able to compare the closely adjacent channels at 9 and 12 $\mu$m, 18 and 25 $\mu$m, 65 and 60 $\mu$m, and 90 and 100 $\mu$m (Fig. 3).

It is clear that whilst the fluxes are similar at 90 and 100 $\mu$m, the other comparisons show systematic differences in emission. For the most part, the IRAS fluxes are larger than those determined by the AKARI all-sky survey.

One possible explanation for these trends is that there is a calibration problem with the AKARI data, although this seems extremely improbable, and can almost certainly be discounted. The AKARI team has used a broad range of standard stars for the IRC calibration, for instance, including sources in the ecliptic polar regions and Large Magellanic Cloud. They have also used the sources employed in calibrating the Infrared Space Observatory (ISO). There is a non-linear relation between input source fluxes and AKARI output; a function which is fitted using a second degree polynomial fit. When this correction is employed, it is clear that residual errors are small; there is certainly little evidence for the biases which are detected in the present work. Similar calibrations are used for the FIS results, where the AKARI team has used stellar, asteroidal, and planetary flux calibrators. Again, non-linear relations are detected between the source fluxes and measured signals, and these trends are subsequently corrected using logarithmic expressions.

We therefore believe that most of the difference can be attributed to the differing mean wavelengths of observation. Thus, where grain temperatures are of the order 110 K, and the grain emissivity function varies as $\varepsilon \propto \lambda^{-\gamma}$, $\gamma \cong 1$, then one obtains similar flux differences to those noted in the 18μm/24μm panel. A somewhat higher temperature (~ 190 K) is required for the 12μm/9μm results; a value which is rather larger than is normally attributed to PNe dust continua. It is possible that these latter results are dominated by warmer components of emission, however, such as has been described in the analyses of 2MASS results (e.g. Phillips & Ramos-Larios 2005, 2006, 2007). It is also worth noting that emission at these shorter wavelengths occurs at the Wien limits of the dust continua, where flux levels tend to be lower. As a consequence, emission from lines and bands have a greater impact upon the results (see Sect. 3.1 for a fuller description of these components), and may contribute to the discrepancies noted above.

Although this interpretation of the trends is plausible, and undoubtedly accounts for much of the disparity in the results, we note that flux differences are larger for fainter sources, and almost disappear where sources are brighter. One likely explanation for this trend arises from the fact that sources having larger apparent fluxes also have lower intrinsic radii. We determine that $F_\nu(18\mu m) = 0.20(D/pc)^{-1.17}$, with a correlation coefficient $r \cong 0.51$. Given that sources with smaller diameters also tend to have higher grain temperatures (Sect. 3.2), this would imply that nebulae having higher fluxes $F_\nu(18\mu m)$ also have higher temperature continua. This, where it is the case, would lead to a narrowing of the flux differences.

## 3. Evolutionary trends in the AKARI Photometry

Previous analyses of IRAS photometry have suggested that the properties of nebular dust evolve with time. It has been suggested for instance that grain temperatures and dust/gas mass ratios both decrease as nebular radii increase. Similarly, it has been proposed that the masses and sizes of the grains decrease with radius (Pottasch et al. 1984; Lenzuni et al. 1989), although such an evolution has been discounted in the study of Stasinska & Szczerba (1999).

Such broad evolutionary trends are also apparent in the present AKARI spectral energy distributions (SEDs), examples of which are illustrated in Fig. 4. In this case, the sources are ordered according to the channel of peak flux, and all of the fluxes have an FQUAL parameter of 3. Sources peaking at 18 μm are shown in the upper panel, whilst those peaking at 65 and 90 μm are illustrated in the middle and lower graphs. We have also inserted histograms showing the distributions of sources with respect to intrinsic diameter, where the latter refers to the size of the ionized (radio/optical) emission envelope. It is clear that as the continua become cooler (i.e. the wavelength of peak FIR flux increases), so the sizes of the nebulae increase; precisely the trend which has been deduced from previous analyses of IRAS photometry. We shall consider this and further evolutionary trends in the following sections.

## 3.1 Line and Band Contributions to the Nebular Fluxes

FIR spectroscopy of PNe shows that most have continua dominated by emission from nebular grains, with temperatures $T_{GR}$ of order ≈ 50-150 K. There is little doubt that these contribute the larger part of the emission in the present PNe, and we shall be assuming that this is the case for much of the analysis below. There are however several other bands and lines which are capable of affecting the results, and their contributions may be quite considerable. Thus for instance, the 9 μm IRC filter covers the regime of the PAH emission bands, and these are likely to contribute appreciable fluxes where C/O > 1. Evidence for these bands has been noted from ISO and Spitzer spectroscopy (e.g. Ramos-Larios et al. 2010, 2011; Cohen & Barlow 2005), and there is evidence that much of the emission arises from PDRs.

The 18 μm filter, by contrast, includes [OIV] $\lambda$25.87 μm, [NeV] $\lambda$24.316 μm, and the $\lambda\lambda$18.713 & 33.482 μm transitions of [SIII]; the 21 and 30 μm emission features; and broad components associated with crystalline and amorphous silicates. The former (21 and 30 μm) features are found in PNe with C/O > 1, whilst silicates are formed in oxygen rich environments.

Perusal of ISO[1] pipeline spectra for 32 PNe reveals that a variety of other transitions affect the longer wave fluxes as well. These include

---

[1] *Based on observations with ISO, an ESA project with instruments funded by ESA Member States (especially the PI countries: France, Germany, the Netherlands and the United Kingdom) and with*

[OIII] $\lambda 51.816$ μm, [NII] $\lambda 57.317$ μm, and [OI] $\lambda 63.184$ μm (in the 65 and 90 μm channels); [OIII] $\lambda 88.355$ μm (in the 90 μm channel); [NII] $\lambda 121.898$ μm and [OI] $\lambda 145.526$ μm (in the 140 μm and 160 μm channels); and [CII] $\lambda 157.741$ μm (in the 160 μm passband).

Several of these transitions are strong, and may measurably affect the present results. Thus a detailed analysis of sources in which the continuum is clearly visible, and the S/N is reasonably high, suggests that [OIV] $\lambda 25.87$ makes the largest average contribution to the 18 μm band (~20 % of the dust continuum flux), whilst total line enhancement is of the order ~40 %. The line contributions to the other channels are less substantial, varying from ~13 % (at 65 μm), to ~7 % (90 μm), ~0.4 % (140 μm) and 4 % (160 μm, where [CII] is strongest). Given the nature of the colour-colour mapping to be discussed in Sect. 3.2, it is relevant to note that mean 18μm/65μm flux ratios would be increased by ~22 % (i.e. points would be shifted by ~0.09 dex), whilst corrections to the other ratios are significantly less.

Finally, of the 32 PNe investigated in this present analysis, nine appear to have strong 30 μm features. These mostly fall between the IRC 18 μm filter on the one hand, and the FIS 65 and 90 μm filters on the other. Their contributions to the AKARI photometry are therefore likely to be modest.

Some care must therefore be taken in interpreting the present photometry - a caveat that also applies to previous such analyses. Thus for instance, the analysis of Pottasch et al. (1984) is similar to that we shall be undertaking below, although the authors appear not to have taken account of colour corrections. Stasinska & Szczerba (1999) have attempted a more sophisticated analysis, in which photometry is corrected using SED evolutionary modeling. The authors ignore several important oxygen- and carbon-rich bands, however, whilst their assumptions of invariant abundances, identical nebular structures, a single grain emissivity function, and various other approximations, may lead to misleading results for many of their sample PNe.

---

*the participation of ISAS and NASA.*

## 3.2 The AKARI Colour-Colour Planes, and Variation of Grain Temperatures

The distribution of PNe within the AKARI colour planes is illustrated in Fig. 5, where the upper panel shows trends with respect to the 65µm/90µm and 90µm/140µm flux ratios, and the lower panel represents the corresponding 18µm/65µm-90µm/140µm ratios. We also indicate the trends to be expected for smooth dust continua having emissivity exponents $\gamma$ = 0, 1 & 2, and temperatures 30 K < $T_{GR}$ < 270 K. The errors in the fluxes have been discussed in Sect. 2.2, where it is noted that PNe with $F_\nu$(18µm) > 1 Jy have S/Ns ~ 15. Similar values apply for the FIS photometric results as well, where combined relative errors of ~10 %, and absolute errors of ~15% lead to overall photometric uncertainties of ~20 %. We have used these estimates in determining the error bars in Fig. 5.

A further correction relates to the broad widths (4.8-52.4 µm) of the AKARI filters, which necessitates a modification to the fluxes depending upon the nature of the SEDs. We have used the colour corrections of Shirahata et al. (2009) and Lorente et al. (2007) to modify the dust continuum trends in Fig. 5, leading to shifts in the loci by as much as ~ 0.1 dex. This leads to an overlapping of loci in the FIS colour plane (upper panel), and a significant narrowing of the dust regime in the IRC/FIS colour plane (lower panel, where we show examples of the corrected and uncorrected loci).

Where the emission in PNe is dominated by grain continua, then the positions of the sources in Fig. 5 will depend upon $T_{GR}$ and $\gamma$. However, it is clear that the majority of nebulae in the upper panel fall well away from the dust emission trends. Some of this scatter is likely to be due to errors in the results, although it is interesting to note the bias in the ratios to larger values of 90µm/140µm. This is rather larger than would be expected from the ISO analysis described in Sect. 3.1. Thus, whilst the shift could be explained in terms of an increase of ≈20 % in the 90 µm flux, this would not be consistent with the smaller mean enhancements expected for lines (~ 7%; see Sect. 3.1). It would also fail to account for the higher levels of line contamination noted for the 65 µm band, which tend to push the PNe to higher positions within this plane.

Similar problems occur for the 18-140 μm trends in the panel below, although in this case, many of the sources lie within the limits of the dust emission loci. We here assume that emission for these sources is dominated by dust continua, and use the positions of the nebulae to determine values for $\gamma$ and $T_{GR}$. The enhancement in 18μm/65μm ratios due to line emission (see Sect. 3.1) may lead to increases in $T_{GR}$ by ~10-20 K – although see also our comments below concerning the partial detection of 18 μm fluxes.

The corresponding variation of $T_{GR}$ with source diameter is illustrated in Fig. 6 (upper panel), where filled circles correspond to sources having $\theta$ < 7 arcsec, and open circles are for nebulae which are larger. It is clear that dust temperatures are significantly higher where the sources have smaller physical dimensions, with $T_{GR}$ approaching ~ 180 K where D is less than ~ 0.08 pc. By contrast, PNe having larger physical dimensions tend to have smaller values of $T_{GR}$, and fall within a narrower range of values $T_{GR}$ ~ 85-120 K.

Stasinska & Szczerba (1999) have represented the variation of dust temperatures against Hβ surface brightnesses S(Hβ). The value of $T_{GR}$ is determined using IRAS 25 & 60 μm fluxes, and by assuming that the emissivity exponents $\gamma$ are uniformly equal to unity. This leads to qualitatively similar trends to those illustrated in the upper panel of Fig. 6, although the nature of their procedure permits them to measure many more PNe. The authors find that temperatures are ~ 60-100 K where $\log(S(H\beta)/\text{erg cm}^{-2} \text{ s}^{-1} \text{ sr}^{-1})$ < -2, and increase rapidly to higher S(Hβ).

A comparable analysis can be achieved using AKARI photometry at 18 and 65 μm, and where one adopts similar simplifying assumptions. The results are illustrated in Fig. 6 (lower panel). It is again clear that there is a marked increase in temperatures for diameters < 0.07 pc; that temperatures are of order ~ 70-110 K for larger nebulae; and that there is a significant overlap of values for angularly small and larger PNe. There is little evidence for the linear fall-off of temperatures noted by Pottasch et al. (1984).

Finally, we note that the FIS beams are relatively large, varying from 37 to 61 arcsec. Nevertheless, a small proportion of nebulae have diameters which are larger than these values. Given that the fluxes for these sources may be being only partially detected, we have excluded such nebulae from the present analyses. By contrast, the 18 μm beam sizes are very much smaller, and this may lead to a quite considerable short-fall in certain of the fluxes. This, where it is the case, would counteract the effects of line and band "contamination" noted in Sect. 3.1. We note however that PNe having warmer dust continua tend to be smaller and unresolved (see Fig. 6), and their fluxes and temperatures are therefore relatively secure. Similarly, sources having lower values of $T_{GR}$ tend to be angularly large, and their shorter wave fluxes may be significantly too low. However, estimates of $T_{GR}$ for such (lower) temperatures are insensitive to such uncertainties, and errors in the trends are likely to be modest. This is consistent with the variations noted in Fig. 6, where we see little inconsistency between the estimated temperatures for large and smaller PNe (Fig. 6).

## 3.3 Variation of Nebular Fluxes and Luminosities with Nebular Evolution

The emission characteristics of dust in PNe are expected to vary as the nebular envelopes evolve. Thus, expansion of the envelopes would be expected to lead to a reduction in stellar radiative flux densities, and decreases in the levels of direct grain heating. Similarly, the roles of changing stellar temperatures; absorption of Lyα radiation; and the transfer of grains from neutral to ionized regimes, will all be important in determining the energy budgets of the grains and, by extension, the temperature of the dust.

A trivial consequence of such evolution is noted in Fig. 7, where we show the variation of intrinsic 18 and 90 μm fluxes as a function of source diameter. It would seem that both of the trends indicate a decrease of flux with D; a variation which may arise through a cooling of the dust continua, such as is noted in Fig. 4.

An even more interesting trend is illustrated in Fig. 8, where we have estimated total fluxes $F_{TOT}(IR)$ for sources within the dust emission regime of Fig. 5 (lower panel). For this case, $F_{TOT}(IR)$ is determined using the values of $\gamma$ and $T_{GR}$ obtained from Fig. 5; observed 90 μm fluxes; and the distances quoted in Table 1. It is again clear that luminosities decline with increasing nebular dimensions, and have values approaching ≈ 1.5 $10^4$ $L_\odot$ where < 0.05 pc – perhaps even larger (~ 4 $10^4$ $L_\odot$) for Hb 12. Since this brightness is typical of the luminosities of PNe central stars, this suggests that a large fraction of the radiation is being absorbed by dust – a situation which would be consistent with the high levels of reddening observed in proto-planetary nebulae (PPNe).

It is finally interesting to adapt the analysis of Ly$\alpha$ heating of Pottasch et al. (1984) to the present AKARI results. The results of such a procedure are illustrated in Fig. 9, where we show the relation between observed integrated IR and 1.4 GHz emission, and between intrinsic (distance corrected) IR and 1.4 GHz results (lower panel). The trends expected for Ly$\alpha$ heating are indicated by the dashed diagonal lines, determined for high and low density nebular regimes. It would appear that the infrared excess (IRE), corresponding to the ratio of the observed flux to Ly$\alpha$ heating, is of order >2 for all of this present sample PNe. It is therefore clear that further heating of the grains is required over and above that expected by Ly$\alpha$ photons.

One would anticipate that values of the IRE would be largest where PNe are more compact, Ly$\alpha$ fluxes are smaller, and direct heating by the stellar radiation field is greatest. Subsequent nebular expansion would lead to further ionization of the shell; increased fluxes of Ly$\alpha$ photons; and the dilution of the stellar radiation field. Under these circumstances, the IRE would be expected to decline rather precipitously, and achieve minimum values of the order of unity.

To investigate whether some such evolution is actually occurring, we have determined FIR luminosities for a broad range of PNe. This involved using the 18 & 65 $\mu$m fluxes to determine $T_{GR}$; assuming a grain emissivity exponent $\gamma$ = 1; and 1.4 GHz fluxes to evaluate Ly$\alpha$ heating. We assume Ly$\alpha$ photon production rates relevant for lower density PNe, whilst the distances and sizes of the sources are taken from Tables 1 & 2. Under these circumstances, appropriate integration of the emission curves show that the luminosity of the source can be represented through

$$L = 3.11 \, 10^{13} \left[ \frac{F_\nu(90\mu m)}{Wm^{-2}Hz} \right] \left[ \frac{d}{kpc} \right]^2 \left[ \sum_{i=0}^{4} a_i T_{GR}^i \right]^{-1} L_{SUN} \quad ........(6)$$

where d is source distance, the polynomial terms [$a_0$, …,$a_4$] are given by [7.162 $10^{10}$, -1.834 $10^9$, 1.868 $10^7$, -8.742 $10^4$, 1.564 $10^2$], and the expression is valid for temperatures 70 < $T_{GR}$/K < 160. Similar expressions may be derived for differing values of $\gamma$.

Note that both here, and in the analysis of the results in Fig. 5 (see above), we have used 90 μm fluxes to scale the results. This is advantageous for a variety of reasons. Not only are these results taken with the most sensitive of the FIS detectors – leading to larger numbers of PNe detections than in any of the other FIS channels – but this wavelength lies close to the continuum peak for most of the PNe SEDs. We have also assessed that levels of line contamination are likely to be modest (Sect. 3.1), although there is evidence that contaminants may be larger than is supposed (see Sect. 3.2 and Fig. 5). It is unlikely, however, that luminosities are very greatly in error, or that mean biases are greater than ~0.1 dex.

The results of this analysis are illustrated in Fig. 10, where the IRE is represented against nebular diameter. It is plain from this that some evolution does indeed appear to be occurring, with values of IRE approaching ~50 for diameters < 0.07 pc. Although the qualitative trend appears consistent with the analysis above, it is interesting to note that the IRE is still relatively large (≈ 5) even where nebulae are evolved – and that the minimum values are larger than found by Pottasch et al. (1984).

**3.4 Radio Infrared Flux Correlation**

It has been noted that the ratio of infrared to radio fluxes in extragalactic systems is reasonably constant, and invariable with redshift z. Appleton et al. (2004) have noted that the parameter $q_{70}$ = log($S_\nu$(70μm)/$S_\nu$(20cm)) is of order ~ 2, whilst $q_{24}$ = log($S_\nu$(24μm)/$S_\nu$(20cm)) is ~ 1. Cohen & Green (2001) have made a similar analysis for 21 Acker et al. (2002) PNe, using 8.3 μm fluxes taken from the Midcourse Science Experiment (MSX; Price et al. 2001). They derived MIR/radio ratios in the region of 12. However, a later analysis for ten MASH PNe (Parker et al. 2006; Miszalski et al. 2008) yielded a ratio of the order of ~ 5 (Cohen et al. 2007). This difference between the less evolved Acker PNe, and more evolved MASH nebulae was attributed to an evolution in the properties of the sources – perhaps a reduction in PAH emission as fluorescent excitation decreased. This mechanism is only applicable however where C/O is greater than or close to unity.

A further investigation of this issue has also been undertaken by Zhang & Kwok (2009), using 8 and 24 μm photometry deriving from the GLIMPSE and MIPSGAL surveys. In this case, the authors note a linear relation between MIR and 1.4 GHz fluxes; a trend they interpret as indicating a constancy in the ratios $q_{24}$ and $q_8$. It is also suggested that the scatter in their figure is attributable to optical thickness at 1.4 GHz – an hypothesis which we will suggest, below, is unlikely to be the case.

We have investigated this question using the present AKARI photometric results, and 1.4 GHz measures deriving from Condon et al. (1998). It should be emphasized that whilst most PNe are likely to be optically thin at 5 GHz, a small proportion may have higher opacities. Thus, we have made a comparison between 5 and 1.4 GHz fluxes in Fig. 11 based upon the data summarized in Tables 1 & 2. It is clear that the vast majority of sources follow the trend for optically thin emission (lower dashed line), although ~20% fall between the optically thick and thin regimes (the former indicated by the upper line). It is therefore clear that whilst certain 5 GHz fluxes derive from partially or fully opaque regimes, the vast majority of 1.4 GHz results are optically thin – and yield fluxes proportional to the total masses of ionised gas.

Phillips (2007) has noted that higher 5GHz/1.4GHz ratios are confined to more compact sources, where density and emission measures are larger. This is also noted for the PNe investigated here (Fig. 11, lower panel), where it is clear that higher 5 GHz opacities arise were diameters are < 0.07 pc; a range similar to that determined for higher values of $T_{GR}$ (Sect. 3.2).

The variation of infrared emission with 1.4 GHz flux is illustrated in Fig. 12, where it is clear that there is a linear relation between the two sets of data. Such trends are strongly affected by a variety of biases, however, and these greatly modify the perceived gradients. In the first place, distance-squared spreading of the fluxes would tend to lead to a one-to-one relation. It is also apparent that source numbers are truncated where $F_\nu(1.4 \text{GHz}) \leq 3$ Jy, corresponding to the sensitivity limit of the Condon et al. (1998) survey.

Even greater biases arise from the limiting sensitivities of the differing AKARI channels, however, and we have indicated these using horizontal dashed lines. We have included values corresponding to 50% source detection rates; a so-called "normal mode" detection limit quoted in the FIS Bright Source Catalogue (Yamamura et al. 2010); and the $5\sigma$ detection limit quoted by Kawada et al. (2007). Most of these limits are mutually consistent, and accord with the fall-offs in the numbers of PNe. The result, for the 9 and 65 $\mu$m results, is for an apparent flattening of the trends. It is apparent however that the 90 and 18 $\mu$m samples are significantly more complete, and trends are close to those expected for unbiased samples of PNe. We shall therefore be using these trends to assess variations in $q_{18}$ and $q_{90}$.

We should finally add that detection limits at 140 and 160 $\mu$m are even higher, whilst PNe fluxes tend to be lower. This leads to extraordinarily strong truncations in the numbers of PNe, and we have excluded these results from the present analysis.

Distance corrected measures of the fluxes are illustrated in Fig. 13. We note that the linear trends in Fig. 12 have now largely disappeared, although the scatter between the nebulae is as large as before. There is still evidence for linearity between the 90 $\mu$m and 1.4 GHz results, although this is consistent with an approximate constancy of IR/radio flux ratios.

It would appear, apart from this, that there is little evidence for evolution or correlation – and certainly nothing like the invariance in q noted for extragalactic systems. The scatter in these planes is huge, and indicates variations of an order of magnitude or more. Most of this scatter likely to be intrinsic to the sources. The combined errors resulting from photometric errors (~20 %) and uncertainties in the distance (~30 %) are likely to be ~60 % for the 90 $\mu$m results, for

instance – and probably somewhat less than that for the 18 μm trends. This is a factor ~15 less than is observed in Fig. 13.

Whilst it is therefore possible to obtain reasonable values of q for large ensembles of PNe, as has previously been undertaken by the authors cited above, there is little evidence that this parameter is invariant between PNe, or indicates any close relation between the mechanisms responsible for radio and IR emission. Whilst it is undoubtedly the case that UV radiation is stimulating the radio and infrared emission, there is little evidence for the levels of correlation noted in extragalactic systems.

## 4. Conclusions

The AKARI all-sky survey has resulted in photometry for $8.77\ 10^5$ sources in the MIR, using the IRC at 9 and 18 μm, and $4.27\ 10^5$ sources in the FIR using the FIS at 65-160 μm. We have interrogated this database to identify the counterparts of Galactic PNe, and undertaken an analysis of the resulting photometry.

Approximately 80 % of the Kerber et al. (2003) sources were detected to quality levels FQUAL=3, of which ~half were observed in both the FIR and MIR regimes. A comparison with photometry deriving from the IRAS survey shows that there are systematic differences between the two sets of flux, although much of the variation is attributable to differences in the wavelengths employed. We also note that the mean AKARI dimensions of the sources are similar to those at radio and optical wavelengths, although the distributions of diameters appear to be radically different. This may arise because of centrally concentrated fluxes due to [OIV] and [NeV], and/or high levels of dust emission in extended PDRs. It is also possible that limiting sensitivities and errors are biasing the distribution of AKARI results.

Analysis of the nebulae in the colour-colour planes suggests that many of the SEDs are consistent with dust continuum emission, although emission bands and ionic transitions also contribute to the results. We find similar evolutionary trends in grain temperatures to those measured in previous studies, with values approaching ~160 K where nebulae are young (D < 0.08 pc), and $T_{GR}$ ~ 60-120 K where nebulae are more evolved.

Fluxes in the individual channels also show an evolution with nebular diameter, much of which may be attributed to cooling of the grains. However, we also note that total FIR emission declines with increasing nebular diameter, from a peak of ≈1.5 $10^4$ $L_\odot$ where D < 0.05 pc, to values ~600 $L_\odot$ in more evolved PNe. The high luminosities of less evolved sources suggests that dust opacities are appreciable. Although up to ~ 0.5 of the grain heating may arise from absorption of Ly$\alpha$ photons, it seems likely that most of the energy budget depends upon absorption of the stellar radiation field. There is evidence for an evolution in the IRE with expansion of the shells, consistent with the variations to be expected in stellar and Ly$\alpha$ heating.

Finally, it has been suggested that PNe may show a consistent IR/radio flux ratio similar to that determined in extragalactic systems, although with some possible evidence of evolution at ~ 8 $\mu$m. Our present results fail to confirm this suggestion. It is plain that whatever the global values of the IR/radio ratio may be, they are highly variable between nebulae, and show little evidence for consistency.

**Acknowledgements**

TABLE 1

Galactic Planetary Nebulae Identified in the AKARI IRC Catalogue

| KN | NAME | R.A. (2000.0) H:M:S | DEC (2000.0) D:M:S | F(9μm) Jy | F(18μm) Jy | θ arcsec | MN_18 arcsec | F(1.4GHz) mJy | F(5GHz) 5GHz | DIST. Kpc | DIAM. pc |
|---|---|---|---|---|---|---|---|---|---|---|---|
| 4 | 010.2+02.7 | 17 58 14.415 | -18 41 26.15 | --- | 0.844 | --- | 3.53 | 18.3 | --- | --- | --- |
| 5 | 021.9+02.7 | 18 21 21.115 | -08 31 42.30 | 0.285 | 4.516 | --- | 4.12 | 36.2 | --- | --- | --- |
| 6 | 050.6+19.7 | 18 08 20.083 | +24 10 43.26 | 2.483 | 13.329 | --- | 4.80 | --- | --- | --- | --- |
| 7 | 099.3-01.9 | 22 04 12.301 | +53 04 01.36 | 0.460 | 11.917 | --- | 4.74 | --- | --- | --- | --- |
| 9 | 19W32 | 17 39 02.870 | -28 56 35.10 | --- | 6.588 | 24 | 4.46 | 13.5 | 21.3 | --- | --- |
| 10 | 341.4-09.0 | 17 35 02.502 | -49 26 26.38 | 5.541 | 77.831 | --- | 6.67 | --- | --- | --- | --- |
| 27 | A 30 | 08 46 53.514 | +17 52 45.47 | 0.548 | 18.039 | 4.47 | 6.04 | --- | 2.3 | 2.55 | 0.055 |
| 31 | A 35 | 12 53 32.792 | -22 52 22.57 | 0.114 | --- | 772 | --- | --- | 255 | 0.74 | 2.770 |
| 42 | A 47 | 18 35 22.569 | -00 13 49.68 | --- | 0.143 | 17 | 2.13 | 7.3 | --- | --- | --- |
| 43 | A 48 | 18 42 46.908 | -03 13 17.30 | --- | 3.342 | 40 | 6.97 | 159.4 | --- | --- | --- |
| 48 | A 53 | 19 06 45.910 | +06 23 52.47 | --- | 1.043 | 31 | 5.16 | 33.6 | 85.7 | 1.68 | 0.252 |
| 53 | A 58 | 19 18 20.476 | +01 46 59.62 | 2.846 | 15.290 | --- | 4.95 | --- | --- | --- | --- |
| 73 | A 78 | 21 35 29.410 | +31 41 45.39 | 0.986 | 30.500 | 107 | 7.23 | 3.7 | 4.9 | 2.22 | 1.152 |
| 83 | Al 2-E | 17 30 14.398 | -27 30 19.41 | --- | 0.873 | --- | 3.44 | 14.7 | --- | --- | --- |
| 85 | Al 2-G | 17 32 22.670 | -28 14 27.32 | --- | 1.013 | --- | 3.39 | --- | --- | --- | --- |
| 90 | Al 2-O | 17 51 44.663 | -31 36 00.18 | --- | 0.775 | 5.7 | 3.18 | 38 | 33 | --- | --- |
| 92 | Al 2-R | 17 53 36.431 | -31 25 25.19 | --- | 0.306 | --- | 3.17 | --- | --- | --- | --- |
| 94 | Ap 1-12 | 18 11 35.088 | -28 22 37.02 | 0.207 | 3.286 | 12 | 4.03 | 6.6 | 18.7 | 5.34 | 0.311 |
| 96 | BD+30 3639 | 19 34 45.235 | +30 30 58.94 | 47.546 | 148.573 | 7.6 | 7.65 | 235.2 | 630 | --- | --- |
| 102 | Bl 2-1 | 22 20 16.638 | +58 14 16.59 | 1.554 | 4.116 | 1.6 | 3.99 | 21.6 | 54 | 6.94 | 0.054 |
| 103 | Bl 3-10 | 17 55 20.537 | -29 57 36.13 | --- | 0.357 | --- | 3.01 | 4.7 | --- | --- | --- |
| 104 | Bl 3-13 | 17 56 02.759 | -29 11 16.21 | --- | 0.809 | --- | 3.31 | 4.7 | 10 | --- | --- |
| 105 | Bl 3-15 | 17 52 35.947 | -29 06 38.97 | --- | 0.372 | --- | 2.67 | --- | --- | --- | --- |
| 106 | Bl B | 17 36 59.836 | -29 40 08.93 | --- | 1.641 | --- | 3.65 | 8.9 | --- | --- | --- |
| 108 | Bl M | 17 53 47.180 | -28 27 16.88 | --- | 1.249 | 4.5 | 3.61 | 9.3 | 17 | 7.03 | 0.153 |
| 109 | Bl O | 17 53 49.717 | -28 59 11.26 | 0.149 | 1.534 | --- | 3.70 | 3.6 | --- | --- | --- |
| 110 | Bl Q | 17 54 34.866 | -28 12 43.04 | 0.195 | 1.650 | --- | 3.76 | 18.2 | --- | --- | --- |
| 112 | CRBB 1 | 20 19 28.702 | -41 31 27.46 | --- | 0.647 | --- | 3.37 | --- | --- | --- | --- |
| 113 | CRL 618 | 04 42 53.672 | +36 06 53.17 | 189.266 | --- | 12 | --- | 4 | 67 | 2.72 | 0.158 |
| 119 | Cn 1-1 | 15 51 15.936 | -48 44 58.67 | 13.024 | 27.810 | --- | 5.27 | --- | 10 | --- | --- |
| 120 | Cn 1-3 | 17 26 12.392 | -44 11 24.52 | 0.525 | --- | --- | --- | --- | --- | --- | --- |
| 122 | Cn 1-5 | 18 29 11.653 | -31 29 59.21 | --- | 4.672 | 7 | 4.16 | 51 | 44 | 4.43 | 0.150 |
| 123 | Cn 2-1 | 17 54 32.978 | -34 22 20.84 | --- | 3.665 | 2.2 | 3.84 | 25.2 | 49 | 6.17 | 0.066 |
| 124 | Cn 3-1 | 18 17 34.133 | +10 09 03.99 | 0.239 | 6.365 | 4.5 | 4.33 | 59.5 | 75 | 4 | 0.087 |
| 126 | DdDm 1 | 16 40 18.156 | +38 42 19.92 | --- | 0.233 | 1 | 2.69 | 5.7 | 6 | 15.85 | 0.077 |
| 150 | Fg 1 | 11 28 36.205 | -52 56 04.01 | 0.416 | 0.920 | 16 | 5.32 | --- | 55 | 2.02 | 0.157 |
| 151 | Fg 2 | 17 39 19.847 | -44 09 37.14 | 0.227 | 1.226 | --- | 3.66 | --- | --- | --- | --- |
| 152 | Fg 3 | 18 00 11.819 | -38 49 52.73 | 1.441 | 16.842 | 2 | 5.10 | 42 | 107 | --- | --- |
| 154 | G001.1+00.8 | 17 45 11.085 | -27 32 38.00 | 1.182 | 4.009 | --- | 4.48 | 73 | --- | --- | --- |
| 155 | G001.5-00.7 | 17 52 08.717 | -28 02 16.56 | 0.394 | 3.804 | --- | 4.32 | 17.2 | --- | --- | --- |
| 158 | G002.5+05.1 | 17 32 12.788 | -24 04 59.56 | 1.138 | 9.218 | --- | 4.56 | --- | --- | --- | --- |
| 163 | G009.3+05.7 | 17 45 14.155 | -17 56 46.57 | --- | 13.562 | --- | 4.75 | --- | --- | --- | --- |
| 164 | G010.2+07.5 | 17 41 00.035 | -16 18 12.50 | --- | 3.076 | --- | 3.83 | --- | --- | --- | --- |

| KN | NAME | R.A. (2000.0) H:M:S | DEC (2000.0) D:M:S | F(9μm) Jy | F(18μm) Jy | θ arcsec | MN_18 arcsec | F(1.4GHz) mJy | F(5GHz) 5GHz | DIST. Kpc | DIAM. pc |
|---|---|---|---|---|---|---|---|---|---|---|---|
| 165 | G011.1-07.9 | 18 40 19.914 | -22 54 29.23 | --- | 2.813 | --- | 3.75 | --- | --- | --- | --- |
| 168 | G013.1-13.2 | 19 04 43.548 | -23 26 08.82 | 7.780 | --- | --- | --- | --- | --- | --- | --- |
| 170 | G014.2+03.8 | 18 02 38.257 | -14 42 03.22 | 0.421 | 7.042 | --- | 4.29 | --- | --- | --- | --- |
| 172 | G014.5-36.1 | 20 42 45.967 | +30 04 06.51 | 3.740 | 37.172 | --- | 5.88 | --- | --- | --- | --- |
| 175 | G017.0+11.1 | 17 42 14.430 | -08 43 18.61 | --- | 2.126 | --- | 4.32 | 23.4 | --- | --- | --- |
| 179 | G019.2-04.4 | 18 42 24.770 | -14 15 11.81 | --- | 0.495 | --- | 3.76 | 7.3 | --- | --- | --- |
| 180 | G020.4+00.6 | 18 25 58.081 | -10 45 28.49 | 0.447 | 5.891 | --- | 4.20 | 48.7 | --- | --- | --- |
| 182 | G032.0-01.7 | 18 56 15.684 | -01 34 00.31 | --- | 0.485 | --- | 2.89 | 10.7 | --- | --- | --- |
| 184 | G033.7-02.0 | 19 00 16.133 | -00 14 32.59 | --- | 0.347 | --- | 2.89 | 3.5 | --- | --- | --- |
| 185 | G034.5-11.7 | 19 36 17.531 | -03 53 25.26 | 0.151 | 3.996 | --- | 3.98 | 8.8 | --- | --- | --- |
| 186 | G038.4+01.8 | 18 54 54.139 | +05 48 11.29 | 0.211 | 2.773 | --- | 3.92 | 20.5 | --- | --- | --- |
| 188 | G039.1-02.2 | 19 10 53.284 | +04 27 28.50 | --- | 0.439 | --- | 3.26 | 11 | --- | --- | --- |
| 191 | G044.1+01.5 | 19 06 32.146 | +10 43 23.93 | --- | 0.815 | --- | 3.24 | 19.4 | --- | --- | --- |
| 193 | G047.2+01.7 | 19 11 35.828 | +13 31 11.58 | 0.780 | 3.869 | --- | 4.07 | --- | --- | --- | --- |
| 195 | G051.5+00.2 | 19 25 40.679 | +16 33 05.56 | --- | 3.863 | --- | 4.38 | 77 | --- | --- | --- |
| 200 | G059.4-00.7 | 19 45 34.244 | +22 58 34.41 | --- | 0.958 | --- | 3.35 | 26.5 | --- | --- | --- |
| 202 | G069.2+01.2 | 20 00 41.998 | +32 27 41.23 | --- | 0.516 | --- | 3.60 | 17 | --- | --- | --- |
| 204 | G076.6-05.7 | 20 48 16.627 | +34 27 24.36 | 0.110 | 6.210 | --- | 4.27 | --- | --- | --- | --- |
| 207 | G095.0-05.5 | 21 56 32.945 | +47 36 13.40 | 0.369 | 1.764 | --- | 3.67 | --- | --- | --- | --- |
| 219 | G199.4+14.3 | 07 20 01.614 | +18 17 26.47 | 0.147 | 0.689 | --- | 3.09 | --- | --- | --- | --- |
| 221 | G222.8-04.2 | 06 54 13.406 | -10 45 38.27 | 0.204 | 1.727 | --- | 4.00 | 2.7 | --- | --- | --- |
| 226 | G255.3-03.6 | 08 06 28.339 | -38 53 23.93 | 0.091 | 1.275 | --- | 3.56 | 2.8 | --- | --- | --- |
| 228 | G260.1+00.2 | 08 37 24.599 | -40 38 07.46 | 0.318 | 2.612 | --- | 4.41 | --- | --- | --- | --- |
| 229 | G270.1-02.9 | 08 59 02.971 | -50 23 39.68 | 0.262 | 1.373 | --- | 3.64 | --- | --- | --- | --- |
| 231 | G277.1-01.5 | 09 37 51.861 | -54 27 08.59 | 0.724 | 3.141 | --- | 3.94 | --- | --- | --- | --- |
| 232 | G279.1-00.4 | 09 53 27.058 | -54 52 39.60 | 0.059 | 0.866 | --- | 3.46 | --- | --- | --- | --- |
| 234 | G282.6-00.4 | 10 13 19.673 | -56 55 32.24 | 0.313 | 3.065 | --- | 3.99 | --- | --- | --- | --- |
| 235 | G285.1-02.7 | 10 19 32.466 | -60 13 29.41 | 3.022 | 17.044 | --- | 5.16 | --- | --- | --- | --- |
| 238 | G301.1-01.4 | 12 34 35.968 | -64 18 16.99 | 2.012 | 17.086 | --- | 5.08 | --- | --- | --- | --- |
| 241 | G308.5-03.5 | 13 46 25.712 | -65 46 24.26 | 0.176 | 2.768 | --- | 3.93 | --- | --- | --- | --- |
| 243 | G313.1+02.1 | 14 16 51.803 | -58 53 10.01 | 0.136 | 1.979 | --- | 3.72 | --- | --- | --- | --- |
| 244 | G313.3+01.1 | 14 15 53.284 | -60 01 37.31 | 0.948 | 9.663 | --- | 4.49 | --- | --- | --- | --- |
| 245 | G314.4+02.2 | 14 21 19.909 | -58 38 22.10 | 0.543 | 5.337 | --- | 4.08 | --- | --- | --- | --- |
| 246 | G316.2+00.8 | 14 38 19.980 | -59 11 46.12 | 1.053 | 10.213 | --- | 4.51 | --- | --- | --- | --- |
| 251 | G328.4-02.8 | 16 09 20.143 | -55 36 09.84 | 0.301 | 2.054 | --- | 3.85 | --- | --- | --- | --- |
| 253 | G331.3-12.1 | 17 16 21.107 | -59 29 23.35 | 0.089 | 2.566 | --- | 3.90 | --- | --- | --- | --- |
| 254 | G336.1+04.1 | 16 15 02.803 | -45 11 54.10 | 0.314 | 2.130 | --- | 3.98 | --- | --- | --- | --- |
| 255 | G340.3-03.2 | 17 03 10.027 | -47 00 27.68 | 18.997 | 158.829 | --- | 7.92 | --- | --- | --- | --- |
| 259 | G342.2-00.3 | 16 56 33.989 | -43 46 13.55 | 0.793 | 9.718 | --- | 4.45 | --- | --- | --- | --- |
| 261 | G343.7-09.6 | 17 45 33.415 | -47 43 50.31 | 2.362 | 2.236 | --- | 3.78 | --- | --- | --- | --- |
| 263 | G344.9-01.9 | 17 12 22.022 | -42 30 41.24 | 0.487 | 9.329 | --- | 4.58 | --- | --- | --- | --- |
| 264 | G347.9-06.0 | 17 40 03.329 | -42 24 05.58 | --- | 0.202 | --- | 3.04 | --- | --- | --- | --- |
| 267 | G351.5-06.5 | 17 52 09.386 | -39 32 14.52 | --- | 0.229 | --- | 2.45 | --- | --- | --- | --- |
| 273 | G353.4-02.4 | 17 39 17.147 | -35 46 58.88 | 0.248 | --- | --- | --- | --- | --- | --- | --- |

| KN | NAME | R.A. (2000.0) H:M:S | DEC (2000.0) D:M:S | F(9μm) Jy | F(18μm) Jy | θ arcsec | MN_18 arcsec | F(1.4GHz) mJy | F(5GHz) 5GHz | DIST. Kpc | DIAM. pc |
|---|---|---|---|---|---|---|---|---|---|---|---|
| 276 | G353.8-01.2 | 17 35 27.372 | -34 47 42.22 | --- | 0.474 | --- | 2.83 | 8.2 | --- | --- | --- |
| 286 | G355.4+02.3 | 17 25 03.470 | -31 28 38.50 | --- | 0.533 | --- | 3.20 | 10.4 | --- | --- | --- |
| 287 | G355.4-01.4 | 17 40 00.347 | -33 35 32.29 | --- | 0.340 | --- | 3.00 | 8.7 | --- | --- | --- |
| 294 | G356.8-03.0 | 17 50 10.734 | -33 14 17.97 | --- | 0.436 | --- | 3.03 | 4.2 | --- | --- | --- |
| 295 | G357.1+01.2 | 17 33 50.871 | -30 42 36.75 | 0.206 | 0.967 | --- | 3.38 | 11.3 | --- | --- | --- |
| 300 | G358.4+01.6 | 17 35 22.745 | -29 22 17.40 | --- | 0.452 | --- | 2.91 | 5.5 | --- | --- | --- |
| 301 | G358.6-02.4 | 17 52 00.342 | -31 17 49.92 | --- | 0.665 | --- | 3.37 | 12 | --- | --- | --- |
| 308 | G359.8+08.9 | 17 11 38.912 | -24 07 32.87 | 0.755 | 6.889 | --- | 4.13 | --- | --- | --- | --- |
| 310 | G000.5+01.9 | 17 39 31.219 | -27 27 46.77 | --- | 0.530 | --- | 3.11 | 6.8 | --- | --- | --- |
| 316 | G001.5+03.6 | 17 35 22.074 | -25 42 46.48 | --- | 0.934 | --- | 3.45 | 12.2 | --- | --- | --- |
| 318 | G001.6+01.5 | 17 43 16.950 | -26 44 17.54 | --- | 0.539 | --- | 3.14 | 9.8 | --- | --- | --- |
| 326 | G003.6+04.9 | 17 35 31.198 | -23 11 47.62 | --- | 0.213 | --- | 2.57 | 4.6 | --- | --- | --- |
| 333 | G070.9+02.2 | 20 00 52.927 | +34 28 22.18 | 0.153 | 3.084 | --- | 4.04 | --- | --- | --- | --- |
| 335 | G000.1-01.2 | 17 50 24.297 | -29 25 18.70 | --- | 2.050 | --- | 3.64 | --- | --- | --- | --- |
| 339 | G000.2-01.4 | 17 51 53.546 | -29 30 53.18 | 3.230 | 3.541 | --- | 3.75 | --- | --- | --- | --- |
| 340 | G000.4+01.1 | 17 42 25.108 | -27 55 35.95 | --- | 1.157 | --- | 4.13 | 23.6 | --- | --- | --- |
| 341 | G000.6-01.0 | 17 51 11.504 | -28 56 26.16 | 0.493 | --- | --- | --- | --- | --- | --- | --- |
| 345 | G000.0+02.0 | 17 37 42.929 | -27 49 09.55 | --- | 1.557 | --- | 3.61 | 6.2 | | --- | --- |
| 359 | H 1-11 | 17 21 17.693 | -22 18 35.32 | --- | 1.044 | 6 | 3.59 | 20.7 | 13 | 7.27 | 0.211 |
| 360 | H 1-12 | 17 26 24.239 | -35 01 41.26 | --- | 30.050 | <25 | 6.40 | 372 | 719 | 1.4 | --- |
| 361 | H 1-13 | 17 28 27.537 | -35 07 31.63 | --- | 27.492 | 14 | 6.16 | 474 | 620 | 1.39 | 0.094 |
| 363 | H 1-15 | 17 28 37.616 | -24 51 06.60 | --- | 1.414 | 4.3 | 3.88 | 13.6 | 13 | 7.9 | 0.165 |
| 364 | H 1-16 | 17 29 23.358 | -26 26 04.39 | 0.272 | --- | 2 | --- | 34.7 | 58 | --- | --- |
| 366 | H 1-18 | 17 29 42.759 | -29 32 50.21 | 0.336 | 2.667 | 1.5 | 3.81 | 17.6 | 33 | 7.37 | 0.054 |
| 368 | H 1-2 | 16 48 54.091 | -35 47 09.01 | 2.425 | 15.806 | 1 | 4.94 | 14.3 | 62 | --- | --- |
| 369 | H 1-20 | 17 30 43.799 | -28 04 06.68 | --- | 1.924 | 3.3 | 3.72 | 26.9 | 44 | 5.12 | 0.082 |
| 371 | H 1-22 | 17 32 22.138 | -37 57 23.81 | --- | 1.517 | <25 | 3.60 | 18.4 | --- | --- | --- |
| 373 | H 1-24 | 17 33 37.565 | -21 46 24.78 | --- | 1.172 | 5 | 3.67 | 5.1 | 15 | 7.32 | 0.177 |
| 374 | H 1-26 | 17 36 29.743 | -39 21 56.95 | 0.385 | 2.345 | <25 | 5.18 | 67.1 | --- | --- | --- |
| 375 | H 1-27 | 17 40 17.926 | -22 19 17.64 | --- | 2.859 | 5.2 | 3.88 | --- | 18 | 10.9 | 0.275 |
| 377 | H 1-29 | 17 44 13.819 | -34 17 33.05 | 0.174 | --- | <10 | --- | 10.6 | --- | --- | --- |
| 379 | H 1-30 | 17 45 06.782 | -38 08 49.48 | --- | 0.556 | <5 | 2.97 | 7 | 81 | 5.21 | --- |
| 381 | H 1-32 | 17 46 06.291 | -34 03 46.25 | 0.266 | --- | 1.1 | --- | 3.4 | 37 | 9.09 | 0.048 |
| 382 | H 1-33 | 17 47 49.387 | -34 08 05.44 | --- | 0.958 | <10 | 3.50 | 15.8 | 12 | --- | --- |
| 383 | H 1-34 | 17 48 07.572 | -22 46 47.33 | 0.352 | --- | 2 | --- | --- | 13 | --- | --- |
| 384 | H 1-35 | 17 49 13.940 | -34 22 52.96 | 1.709 | 19.148 | 2 | 4.86 | 18.4 | 72 | 6.03 | 0.058 |
| 385 | H 1-36 | 17 49 48.172 | -37 01 29.47 | --- | 20.615 | 0.8 | 5.25 | 11.8 | 50 | --- | --- |
| 386 | H 1-37 | 17 50 44.569 | -39 17 25.98 | --- | 0.508 | --- | 3.56 | 8.5 | --- | --- | --- |
| 388 | H 1-39 | 17 53 21.003 | -33 55 58.42 | --- | 1.308 | 1.7 | 3.62 | 10.8 | 13 | 10.04 | 0.083 |
| 390 | H 1-40 | 17 55 36.049 | -30 33 32.08 | 1.255 | 10.494 | 3 | 4.48 | 8.1 | 31 | 7.84 | 0.114 |
| 391 | H 1-41 | 17 57 19.145 | -34 09 49.12 | --- | 0.557 | 9.6 | 3.33 | 16.6 | 12 | 6.6 | 0.307 |
| 392 | H 1-42 | 17 57 25.169 | -33 35 42.94 | --- | 1.769 | 5.8 | 3.77 | 34.5 | 40 | 4.82 | 0.136 |
| 393 | H 1-43 | 17 58 14.430 | -33 47 37.51 | 0.309 | 3.571 | 3 | 4.03 | 4 | 6 | 11.56 | 0.168 |
| 394 | H 1-44 | 17 58 10.641 | -31 42 56.08 | --- | 0.484 | <5 | 2.98 | 8 | --- | --- | --- |

| KN | NAME | R.A. (2000.0) H:M:S | DEC (2000.0) D:M:S | F(9μm) Jy | F(18μm) Jy | θ arcsec | MN_18 arcsec | F(1.4GHz) mJy | F(5GHz) 5GHz | DIST. Kpc | DIAM. pc |
|---|---|---|---|---|---|---|---|---|---|---|---|
| 395 | H 1-45 | 17 58 21.865 | -28 14 52.20 | 5.230 | 2.443 | <10 | 3.94 | --- | 23.5 | 8.2 | --- |
| 396 | H 1-46 | 17 59 02.490 | -32 21 43.42 | 0.303 | 3.205 | 1.2 | 3.94 | 18.3 | 43 | 7.11 | 0.041 |
| 397 | H 1-47 | 18 00 37.596 | -29 21 50.22 | 0.137 | 2.257 | 2.5 | 3.88 | 8.1 | 10 | 10.01 | 0.121 |
| 398 | H 1-5 | 16 57 23.740 | -41 37 57.88 | 1.631 | --- | --- | --- | --- | --- | --- | --- |
| 399 | H 1-50 | 18 03 53.461 | -32 41 42.15 | 0.189 | 2.078 | 1.4 | 3.69 | 19.6 | 31 | 7.68 | 0.052 |
| 402 | H 1-53 | 18 05 57.404 | -26 29 41.50 | --- | 0.651 | <25 | 3.15 | 7.2 | --- | --- | --- |
| 403 | H 1-54 | 18 07 07.239 | -29 13 06.35 | 0.192 | 3.092 | 2.5 | 3.87 | 12.7 | 31 | 8.01 | 0.097 |
| 404 | H 1-55 | 18 07 14.542 | -29 41 24.51 | 0.176 | 0.518 | 10 | 3.06 | 2.5 | 5.3 | --- | --- |
| 405 | H 1-56 | 18 07 53.880 | -29 44 34.26 | --- | 0.894 | 3 | 3.37 | 8.3 | 7.8 | 10.98 | 0.160 |
| 407 | H 1-58 | 18 09 13.835 | -26 02 28.78 | 0.638 | 5.951 | --- | 4.24 | 2.7 | --- | --- | --- |
| 408 | H 1-59 | 18 11 29.263 | -27 46 15.69 | --- | 0.317 | 6 | 2.99 | 3.4 | 4.2 | 11.12 | 0.323 |
| 410 | H 1-60 | 18 12 25.155 | -27 29 12.96 | --- | 0.411 | 3.46 | 2.73 | --- | 11.3 | 8.59 | 0.144 |
| 411 | H 1-61 | 18 12 33.977 | -24 50 00.28 | 0.427 | --- | <25 | --- | 4.7 | --- | --- | --- |
| 412 | H 1-62 | 18 13 17.951 | -32 19 42.74 | 0.123 | 1.979 | --- | 3.78 | 11.9 | --- | --- | --- |
| 413 | H 1-63 | 18 16 19.336 | -30 07 35.98 | 0.578 | 4.882 | --- | 4.02 | 2.5 | 9 | --- | --- |
| 414 | H 1-64 | 18 18 23.908 | -23 24 57.15 | --- | 0.297 | <25 | 2.71 | 8.1 | --- | --- | --- |
| 415 | H 1-65 | 18 20 08.847 | -24 15 05.09 | 0.195 | 2.520 | 2.5 | 3.75 | 5.9 | 10 | 10.23 | 0.124 |
| 416 | H 1-66 | 18 24 57.539 | -25 41 55.76 | --- | 0.419 | <10 | 3.53 | 10.4 | 6 | 9.5 | --- |
| 417 | H 1-67 | 18 25 04.976 | -22 34 52.64 | 0.100 | --- | 6 | --- | 12.2 | 11.9 | 7.51 | 0.218 |
| 418 | H 1-7 | 17 10 27.389 | -41 52 49.42 | 1.893 | 13.029 | --- | 4.72 | --- | --- | --- | --- |
| 419 | H 1-8 | 17 14 42.902 | -33 24 47.21 | 0.292 | 2.295 | 3.4 | 3.99 | 22.2 | --- | --- | --- |
| 421 | H 2-1 | 17 04 36.257 | -33 59 18.75 | 0.967 | 8.184 | 2.4 | 4.23 | 51.5 | 61 | 5.84 | 0.068 |
| 422 | H 2-10 | 17 27 32.848 | -28 31 06.90 | --- | 1.045 | 2 | 3.44 | 13.4 | 20 | 8.21 | 0.080 |
| 423 | H 2-11 | 17 29 25.947 | -25 49 06.64 | 0.248 | --- | 1.5 | --- | 11.5 | 27.7 | --- | --- |
| 425 | H 2-13 | 17 31 08.084 | -30 10 28.00 | --- | 1.052 | --- | 3.38 | 17.2 | --- | --- | --- |
| 426 | H 2-14 | 17 32 20.093 | -39 51 25.60 | --- | 0.326 | 15 | 3.70 | 5.2 | 4.9 | --- | --- |
| 430 | H 2-18 | 17 43 28.751 | -21 09 51.29 | --- | 0.428 | 3.5 | 3.26 | 10.1 | 11 | --- | --- |
| 431 | H 2-20 | 17 45 39.767 | -25 40 00.04 | --- | 1.474 | 3.5 | 3.66 | 13.1 | 16.3 | 7.5 | 0.127 |
| 432 | H 2-22 | 17 47 33.926 | -21 47 23.11 | --- | 0.236 | <8 | 2.30 | 5.4 | --- | --- | --- |
| 433 | H 2-23 | 17 48 57.994 | -34 21 53.28 | --- | 0.269 | --- | 2.48 | 5.3 | --- | --- | --- |
| 434 | H 2-24 | 17 48 36.494 | -24 16 34.26 | --- | 3.535 | 4.90 | 4.00 | 4.5 | 13 | 7.77 | 0.185 |
| 437 | H 2-27 | 17 51 50.576 | -33 47 35.59 | 0.098 | 0.541 | --- | 3.00 | 7.9 | --- | --- | --- |
| 440 | H 2-31 | 17 56 02.350 | -28 14 11.36 | --- | 0.877 | --- | 3.37 | 5.3 | --- | --- | --- |
| 442 | H 2-33 | 17 58 12.535 | -31 08 05.99 | --- | 0.393 | 8 | 3.25 | 8.1 | 5 | 9.71 | 0.377 |
| 446 | H 2-39 | 18 08 05.759 | -28 26 10.50 | --- | 0.208 | --- | 2.43 | 4.9 | --- | --- | --- |
| 450 | H 2-43 | 18 12 47.953 | -28 19 59.67 | 0.802 | 0.702 | 9 | 3.28 | --- | 26.4 | 5.03 | 0.219 |
| 452 | H 2-45 | 18 14 28.844 | -24 43 38.35 | --- | 0.317 | 4.447 | 2.82 | 13.2 | --- | --- | --- |
| 454 | H 2-48 | 18 46 35.149 | -23 26 48.24 | 0.650 | 9.835 | 2 | 4.26 | 31.6 | 66 | 6.23 | 0.060 |
| 455 | H 2-7 | 17 23 24.936 | -28 59 06.05 | --- | 0.702 | 4.2 | 3.47 | 11.3 | 9 | 9.18 | 0.187 |
| 457 | H 3-29 | 04 37 23.484 | +25 02 40.94 | 0.087 | 0.711 | 21 | 4.06 | 18.3 | 18 | 3.4 | 0.346 |
| 462 | HaTr 11 | 19 02 59.373 | +03 02 20.77 | --- | 0.566 | --- | 3.74 | 17 | 83 | --- | --- |
| 465 | HaTr 2 | 15 30 18.542 | -61 01 38.79 | --- | 0.199 | --- | 2.95 | --- | --- | --- | --- |
| 474 | Hb 12 | 23 26 14.814 | +58 10 54.65 | 11.885 | 40.599 | 0.8 | 6.06 | --- | 45 | 10.46 | 0.041 |
| 476 | Hb 5 | 17 47 56.187 | -29 59 41.91 | --- | 41.232 | 3.4 | 6.43 | 179.5 | 548 | 1.32 | 0.022 |

| KN | NAME | R.A. (2000.0) H:M:S | DEC (2000.0) D:M:S | F(9μm) Jy | F(18μm) Jy | θ arcsec | MN_18 arcsec | F(1.4GHz) mJy | F(5GHz) 5GHz | DIST. Kpc | DIAM. pc |
|---|---|---|---|---|---|---|---|---|---|---|---|
| 477 | Hb 6 | 17 55 07.023 | -21 44 39.98 | 1.144 | --- | 6 | --- | 190.5 | 243 | 2.45 | 0.071 |
| 478 | Hb 7 | 18 55 37.950 | -32 15 47.07 | 0.161 | 2.143 | 4 | 3.78 | 25.4 | 30 | 5.9 | 0.114 |
| 479 | He 1-1 | 19 23 46.875 | +21 06 38.60 | --- | 0.611 | 8 | 3.40 | 16 | 14 | --- | --- |
| 480 | He 1-2 | 19 26 37.757 | +21 09 27.04 | 0.125 | 2.281 | 4.7 | 4.01 | 14.6 | 15 | --- | --- |
| 481 | He 1-3 | 19 48 26.422 | +22 08 37.62 | 0.120 | --- | 8 | --- | 53.5 | --- | --- | --- |
| 483 | He 1-5 | 20 11 56.057 | +20 20 04.43 | 14.057 | 5.460 | 36 | 4.07 | 4.8 | 5.2 | 6.71 | 1.171 |
| 485 | He 2-101 | 13 54 55.721 | -58 27 16.46 | 1.096 | 1.054 | <10 | 3.55 | --- | --- | --- | --- |
| 486 | He 2-102 | 13 58 13.866 | -58 54 31.78 | 0.215 | 1.276 | 9 | 4.09 | --- | 33 | 3.77 | 0.164 |
| 487 | He 2-104 | 14 11 52.077 | -51 26 24.18 | 7.077 | 7.358 | 5 | 4.24 | --- | 15 | --- | --- |
| 489 | He 2-107 | 14 18 43.339 | -63 07 09.48 | 0.155 | 4.631 | 10 | 4.49 | --- | 65 | 2.97 | 0.144 |
| 490 | He 2-108 | 14 18 08.889 | -52 10 39.76 | 0.179 | 4.040 | 11 | 4.76 | --- | 32 | 3.49 | 0.186 |
| 494 | He 2-112 | 14 40 30.924 | -52 34 56.66 | 0.271 | 3.300 | 14.6 | 3.93 | --- | 82 | 2.36 | 0.167 |
| 495 | He 2-113 | 14 59 53.523 | -54 18 07.20 | 62.991 | 209.197 | --- | 7.79 | --- | 115 | --- | --- |
| 499 | He 2-117 | 15 05 59.193 | -55 59 16.53 | 1.558 | 20.941 | 5 | 5.32 | --- | 267 | 2.69 | 0.065 |
| 500 | He 2-118 | 15 06 13.731 | -42 59 56.46 | --- | 0.870 | 5 | 3.50 | --- | 10 | --- | --- |
| 502 | He 2-120 | 15 11 56.388 | -55 39 46.67 | --- | 0.567 | 27 | 4.07 | --- | 26 | 2.51 | 0.329 |
| 503 | He 2-123 | 15 22 19.360 | -54 08 12.77 | 0.419 | 2.327 | 4.6 | 3.88 | --- | 110 | 3.59 | 0.080 |
| 504 | He 2-125 | 15 23 36.326 | -53 51 27.95 | 0.109 | 1.870 | 3 | 3.84 | --- | 20.4 | 6.96 | 0.101 |
| 505 | He 2-128 | 15 25 07.841 | -51 19 42.29 | 0.201 | 2.529 | 5 | 3.86 | --- | 40 | --- | --- |
| 506 | He 2-129 | 15 25 32.677 | -52 50 37.96 | 0.143 | 1.474 | 1.6 | 3.59 | --- | 35 | 7.85 | 0.061 |
| 507 | He 2-131 | 15 37 11.210 | -71 54 52.89 | 2.586 | 50.505 | 6 | 6.43 | --- | 325 | 2.2 | 0.064 |
| 508 | He 2-132 | 15 38 01.182 | -58 44 42.07 | 0.232 | 1.382 | 17.8 | 4.82 | --- | 25 | 3.02 | 0.261 |
| 509 | He 2-133 | 15 41 58.788 | -56 36 25.72 | 1.254 | 16.766 | <10 | 5.03 | --- | 210 | 3.18 | --- |
| 510 | He 2-136 | 15 52 10.666 | -62 30 46.94 | 0.137 | 0.588 | 10 | 3.14 | --- | 23 | --- | --- |
| 511 | He 2-138 | 15 56 01.694 | -66 09 09.23 | 1.015 | 21.099 | 7 | 5.50 | --- | 76 | 3.61 | 0.123 |
| 512 | He 2-140 | 15 58 08.063 | -55 41 50.13 | 0.309 | 6.339 | 2.6 | 4.32 | --- | 80 | 5.03 | 0.063 |
| 513 | He 2-141 | 15 59 08.762 | -58 23 53.15 | 0.151 | 1.255 | 13.8 | 4.26 | --- | 51 | 2.77 | 0.185 |
| 514 | He 2-142 | 15 59 57.608 | -55 55 32.89 | 4.033 | 15.811 | 3.6 | 4.89 | --- | 65 | 4.75 | 0.083 |
| 515 | He 2-143 | 16 00 59.125 | -55 05 39.74 | 0.854 | 8.233 | 5.2 | 4.34 | --- | 120 | 3.32 | 0.084 |
| 516 | He 2-145 | 16 08 58.902 | -51 01 57.74 | --- | 0.713 | <25 | 5.98 | --- | --- | --- | --- |
| 518 | He 2-147 | 16 14 01.045 | -56 59 27.78 | 5.208 | 2.678 | --- | 3.65 | --- | --- | --- | --- |
| 519 | He 2-149 | 16 14 24.266 | -54 47 38.82 | --- | 0.542 | 3 | 2.96 | --- | 10 | 8.52 | 0.124 |
| 520 | He 2-15 | 08 53 30.701 | -40 03 42.08 | 0.227 | 1.209 | 23.8 | 4.64 | 94.5 | 105 | 1.92 | 0.222 |
| 521 | He 2-151 | 16 15 42.269 | -59 54 00.96 | 0.616 | 12.148 | 5 | 4.62 | --- | 10 | --- | --- |
| 522 | He 2-152 | 16 15 20.031 | -49 13 20.76 | 1.234 | 7.995 | 11 | 4.64 | --- | 196 | 2.09 | 0.111 |
| 524 | He 2-155 | 16 19 23.101 | -42 15 35.98 | 0.223 | 0.721 | 14.6 | 3.72 | --- | 70 | 2.47 | 0.175 |
| 525 | He 2-157 | 16 22 14.264 | -53 40 54.09 | --- | 1.951 | <5 | 3.89 | --- | 30 | 6.24 | --- |
| 526 | He 2-158 | 16 23 30.605 | -58 19 22.64 | --- | 0.455 | 2 | 3.37 | --- | 2.7 | 14.71 | 0.143 |
| 527 | He 2-159 | 16 24 21.417 | -54 36 02.82 | --- | 0.303 | 10 | 2.52 | --- | 25 | 3.9 | 0.189 |
| 528 | He 2-161 | 16 24 37.786 | -53 22 34.14 | 0.113 | 0.717 | 10 | 3.43 | --- | 32 | 3.64 | 0.176 |
| 529 | He 2-162 | 16 27 50.914 | -54 01 28.36 | 0.250 | 0.927 | <5 | 3.78 | --- | 28 | --- | --- |
| 531 | He 2-164 | 16 29 53.255 | -53 23 15.37 | 0.226 | 1.529 | 16 | 4.60 | --- | 97 | 2.17 | 0.168 |
| 533 | He 2-169 | 16 34 13.328 | -49 21 13.20 | 0.338 | 1.699 | 21.8 | 4.60 | --- | 128 | 2.7 | 0.285 |
| 534 | He 2-170 | 16 35 21.170 | -53 50 11.11 | 0.160 | 1.675 | 5 | 3.80 | --- | 15 | --- | --- |

| KN | NAME | R.A. (2000.0) H:M:S | DEC (2000.0) D:M:S | F(9μm) Jy | F(18μm) Jy | θ arcsec | MN_18 arcsec | F(1.4GHz) mJy | F(5GHz) 5GHz | DIST. Kpc | DIAM. pc |
|---|---|---|---|---|---|---|---|---|---|---|---|
| 535 | He 2-171 | 16 34 04.240 | -35 05 26.93 | 4.775 | 3.186 | 10 | 3.75 | --- | 10 | --- | --- |
| 536 | He 2-175 | 16 39 28.112 | -36 34 16.41 | 0.183 | --- | 6.6 | --- | 19.2 | 26.8 | 4.58 | 0.147 |
| 538 | He 2-182 | 16 54 35.167 | -64 14 28.43 | 0.365 | 4.408 | 10 | 4.21 | --- | 62 | --- | --- |
| 539 | He 2-185 | 17 01 17.254 | -70 06 03.35 | --- | 0.506 | 10 | 3.17 | --- | 18 | --- | --- |
| 540 | He 2-186 | 16 59 36.064 | -51 42 06.46 | 0.091 | 0.752 | 3 | 3.44 | --- | 21 | 6.9 | 0.100 |
| 541 | He 2-187 | 17 01 36.976 | -50 22 57.49 | --- | 0.549 | 6 | 3.65 | --- | --- | --- | --- |
| 543 | He 2-21 | 09 13 52.823 | -55 28 16.80 | --- | 0.237 | <10 | 2.46 | --- | 16 | 8.21 | --- |
| 545 | He 2-25 | 09 18 01.308 | -54 39 29.01 | 1.781 | 3.827 | 4.4 | 3.95 | --- | 827.9 | 2.07 | 0.044 |
| 546 | He 2-250 | 17 34 54.710 | -26 35 56.92 | --- | 0.728 | 5 | 3.40 | 15.3 | 15 | 7.01 | 0.170 |
| 547 | He 2-26 | 09 19 27.476 | -59 12 00.39 | 0.171 | 1.109 | <10 | 3.54 | --- | 40 | 1.91 | --- |
| 548 | He 2-262 | 17 40 12.814 | -26 44 21.37 | --- | 1.082 | <10 | 3.46 | 24.1 | 26 | 6.64 | --- |
| 549 | He 2-28 | 09 22 06.825 | -54 09 38.60 | --- | 0.123 | 10 | 1.95 | --- | 20 | 4.15 | 0.201 |
| 550 | He 2-29 | 09 24 45.834 | -54 36 15.45 | --- | 0.326 | 14 | 3.07 | --- | 24 | 3.41 | 0.231 |
| 551 | He 2-306 | 17 56 33.706 | -43 03 18.90 | --- | 0.830 | --- | 3.32 | --- | --- | --- | --- |
| 553 | He 2-34 | 09 41 13.997 | -49 22 47.18 | 8.941 | 6.160 | <10 | 4.19 | --- | --- | --- | --- |
| 554 | He 2-35 | 09 41 37.504 | -49 57 58.71 | --- | 1.009 | 5 | 3.61 | --- | 20 | 5.61 | 0.136 |
| 555 | He 2-36 | 09 43 25.538 | -57 16 55.44 | 0.192 | 1.686 | 22 | 4.25 | --- | 90 | 3.1 | 0.331 |
| 559 | He 2-41 | 10 07 23.611 | -63 54 30.06 | --- | 0.393 | <10 | 2.94 | --- | 41 | 3.39 | --- |
| 561 | He 2-428 | 19 13 05.239 | +15 46 39.80 | --- | 0.331 | 8 | 2.99 | 26.7 | 7 | 6.16 | 0.239 |
| 562 | He 2-429 | 19 13 38.422 | +14 59 19.21 | 0.271 | 2.240 | 4.2 | 3.96 | 58.5 | 64.9 | 4.33 | 0.088 |
| 563 | He 2-430 | 19 14 04.197 | +17 31 32.92 | 0.255 | --- | 1.7 | --- | 18.2 | 40 | 7.36 | 0.061 |
| 564 | He 2-432 | 19 23 24.821 | +21 08 00.50 | 0.152 | 1.286 | 2.3 | 3.61 | 21.3 | 32 | 6.88 | 0.077 |
| 565 | He 2-434 | 19 33 49.420 | -74 32 58.66 | --- | 1.035 | 10 | 3.42 | --- | --- | --- | --- |
| 566 | He 2-436 | 19 32 06.701 | -34 12 57.43 | 0.327 | 0.457 | 10 | 2.86 | --- | 23 | --- | --- |
| 567 | He 2-437 | 19 32 57.657 | +26 52 43.35 | 2.839 | 5.478 | --- | 4.11 | 6.6 | --- | --- | --- |
| 568 | He 2-440 | 19 38 08.403 | +25 15 40.98 | 0.233 | 2.995 | 2.2 | 3.98 | 26.4 | 43 | 6.45 | 0.069 |
| 569 | He 2-442 | 19 39 43.376 | +26 29 33.05 | 14.020 | 7.485 | <10 | 4.24 | --- | 6 | --- | --- |
| 570 | He 2-447 | 19 45 22.164 | +21 20 04.03 | 0.951 | 6.286 | 1.2 | 4.21 | 22.9 | 60 | 7.63 | 0.044 |
| 571 | He 2-459 | 20 13 57.898 | +29 33 55.94 | 2.042 | 17.276 | 1.3 | 5.12 | 13.9 | 64 | 7.24 | 0.046 |
| 572 | He 2-47 | 10 23 09.143 | -60 32 42.21 | 1.400 | 22.347 | 5 | 5.53 | --- | 170 | --- | --- |
| 574 | He 2-5 | 07 47 20.022 | -51 15 03.42 | 0.080 | 1.011 | 3 | 3.51 | --- | 29 | 6.3 | 0.092 |
| 577 | He 2-55 | 10 48 43.168 | -56 03 10.21 | --- | 0.324 | <25 | 3.58 | --- | --- | --- | --- |
| 579 | He 2-62 | 11 17 43.157 | -70 49 32.41 | 0.075 | 0.692 | <10 | 3.36 | --- | --- | --- | --- |
| 582 | He 2-67 | 11 28 47.369 | -60 06 37.28 | 0.224 | 1.375 | 5 | 3.64 | --- | 41 | --- | --- |
| 583 | He 2-68 | 11 31 45.427 | -65 58 13.67 | 0.369 | 2.549 | 10 | 3.90 | --- | 34 | --- | --- |
| 584 | He 2-7 | 08 11 31.890 | -48 43 16.71 | 0.102 | 0.596 | 44.6 | 3.27 | --- | 47 | 2.91 | 0.629 |
| 586 | He 2-71 | 11 39 11.198 | -68 52 09.14 | 0.298 | 3.484 | 5 | 3.95 | --- | 12 | --- | --- |
| 588 | He 2-73 | 11 48 38.191 | -65 08 37.33 | 0.378 | 3.875 | 4 | 3.99 | --- | 76 | 4.23 | 0.082 |
| 589 | He 2-76 | 12 08 25.433 | -64 12 09.32 | --- | 0.467 | --- | 3.43 | --- | --- | --- | --- |
| 590 | He 2-78 | 12 09 10.196 | -58 42 37.44 | --- | 0.717 | --- | 3.40 | --- | --- | --- | --- |
| 591 | He 2-81 | 12 23 01.237 | -64 01 45.96 | --- | 0.308 | <25 | 2.70 | --- | --- | --- | --- |
| 593 | He 2-83 | 12 28 43.997 | -62 05 35.05 | 0.357 | 6.569 | --- | 4.27 | --- | --- | --- | --- |
| 594 | He 2-84 | 12 28 46.822 | -63 44 37.22 | --- | 0.261 | --- | 2.56 | --- | --- | --- | --- |
| 595 | He 2-85 | 12 30 07.567 | -63 53 00.32 | 0.463 | 3.290 | 10.2 | 4.26 | --- | 111.7 | 2.53 | 0.125 |

Table 1 (cont.)

| KN | NAME | R.A. (2000.0) H:M:S | DEC (2000.0) D:M:S | F(9μm) Jy | F(18μm) Jy | θ arcsec | MN_18 arcsec | F(1.4GHz) mJy | F(5GHz) 5GHz | DIST. Kpc | DIAM. pc |
|---|---|---|---|---|---|---|---|---|---|---|---|
| 596 | He 2-86 | 12 30 30.426 | -64 52 05.58 | 1.102 | 12.943 | 3.6 | 4.81 | --- | 125 | 3.85 | 0.067 |
| 598 | He 2-9 | 08 28 27.988 | -39 23 40.27 | 0.754 | 6.989 | 4.4 | 4.37 | 135.6 | 194.8 | 3.11 | 0.066 |
| 599 | He 2-90 | 13 09 36.252 | -61 19 35.96 | 32.451 | 62.806 | 10 | 6.59 | --- | 25 | --- | --- |
| 600 | He 2-96 | 13 42 36.150 | -61 22 29.18 | 0.762 | 7.397 | --- | 4.25 | --- | --- | --- | --- |
| 601 | He 2-97 | 13 45 22.394 | -71 28 56.13 | 0.419 | 4.989 | 5 | 4.03 | --- | 30 | --- | --- |
| 602 | He 2-99 | 13 52 30.683 | -66 23 26.49 | 0.366 | 5.097 | 17 | 5.04 | --- | 18 | 3.4 | 0.280 |
| 603 | He 3-1333 | 17 09 00.932 | -56 54 48.25 | 118.338 | 165.011 | --- | 7.72 | --- | 26 | --- | --- |
| 605 | Hf 2-2 | 18 32 30.935 | -28 43 20.47 | --- | 0.300 | --- | 2.31 | 4.4 | --- | --- | --- |
| 606 | Hf 39 | 10 53 59.586 | -60 26 44.31 | 1.092 | 8.060 | --- | 10.32 | --- | --- | --- | --- |
| 607 | Hf 48 | 11 03 55.979 | -60 36 04.57 | --- | 0.416 | --- | 3.26 | --- | --- | --- | --- |
| 608 | Hu 1-1 | 00 28 15.614 | +55 57 54.71 | 0.200 | 0.665 | 10 | 3.59 | 28 | 26 | 3.86 | 0.187 |
| 609 | Hu 1-2 | 21 33 08.349 | +39 38 09.57 | 0.211 | 1.849 | 6.5 | 4.21 | 107 | 155 | 2.53 | 0.080 |
| 610 | Hu 2-1 | 18 49 47.562 | +20 50 39.46 | 0.782 | 5.679 | 1.8 | 4.24 | 42.6 | 110 | 4.52 | 0.039 |
| 612 | IC 1297 | 19 17 23.459 | -39 36 46.40 | 0.320 | 1.450 | 7 | 3.93 | 59.4 | 68.9 | 3.75 | 0.127 |
| 614 | IC 1747 | 01 57 35.896 | +63 19 19.36 | 0.499 | 2.162 | 13 | 4.28 | 85.8 | 124 | 2.23 | 0.141 |
| 615 | IC 2003 | 03 56 21.984 | +33 52 30.59 | 0.245 | 1.539 | 10 | 3.93 | 54.8 | 30 | 3.89 | 0.189 |
| 616 | IC 2149 | 05 56 23.908 | +46 06 17.32 | 0.786 | 9.271 | 8.4 | 5.48 | 177 | 280 | 2.03 | 0.083 |
| 617 | IC 2165 | 06 21 42.775 | -12 59 13.96 | 0.485 | 4.793 | 9 | 4.21 | 180 | 188 | 2.47 | 0.108 |
| 618 | IC 2448 | 09 07 06.261 | -69 56 30.74 | 0.417 | 1.940 | 10 | 5.02 | --- | 73 | 3.41 | 0.165 |
| 619 | IC 2501 | 09 38 47.213 | -60 05 30.92 | 1.939 | 13.312 | 2 | 4.99 | --- | 261 | 2.09 | 0.020 |
| 620 | IC 2553 | 10 09 20.856 | -62 36 48.40 | 0.817 | 5.105 | 9 | 4.38 | --- | 92 | 2.85 | 0.124 |
| 621 | IC 2621 | 11 00 20.111 | -65 14 57.77 | 2.614 | 15.350 | 5 | 4.89 | --- | 195 | 2.94 | 0.071 |
| 622 | IC 289 | 03 10 19.273 | +61 19 00.91 | --- | 5.700 | 36.8 | 7.95 | 152 | 170 | 1.18 | 0.211 |
| 623 | IC 351 | 03 47 33.143 | +35 02 48.50 | --- | 0.637 | 7 | 3.51 | 31.9 | 27 | 4.45 | 0.151 |
| 624 | IC 3568 | 12 33 06.834 | +82 33 50.29 | 0.340 | 3.570 | 18 | 5.27 | 94.1 | 95 | 2.47 | 0.216 |
| 626 | IC 4191 | 13 08 47.343 | -67 38 37.67 | 0.961 | 9.494 | 14 | 4.67 | --- | 172 | 1.95 | 0.132 |
| 627 | IC 4406 | 14 22 26.278 | -44 09 04.35 | --- | 1.841 | 20 | 5.83 | --- | 110 | 1.5 | 0.145 |
| 628 | IC 4593 | 16 11 44.545 | +12 04 17.06 | 0.375 | 9.950 | 12.8 | 5.16 | 90.6 | 92 | 2.42 | 0.150 |
| 629 | IC 4634 | 17 01 33.572 | -21 49 32.77 | 0.505 | 6.146 | 5.5 | 4.38 | 116 | 100 | 3.47 | 0.093 |
| 630 | IC 4637 | 17 05 10.506 | -40 53 08.44 | --- | 9.044 | 18.6 | 5.70 | --- | 132.5 | 2.29 | 0.207 |
| 631 | IC 4642 | 17 11 45.025 | -55 24 01.47 | --- | 2.797 | 16.6 | 5.51 | --- | 60 | 2.52 | 0.203 |
| 633 | IC 4673 | 18 03 18.408 | -27 06 22.61 | 0.692 | 3.057 | 16 | 5.02 | 53.4 | 62 | 3.12 | 0.242 |
| 634 | IC 4699 | 18 18 32.024 | -45 59 01.70 | --- | 0.283 | 5 | 2.86 | --- | 20 | 4.91 | 0.119 |
| 635 | IC 4732 | 18 33 54.633 | -22 38 40.91 | 0.203 | --- | 3 | --- | 12.2 | 49 | 4.79 | 0.070 |
| 636 | IC 4846 | 19 16 28.220 | -09 02 36.57 | 0.131 | 2.038 | 2.9 | 3.85 | 38.7 | 43 | 5.72 | 0.080 |
| 637 | IC 4997 | 20 20 08.741 | +16 43 53.71 | 1.143 | 16.078 | 2 | 4.97 | 30.1 | 127 | 5.45 | 0.053 |
| 638 | IC 5117 | 21 32 31.027 | +44 35 48.53 | 6.448 | 26.937 | 1.5 | 5.60 | 34.1 | 210 | 5.01 | 0.036 |
| 639 | IC 5217 | 22 23 55.725 | +50 58 00.43 | 0.278 | 1.772 | 6.5 | 3.89 | 50.4 | 163 | 2.71 | 0.085 |
| 643 | J 320 | 05 05 34.313 | +10 42 22.73 | --- | 0.545 | 7.1 | 3.32 | 29.6 | 23 | 4.63 | 0.159 |
| 644 | J 900 | 06 25 57.275 | +17 47 27.19 | 1.467 | 4.976 | 6 | 4.24 | 107.8 | 100 | 3.29 | 0.096 |
| 653 | KFL 15 | 18 14 19.324 | -25 20 51.22 | --- | 0.451 | 8.5 | 3.31 | 11 | 11.2 | --- | --- |
| 681 | K 2-16 | 16 44 49.050 | -28 04 04.56 | 2.535 | --- | --- | --- | --- | --- | --- | --- |
| 687 | K 3-11 | 18 41 07.311 | -08 55 58.97 | 0.154 | 3.265 | 3 | 3.99 | 14.5 | 17 | 7.33 | 0.107 |
| 688 | K 3-13 | 18 45 24.582 | +02 01 23.85 | --- | 1.335 | 3.7 | 3.54 | 30.6 | 34 | 5.5 | 0.099 |

| KN | NAME | R.A. (2000.0) H:M:S | DEC (2000.0) D:M:S | F(9μm) Jy | F(18μm) Jy | θ arcsec | MN_18 arcsec | F(1.4GHz) mJy | F(5GHz) 5GHz | DIST. Kpc | DIAM. pc |
|---|---|---|---|---|---|---|---|---|---|---|---|
| 689 | K 3-14 | 18 48 32.821 | +10 35 50.74 | --- | 0.667 | 1 | 3.31 | --- | 4.8 | 16.89 | 0.082 |
| 690 | K 3-15 | 18 51 41.522 | +09 54 53.06 | 0.478 | --- | 0.3 | --- | --- | 4 | --- | --- |
| 692 | K 3-17 | 18 56 18.171 | +07 07 26.31 | 1.967 | 18.124 | 8 | 5.47 | 319.5 | 345 | 2.04 | 0.079 |
| 693 | K 3-18 | 19 00 34.829 | -02 11 57.62 | --- | 9.729 | 3.5 | 4.82 | --- | 11 | 13.98 | 0.235 |
| 694 | K 3-19 | 19 01 36.575 | -01 19 07.85 | 0.131 | 1.262 | 1.2 | 3.56 | 11.8 | 23 | 8.93 | 0.052 |
| 695 | K 3-2 | 18 25 00.576 | -01 30 52.64 | 0.082 | 2.431 | 2.8 | 3.77 | 25.9 | 31 | 6.37 | 0.086 |
| 698 | K 3-22 | 19 09 26.658 | +12 00 44.05 | 0.832 | 0.715 | --- | 3.21 | --- | --- | --- | --- |
| 699 | K 3-24 | 19 12 05.820 | +15 09 04.47 | --- | 0.268 | 6.2 | 2.70 | 8.5 | --- | --- | --- |
| 700 | K 3-26 | 19 14 39.178 | +00 13 36.29 | --- | 0.286 | --- | 3.03 | 6.4 | 0.5 | --- | --- |
| 702 | K 3-29 | 19 15 30.561 | +14 03 49.83 | 0.869 | 7.708 | 1 | 4.29 | 13.9 | 65 | 8.07 | 0.039 |
| 703 | K 3-3 | 18 27 09.338 | +01 14 26.88 | 0.321 | 1.614 | 10 | 4.05 | 35.9 | 34 | 3.43 | 0.166 |
| 704 | K 3-30 | 19 16 27.691 | +05 13 19.47 | 0.110 | 0.662 | 3.3 | 3.23 | 12.1 | 23 | 6.46 | 0.103 |
| 705 | K 3-31 | 19 19 02.666 | +19 02 20.85 | 0.269 | 2.394 | 1.5 | 3.86 | 16.8 | 39 | 7.83 | 0.057 |
| 706 | K 3-33 | 19 22 26.670 | +10 41 21.39 | --- | 1.613 | 1.1 | 3.83 | 8.6 | 17 | 11.33 | 0.060 |
| 708 | K 3-35 | 19 27 44.036 | +21 30 03.83 | 0.370 | 6.017 | 1.7 | 4.35 | 14.5 | 60 | 6.56 | 0.054 |
| 709 | K 3-36 | 19 32 39.557 | +07 27 51.57 | --- | 0.164 | --- | 1.88 | 3.1 | 0.2 | --- | --- |
| 710 | K 3-37 | 19 33 46.749 | +24 32 27.09 | --- | 0.466 | 2.5 | 2.97 | 14.4 | 17 | 7.93 | 0.096 |
| 711 | K 3-38 | 19 35 18.354 | +17 13 00.71 | 0.236 | 1.362 | 5 | 3.81 | 28.7 | 31 | 5.09 | 0.123 |
| 712 | K 3-39 | 19 35 54.469 | +24 54 48.20 | 0.393 | 4.122 | 1 | 4.15 | 3.6 | 11 | 13.35 | 0.065 |
| 713 | K 3-4 | 18 31 00.227 | +02 25 27.35 | --- | 0.180 | 21.0 | 2.84 | 17 | 21 | --- | --- |
| 714 | K 3-40 | 19 36 21.828 | +23 39 47.93 | 0.216 | 1.597 | 4 | 3.76 | 17.1 | 20 | 6.18 | 0.120 |
| 720 | K 3-49 | 19 54 00.702 | +33 22 12.95 | 0.227 | 2.631 | --- | 3.87 | 5.5 | 7 | 34.5 | --- |
| 721 | K 3-5 | 18 31 45.833 | +04 05 09.12 | 0.109 | 0.276 | 10 | 2.86 | 6.7 | 3 | 7.11 | 0.345 |
| 723 | K 3-52 | 20 03 11.435 | +30 32 34.15 | 0.877 | 7.924 | 0.7 | 4.28 | 16.9 | 65 | 9.42 | 0.032 |
| 724 | K 3-53 | 20 03 22.475 | +27 00 54.73 | 1.360 | 7.328 | 0.9 | 4.31 | 10 | 69 | 9.27 | 0.040 |
| 725 | K 3-54 | 20 04 58.639 | +25 26 37.36 | 0.155 | 0.638 | 0.8 | 3.29 | 4.8 | 7.5 | 16.39 | 0.064 |
| 726 | K 3-55 | 20 06 56.210 | +32 16 33.60 | 0.325 | 2.551 | 8.2 | 4.05 | 86.6 | 90 | 2.96 | 0.118 |
| 728 | K 3-57 | 20 12 47.679 | +34 20 32.52 | 0.245 | 1.861 | 6.3 | 3.96 | 46.8 | 60 | 3.72 | 0.114 |
| 731 | K 3-60 | 21 27 26.477 | +57 39 06.45 | 0.962 | 3.113 | 1.9 | 3.92 | 28 | 43 | 6.87 | 0.063 |
| 732 | K 3-61 | 21 30 00.710 | +54 27 27.45 | 0.118 | 0.606 | 6 | 3.25 | 16.6 | 14 | 5.73 | 0.167 |
| 733 | K 3-62 | 21 31 50.203 | +52 33 51.64 | 0.748 | 6.410 | 2.5 | 4.29 | 59.5 | 115 | 4.62 | 0.056 |
| 734 | K 3-63 | 21 39 11.976 | +55 46 03.94 | --- | 0.443 | 7 | 3.18 | 8.5 | 29 | --- | --- |
| 737 | K 3-66 | 04 36 37.243 | +33 39 29.87 | --- | 0.910 | 2.1 | 3.24 | 15.4 | 18 | 8.42 | 0.086 |
| 738 | K 3-67 | 04 39 47.905 | +36 45 42.85 | 0.119 | 1.456 | 2.2 | 3.74 | 33.3 | 42 | 6.49 | 0.069 |
| 740 | K 3-69 | 05 41 22.147 | +39 15 08.09 | 0.162 | 0.838 | 0.46 | 3.40 | 2.9 | 0.4 | 13.87 | 0.031 |
| 741 | K 3-7 | 18 34 13.601 | -02 27 36.41 | 0.287 | 1.770 | 6.3 | 3.85 | 29.5 | 30 | 4.52 | 0.138 |
| 746 | K 3-76 | 20 25 04.865 | +33 34 50.70 | --- | 0.539 | 0.2 | 3.76 | 5.1 | 12 | --- | --- |
| 747 | K 3-78 | 20 45 22.710 | +50 22 39.92 | 0.146 | 0.911 | 3.8 | 3.55 | 14.9 | 15 | 6.85 | 0.126 |
| 750 | K 3-82 | 21 30 51.636 | +50 00 06.96 | --- | 0.951 | 24 | 4.73 | 35.6 | 30 | 2.49 | 0.290 |
| 751 | K 3-83 | 21 35 43.877 | +50 54 16.94 | --- | 0.172 | 5 | 1.92 | 4.7 | 6.5 | 7.71 | 0.187 |
| 753 | K 3-87 | 22 55 06.987 | +56 42 31.14 | --- | 0.181 | 6 | 2.45 | 10.6 | 4.5 | 7.91 | 0.230 |
| 755 | K 3-90 | 01 24 58.628 | +65 38 36.14 | --- | 0.339 | 8.5 | 3.25 | 12.4 | 13.9 | 4.6 | 0.190 |
| 761 | K 4-19 | 19 13 22.624 | +03 25 00.30 | --- | 0.708 | --- | 3.33 | --- | --- | --- | --- |
| 762 | K 4-28 | 19 30 16.668 | +14 47 21.84 | 0.129 | 2.229 | 0.6 | 3.82 | 5 | 19 | 14.27 | 0.042 |

| KN | NAME | R.A. (2000.0) H:M:S | DEC (2000.0) D:M:S | F(9μm) Jy | F(18μm) Jy | θ arcsec | MN_18 arcsec | F(1.4GHz) mJy | F(5GHz) 5GHz | DIST. Kpc | DIAM. pc |
|---|---|---|---|---|---|---|---|---|---|---|---|
| 764 | K 4-41 | 19 56 34.027 | +32 22 12.96 | --- | 0.930 | 3 | 3.58 | 11.5 | 15 | 7.59 | 0.110 |
| 765 | K 4-47 | 04 20 45.157 | +56 18 11.57 | 0.264 | 1.525 | 7.8 | 3.70 | --- | 7.7 | 22.46 | 0.849 |
| 766 | K 4-48 | 06 39 55.879 | +11 06 30.69 | --- | 0.707 | 2.2 | 3.11 | 11.8 | 14 | 8.86 | 0.095 |
| 770 | K 4-57 | 22 48 34.363 | +58 29 08.18 | 1.460 | 1.081 | --- | 3.41 | --- | --- | --- | --- |
| 771 | K 4-8 | 18 54 20.003 | -08 47 33.17 | --- | 0.229 | --- | 2.91 | 3.1 | --- | --- | --- |
| 773 | KjPn 2 | 20 15 22.209 | +40 34 44.76 | 0.093 | 0.661 | --- | 3.36 | 7.3 | --- | --- | --- |
| 781 | LoTr 5 | 12 55 33.745 | +25 53 30.60 | 0.137 | --- | 10.6 | --- | --- | 5.6 | 8.69 | 0.447 |
| 798 | MA 13 | 18 30 30.388 | -07 27 38.29 | 0.182 | --- | --- | --- | 28.7 | --- | --- | --- |
| 800 | MA 3 | 18 17 49.379 | -06 48 21.55 | --- | 0.564 | --- | 3.19 | 6.3 | --- | --- | --- |
| 801 | MGP 1 | 16 48 48.548 | -35 00 57.39 | --- | 0.601 | 9 | 3.26 | 14.7 | 10.2 | --- | --- |
| 802 | M 1-1 | 01 37 19.430 | +50 28 11.50 | --- | 0.433 | 4.5 | 2.92 | 14 | 8 | 7.27 | 0.159 |
| 803 | M 1-11 | 07 11 16.693 | -19 51 02.87 | 6.754 | 31.786 | 2.2 | 5.52 | 25.9 | 113 | 4.9 | 0.052 |
| 804 | M 1-12 | 07 19 21.471 | -21 43 55.46 | 1.030 | 5.606 | 1.8 | 4.21 | 21.9 | 41 | 7.13 | 0.062 |
| 805 | M 1-13 | 07 21 14.952 | -18 08 36.91 | --- | 0.502 | 10 | 3.31 | 18.9 | 15 | 4.51 | 0.219 |
| 807 | M 1-16 | 07 37 18.955 | -09 38 49.67 | 0.195 | 1.148 | 3.6 | 3.67 | 31.3 | 31 | 5.71 | 0.100 |
| 808 | M 1-17 | 07 40 22.206 | -11 32 29.81 | 0.204 | 0.982 | 2.5 | 3.57 | 18.3 | 17 | 8.23 | 0.100 |
| 810 | M 1-19 | 17 03 46.809 | -33 29 43.75 | --- | 1.881 | 2.8 | 3.80 | 23.2 | 20 | --- | --- |
| 811 | M 1-2 | 01 58 49.675 | +52 53 48.57 | 0.861 | 1.862 | 0.5 | 3.64 | --- | 10 | --- | --- |
| 814 | M 1-24 | 17 38 11.588 | -19 37 37.64 | 0.243 | 3.627 | <10 | 4.34 | 43.5 | --- | --- | --- |
| 815 | M 1-25 | 17 38 30.307 | -22 08 38.88 | 0.237 | --- | 3.2 | --- | 39.9 | 57 | 4.93 | 0.076 |
| 816 | M 1-26 | 17 45 57.653 | -30 12 00.58 | --- | 104.933 | 3.2 | 7.32 | 66 | 400 | 2.42 | 0.038 |
| 817 | M 1-27 | 17 46 45.450 | -33 08 35.06 | 0.446 | --- | 8 | --- | 64.9 | 63 | 3.74 | 0.145 |
| 818 | M 1-29 | 17 50 18.003 | -30 34 54.90 | --- | 4.343 | 8.2 | 4.21 | 91.5 | 97 | 3.3 | 0.131 |
| 820 | M 1-31 | 17 52 41.440 | -22 21 56.83 | 0.435 | 5.727 | 2.3 | 4.23 | 28.8 | 60.5 | --- | --- |
| 821 | M 1-32 | 17 56 19.980 | -16 29 03.95 | 1.481 | 8.938 | 9 | 4.63 | 70.1 | 61 | 3.17 | 0.138 |
| 822 | M 1-33 | 17 58 58.794 | -15 32 14.79 | 0.495 | 4.194 | 4 | 4.03 | 49 | 60 | 4.63 | 0.090 |
| 823 | M 1-34 | 18 01 22.193 | -33 17 43.08 | --- | 0.575 | 11.2 | 3.68 | 13.6 | 14.7 | 5.95 | 0.323 |
| 824 | M 1-35 | 18 03 39.255 | -26 43 33.47 | 0.514 | 3.572 | <25 | 4.06 | 54 | 54 | 4.63 | --- |
| 825 | M 1-37 | 18 05 25.801 | -28 22 04.19 | 0.208 | 3.508 | 2.5 | 4.06 | 10.9 | 15 | 8.44 | 0.102 |
| 826 | M 1-38 | 18 06 05.765 | -28 40 29.28 | --- | 3.049 | 3.5 | 4.06 | 14.9 | 24 | 6.58 | 0.112 |
| 827 | M 1-39 | 18 07 30.699 | -13 28 47.61 | 0.692 | --- | 4 | --- | 58.2 | 100 | 3.92 | 0.076 |
| 828 | M 1-4 | 03 41 43.428 | +52 17 00.28 | 0.284 | 1.806 | 6 | 3.89 | 77.8 | 90 | 3.39 | 0.099 |
| 829 | M 1-40 | 18 08 25.994 | -22 16 53.25 | 1.187 | 11.344 | 4.5 | 4.75 | 163.3 | 208 | 2.83 | 0.062 |
| 831 | M 1-42 | 18 11 05.028 | -28 58 59.33 | --- | 0.900 | 9 | 3.53 | 28.3 | 28.5 | 4.89 | 0.213 |
| 833 | M 1-44 | 18 16 17.365 | -27 04 32.47 | 0.336 | 0.841 | 4 | 3.37 | 9.4 | 9 | 9.24 | 0.179 |
| 834 | M 1-45 | 18 23 07.984 | -19 17 05.26 | --- | 3.090 | 2.5 | 4.04 | 15.1 | 19 | 7.43 | 0.090 |
| 835 | M 1-46 | 18 27 56.339 | -15 32 54.43 | 0.327 | --- | 11.5 | --- | 78.1 | 81 | 2.58 | 0.144 |
| 838 | M 1-5 | 05 46 50.000 | +24 22 02.32 | 0.860 | 4.338 | 2.3 | 3.98 | 41.1 | 71 | 5.49 | 0.061 |
| 839 | M 1-50 | 18 33 20.886 | -18 16 36.86 | 0.285 | 2.153 | 5.6 | 3.95 | 35.4 | 50 | 4.12 | 0.112 |
| 841 | M 1-52 | 18 33 58.542 | -14 52 25.02 | --- | 0.303 | 43.1 | 3.17 | 7.8 | 319 | 5.93 | 1.240 |
| 842 | M 1-53 | 18 35 48.267 | -17 36 08.71 | 0.126 | 1.422 | 6 | 4.16 | 19.6 | 8 | 3.93 | 0.114 |
| 843 | M 1-54 | 18 36 08.357 | -16 59 57.02 | 0.215 | 0.933 | 13 | 3.57 | 35.5 | 53 | 3.09 | 0.195 |
| 844 | M 1-55 | 18 36 42.550 | -21 48 59.09 | 1.613 | 0.475 | <25 | 3.07 | --- | 38 | --- | --- |
| 845 | M 1-56 | 18 37 46.250 | -17 05 46.55 | 0.195 | 2.641 | 1.5 | 3.84 | 14.6 | 22 | 9.48 | 0.069 |

Table 1 (cont.)

| KN | NAME | R.A. (2000.0) H:M:S | DEC (2000.0) D:M:S | F(9μm) Jy | F(18μm) Jy | θ arcsec | MN_18 arcsec | F(1.4GHz) mJy | F(5GHz) 5GHz | DIST. Kpc | DIAM. pc |
|---|---|---|---|---|---|---|---|---|---|---|---|
| 846 | M 1-57 | 18 40 20.248 | -10 39 47.17 | 0.413 | 5.209 | 8 | 4.37 | 53.9 | 70 | 3.21 | 0.125 |
| 847 | M 1-58 | 18 42 56.979 | -11 06 53.10 | 0.229 | --- | 6.4 | --- | 37.7 | 60 | 3.69 | 0.114 |
| 848 | M 1-59 | 18 43 20.178 | -09 04 48.64 | 0.899 | 3.844 | 4.8 | 4.03 | 90.1 | 108 | 3.54 | 0.082 |
| 849 | M 1-6 | 06 35 45.140 | -00 05 37.37 | 1.139 | 5.895 | 3 | 4.25 | 54.4 | 86 | 4.7 | 0.068 |
| 851 | M 1-61 | 18 45 55.072 | -14 27 37.93 | --- | 13.886 | 1.8 | 4.81 | 32.5 | 97 | 5.59 | 0.049 |
| 852 | M 1-62 | 18 50 26.030 | -22 34 22.54 | --- | 0.230 | 1.5 | 2.39 | 13.5 | 12.8 | --- | --- |
| 855 | M 1-65 | 18 56 33.639 | +10 52 10.05 | --- | 1.665 | 4 | 3.79 | 21 | 23 | 5.94 | 0.115 |
| 856 | M 1-66 | 18 58 26.247 | -01 03 45.70 | 0.215 | --- | 2.7 | --- | 45.7 | 59 | 5.39 | 0.071 |
| 857 | M 1-67 | 19 11 30.881 | +16 51 38.21 | 0.955 | --- | 120.4 | --- | 217.6 | 250 | 0.69 | 0.403 |
| 858 | M 1-69 | 19 13 53.961 | +03 37 41.90 | 0.230 | 1.813 | --- | 3.85 | 31.4 | --- | --- | --- |
| 859 | M 1-7 | 06 37 20.955 | +24 00 35.38 | --- | 0.422 | 11 | 3.18 | 18.5 | 13 | 4.5 | 0.240 |
| 860 | M 1-71 | 19 36 26.927 | +19 42 23.99 | 1.507 | 15.253 | 3 | 4.89 | 83.8 | 204 | --- | --- |
| 861 | M 1-72 | 19 41 33.975 | +17 45 17.71 | 0.547 | 6.554 | 0.7 | 4.25 | 5.3 | 30 | 12.54 | 0.043 |
| 862 | M 1-73 | 19 41 09.323 | +14 56 59.37 | 0.160 | 3.830 | 6 | 4.27 | 47.7 | 43 | 4.17 | 0.121 |
| 863 | M 1-74 | 19 42 18.874 | +15 09 08.16 | 0.292 | 3.285 | 1 | 3.99 | 7.9 | 29 | 10.15 | 0.049 |
| 864 | M 1-75 | 20 04 44.086 | +31 27 24.42 | 0.187 | 0.739 | 14 | 3.81 | 36.5 | 26.2 | 3.32 | 0.225 |
| 865 | M 1-77 | 21 19 07.360 | +46 18 47.24 | 1.874 | 6.154 | 8 | 5.23 | 29.4 | 25 | 4.29 | 0.166 |
| 866 | M 1-78 | 21 20 44.809 | +51 53 27.88 | 19.514 | 226.731 | 6 | 8.41 | 363.5 | 1103.9 | 1.62 | 0.047 |
| 868 | M 1-8 | 06 53 33.795 | +03 08 26.96 | --- | 0.303 | 18.4 | 2.92 | 16.3 | 23 | 3.06 | 0.273 |
| 869 | M 1-80 | 22 56 19.806 | +57 09 20.68 | --- | 0.336 | 8 | 2.96 | 17.9 | 25 | 4.29 | 0.166 |
| 870 | M 1-9 | 07 05 19.204 | +02 46 59.16 | 0.113 | 0.941 | 2.3 | 3.34 | 22.4 | 27 | 7.22 | 0.081 |
| 873 | M 2-12 | 17 24 01.451 | -25 59 23.16 | --- | 2.228 | 4.4 | 3.85 | 10.3 | 12.9 | --- | --- |
| 874 | M 2-13 | 17 28 34.188 | -13 26 20.82 | --- | 0.888 | 1.5 | 3.35 | 9.7 | 13 | 10.68 | 0.078 |
| 876 | M 2-15 | 17 46 54.449 | -16 17 24.80 | --- | 0.870 | 7.4 | 3.64 | 18.7 | 12.5 | 5.37 | 0.193 |
| 877 | M 2-16 | 17 52 34.362 | -32 45 51.10 | --- | 1.380 | 2.5 | 3.56 | 20.4 | 24.8 | 5.86 | 0.071 |
| 878 | M 2-17 | 17 52 04.896 | -17 36 05.23 | 0.107 | --- | 8 | --- | 10.6 | 10 | 5.57 | 0.216 |
| 879 | M 2-18 | 17 53 37.844 | -32 58 47.93 | --- | 1.349 | 1.5 | 3.69 | 11.1 | 17 | 9.4 | 0.068 |
| 880 | M 2-19 | 17 53 45.619 | -29 43 46.34 | 0.182 | 0.520 | 7.5 | 3.27 | 10.5 | 14 | 7.4 | 0.269 |
| 881 | M 2-2 | 04 13 15.045 | +56 56 58.08 | 0.310 | 1.828 | 6.5 | 4.57 | 52 | 54 | 3.66 | 0.115 |
| 882 | M 2-20 | 17 54 25.422 | -29 36 08.20 | 0.127 | 1.934 | 16.4 | 3.83 | 22.2 | 3.2 | --- | --- |
| 883 | M 2-21 | 17 58 09.565 | -29 44 20.08 | 0.305 | 0.789 | 3 | 3.42 | 20.3 | 23 | 7.01 | 0.102 |
| 884 | M 2-22 | 17 58 32.634 | -33 28 36.60 | --- | 0.490 | 5.2 | 2.92 | 9.1 | 6.5 | 9.87 | 0.249 |
| 885 | M 2-23 | 18 01 42.610 | -28 25 43.90 | --- | 6.155 | 8.8 | 3.98 | 4.2 | 41 | 8.38 | 0.358 |
| 886 | M 2-24 | 18 02 02.881 | -34 27 47.07 | --- | 3.190 | 6.8 | 3.94 | 3.2 | 3 | 11.79 | 0.389 |
| 888 | M 2-26 | 18 03 11.844 | -26 58 30.23 | --- | 0.398 | 7 | 2.76 | 7.9 | 5 | 9.43 | 0.320 |
| 889 | M 2-27 | 18 03 52.588 | -31 17 46.54 | --- | 3.369 | 3.8 | 4.00 | 22.5 | 50 | 5.65 | 0.104 |
| 890 | M 2-28 | 18 05 02.695 | -30 58 17.39 | --- | 0.508 | 5.3 | 3.22 | 11.3 | 10 | 8.39 | 0.216 |
| 891 | M 2-29 | 18 06 40.862 | -26 54 55.95 | --- | 1.179 | <5 | 4.08 | 9.4 | 8 | 9.98 | --- |
| 892 | M 2-30 | 18 12 34.414 | -27 58 11.59 | 0.131 | 0.924 | 3.5 | 3.40 | 13.9 | 14 | 8.1 | 0.137 |
| 893 | M 2-31 | 18 13 16.026 | -25 30 04.97 | 0.223 | 1.653 | 4 | 3.66 | 39.6 | 51 | 4.91 | 0.095 |
| 894 | M 2-32 | 18 14 50.617 | -32 36 55.21 | --- | 0.585 | <10 | 3.29 | 9.5 | --- | --- | --- |
| 895 | M 2-33 | 18 15 06.534 | -30 15 32.89 | --- | 1.138 | 5 | 3.77 | 11.9 | 12 | 8.52 | 0.207 |
| 898 | M 2-36 | 18 17 41.418 | -29 08 19.59 | 0.218 | --- | 6.8 | --- | 22.6 | 13 | 6.99 | 0.230 |
| 900 | M 2-38 | 18 19 25.162 | -26 35 19.92 | --- | 0.522 | 8 | 3.20 | 7.5 | 8 | 8.13 | 0.315 |

| KN | NAME | R.A. (2000.0) H:M:S | DEC (2000.0) D:M:S | F(9μm) Jy | F(18μm) Jy | θ arcsec | MN_18 arcsec | F(1.4GHz) mJy | F(5GHz) 5GHz | DIST. Kpc | DIAM. pc |
|---|---|---|---|---|---|---|---|---|---|---|---|
| 901 | M 2-39 | 18 22 01.148 | -24 10 40.18 | 0.211 | 0.802 | 3.2 | 3.37 | 7.2 | 8 | 10.21 | 0.158 |
| 902 | M 2-4 | 17 01 06.231 | -34 49 38.58 | 0.194 | 2.996 | 2 | 3.87 | 24.5 | 32 | 7.15 | 0.069 |
| 903 | M 2-40 | 18 21 23.851 | -06 01 55.79 | --- | 3.620 | 5.5 | 4.19 | 45.5 | 33 | 4.67 | 0.125 |
| 905 | M 2-42 | 18 22 32.020 | -24 09 28.40 | --- | 0.455 | 3.9 | 3.08 | 9.6 | 14 | 7.83 | 0.148 |
| 906 | M 2-43 | 18 26 40.048 | -02 42 57.63 | 9.659 | 32.600 | 1.5 | 5.68 | 20.4 | 237 | 5 | 0.036 |
| 907 | M 2-44 | 18 37 36.908 | -03 05 55.96 | 0.385 | 2.105 | 8 | 3.89 | 48.3 | 54 | 3.45 | 0.134 |
| 908 | M 2-45 | 18 39 21.837 | -04 19 50.90 | 0.760 | 6.161 | 6.4 | 4.28 | 107 | 154 | 2.64 | 0.082 |
| 909 | M 2-46 | 18 46 34.620 | -08 28 01.85 | 0.111 | 0.455 | 4.4 | 3.12 | 13.1 | 12 | 6.48 | 0.138 |
| 910 | M 2-47 | 19 13 34.559 | +04 38 04.45 | 0.180 | 2.890 | 6 | 4.04 | 38.3 | 45 | 4.12 | 0.120 |
| 911 | M 2-48 | 19 50 28.543 | +25 54 30.21 | --- | 0.376 | 4 | 2.79 | 16.9 | 19 | 7 | 0.136 |
| 912 | M 2-49 | 21 43 17.624 | +50 25 14.57 | 0.408 | 1.202 | 2.5 | 3.50 | 28.2 | 35 | 6.47 | 0.078 |
| 913 | M 2-5 | 17 02 19.069 | -33 10 05.01 | 0.202 | --- | 5 | --- | 14.1 | 12 | 7.69 | 0.186 |
| 914 | M 2-50 | 21 57 41.814 | +51 41 39.01 | --- | 0.183 | 4.6 | 2.45 | 8.2 | 6.5 | 8.07 | 0.180 |
| 916 | M 2-52 | 22 20 30.745 | +57 36 21.62 | --- | 0.133 | 13.5 | 2.39 | 14.9 | 14 | 3.97 | 0.260 |
| 917 | M 2-53 | 22 32 17.720 | +56 10 26.12 | --- | 0.291 | 18 | 2.74 | 14.6 | 11 | 3.64 | 0.318 |
| 918 | M 2-54 | 22 51 38.923 | +51 50 42.50 | 0.220 | 6.060 | --- | 4.19 | 5.4 | 8 | 14.61 | --- |
| 920 | M 2-6 | 17 04 18.326 | -30 53 28.71 | --- | 0.751 | 2 | 3.15 | 13.6 | 17 | --- | --- |
| 922 | M 2-8 | 17 05 30.673 | -32 32 08.30 | 0.231 | 0.921 | 4 | 3.50 | 16.2 | 18 | 7.27 | 0.141 |
| 923 | M 2-9 | 17 05 37.952 | -10 08 34.58 | 38.070 | 73.255 | 17.5 | 6.93 | 37.1 | 36 | 3.01 | 0.255 |
| 924 | M 3-1 | 07 02 49.980 | -31 35 31.50 | --- | 0.210 | 11.2 | 2.71 | 25.7 | 24 | 3.78 | 0.205 |
| 925 | M 3-10 | 17 27 20.152 | -28 27 51.15 | 0.311 | 2.611 | 3.1 | 3.81 | 35.2 | 29 | 6.33 | 0.095 |
| 926 | M 3-12 | 17 36 22.634 | -21 31 12.22 | --- | 0.616 | 7.5 | 3.01 | 11.2 | 12.5 | 6.98 | 0.254 |
| 927 | M 3-13 | 17 41 36.594 | -22 13 02.46 | 0.862 | 1.893 | <7 | 3.73 | --- | --- | --- | --- |
| 928 | M 3-14 | 17 44 20.598 | -34 06 40.59 | 0.344 | 1.890 | 4 | 3.91 | 22.6 | 30 | 6.47 | 0.125 |
| 930 | M 3-17 | 17 56 25.640 | -31 04 16.82 | --- | 1.075 | <5 | 3.43 | 10 | 12 | 9 | --- |
| 931 | M 3-19 | 17 58 19.337 | -30 00 39.32 | --- | 0.800 | 7 | 3.50 | 7.9 | 7.4 | 8.62 | 0.293 |
| 933 | M 3-20 | 17 59 19.318 | -28 13 48.05 | 0.091 | --- | 4 | --- | 16.2 | 40 | 5.3 | 0.103 |
| 934 | M 3-21 | 18 02 32.324 | -36 39 12.24 | 0.265 | 2.617 | <5 | 3.92 | 15.6 | 30 | 5.57 | --- |
| 935 | M 3-22 | 18 02 19.238 | -30 14 25.38 | --- | 0.451 | 6.4 | 3.15 | 10.1 | 8.7 | 8.45 | 0.262 |
| 936 | M 3-23 | 18 07 06.148 | -30 34 16.96 | 0.231 | 1.245 | 11.5 | 3.85 | 30 | 28 | 4.58 | 0.255 |
| 937 | M 3-24 | 18 07 53.914 | -25 24 02.71 | --- | 0.887 | <25 | 3.40 | 17.4 | --- | --- | --- |
| 938 | M 3-25 | 18 15 16.967 | -10 10 09.47 | 0.831 | --- | 1.5 | --- | 25 | 76 | 6.89 | 0.050 |
| 939 | M 3-26 | 18 16 11.405 | -27 14 57.46 | --- | 0.479 | 7 | 2.96 | 8.1 | 8 | 8.4 | 0.285 |
| 940 | M 3-27 | 18 27 48.273 | +14 29 06.06 | 0.210 | 0.996 | 1 | 3.55 | --- | 53.2 | 8.55 | 0.041 |
| 941 | M 3-28 | 18 32 41.288 | -10 05 50.03 | --- | 1.446 | 9 | 5.11 | --- | 33 | 4.87 | 0.212 |
| 945 | M 3-31 | 18 44 01.767 | -19 54 52.52 | --- | 1.205 | <10 | 3.44 | --- | 9 | 14.79 | --- |
| 946 | M 3-32 | 18 44 43.127 | -25 21 33.85 | 0.159 | 0.705 | 7.5 | 3.27 | 12.2 | 12 | 7.06 | 0.257 |
| 947 | M 3-33 | 18 48 12.131 | -25 28 52.39 | --- | 0.470 | 6 | 3.32 | 9.8 | 7.5 | 8.94 | 0.260 |
| 948 | M 3-34 | 19 27 01.897 | -06 35 04.63 | 0.154 | 1.764 | 8 | 3.91 | 29.6 | 29 | 4.12 | 0.160 |
| 949 | M 3-35 | 20 21 03.769 | +32 29 23.86 | 3.437 | 16.454 | 1.5 | 4.97 | 29.5 | 140 | 5.62 | 0.041 |
| 950 | M 3-36 | 17 12 39.152 | -25 43 37.39 | --- | 0.577 | 3.2 | 3.22 | 4.9 | 3.5 | 13.92 | 0.216 |
| 953 | M 3-39 | 17 21 11.505 | -27 11 38.13 | 1.177 | --- | 19 | --- | 279.4 | 280 | 1.7 | 0.157 |
| 956 | M 3-41 | 17 25 59.784 | -29 21 50.15 | --- | 1.007 | <25 | 3.35 | 11.3 | 75 | 4.15 | --- |
| 958 | M 3-43 | 17 50 24.297 | -29 25 18.70 | --- | 2.050 | 3.8 | 3.64 | --- | 27.2 | 6.2 | 0.114 |

| KN | NAME | R.A. (2000.0) H:M:S | DEC (2000.0) D:M:S | F(9μm) Jy | F(18μm) Jy | θ arcsec | MN_18 arcsec | F(1.4GHz) mJy | F(5GHz) 5GHz | DIST. Kpc | DIAM. pc |
|---|---|---|---|---|---|---|---|---|---|---|---|
| 959 | M 3-44 | 17 51 18.894 | -30 23 52.99 | 0.342 | 6.899 | 4.2 | 4.35 | 24.3 | 35 | --- | --- |
| 960 | M 3-45 | 17 52 05.927 | -30 05 13.81 | --- | 0.614 | 3.5 | 3.00 | 20.4 | 24.3 | --- | --- |
| 965 | M 3-5 | 08 02 28.932 | -27 41 55.44 | --- | 0.286 | 7 | 2.91 | 11 | 10 | 5.9 | 0.200 |
| 969 | M 3-53 | 18 24 07.890 | -11 06 42.08 | 0.233 | 1.240 | --- | 3.61 | 19.5 | --- | --- | --- |
| 970 | M 3-54 | 18 33 03.745 | -13 44 19.95 | --- | 0.422 | 5.6 | 3.01 | 8.6 | 8 | 6.92 | 0.188 |
| 972 | M 3-6 | 08 40 40.205 | -32 22 33.22 | 0.407 | 5.983 | 11 | 4.95 | 77.4 | 75 | 2.74 | 0.146 |
| 973 | M 3-7 | 17 24 34.429 | -29 24 19.47 | 0.096 | 1.553 | 5 | 3.08 | 32.9 | 28 | 5.41 | 0.131 |
| 974 | M 3-8 | 17 24 52.152 | -28 05 54.61 | 0.297 | --- | 3.2 | --- | 18 | 20 | 7.45 | 0.116 |
| 975 | M 3-9 | 17 25 43.364 | -26 11 55.47 | 0.364 | 2.220 | 16 | 3.81 | 37.3 | 35 | 3.88 | 0.301 |
| 976 | M 4-10 | 18 34 13.821 | -13 12 24.57 | --- | 1.920 | 1.2 | 3.81 | 12.1 | 33 | 9.04 | 0.053 |
| 979 | M 4-17 | 20 09 01.927 | +43 43 43.54 | --- | 0.447 | --- | 3.36 | 20 | --- | --- | --- |
| 980 | M 4-18 | 04 25 50.831 | +60 07 12.72 | 1.687 | 6.346 | 3.75 | 4.22 | 18.5 | 22 | 6.37 | 0.116 |
| 981 | M 4-2 | 07 28 53.808 | -35 45 13.92 | --- | 0.481 | 6 | 3.25 | 24.2 | 19 | 5.26 | 0.153 |
| 983 | M 4-4 | 17 28 50.344 | -30 07 44.47 | 0.088 | --- | --- | --- | 13.9 | 9.8 | --- | --- |
| 985 | M 4-7 | 17 51 44.663 | -31 36 00.18 | --- | 0.775 | 5.7 | 3.18 | 38 | 33 | 5.2 | 0.144 |
| 986 | M 4-8 | 18 12 09.587 | -10 42 58.30 | 0.117 | 2.155 | 1.4 | 3.67 | 10.7 | 19 | 10.21 | 0.069 |
| 989 | MaC 1-11 | 18 14 50.890 | -22 43 55.43 | --- | 0.399 | --- | 3.01 | 12.2 | --- | --- | --- |
| 990 | MaC 1-13 | 18 28 35.242 | -08 43 22.82 | --- | 0.683 | --- | 3.50 | 30.4 | --- | --- | --- |
| 1000 | MeWe 1-7 | 16 47 57.070 | -50 42 48.32 | --- | 0.356 | --- | 3.30 | --- | --- | --- | --- |
| 1003 | Me 1-1 | 19 39 09.813 | +15 56 48.13 | 0.450 | 0.931 | 4.7 | 3.46 | 36 | 45.1 | 4.58 | 0.104 |
| 1004 | Me 2-1 | 15 22 19.274 | -23 37 31.34 | --- | 0.930 | 7 | 3.55 | 33.6 | 30 | 4.32 | 0.147 |
| 1005 | Me 2-2 | 22 31 43.686 | +47 48 03.96 | 0.606 | 2.556 | 1.2 | 3.85 | 16 | 40 | 8.56 | 0.050 |
| 1006 | MyCn 18 | 13 39 35.116 | -67 22 51.45 | 0.862 | 10.869 | 12.6 | 4.66 | --- | 106 | 3.23 | 0.197 |
| 1007 | My 60 | 10 31 33.390 | -55 20 50.87 | 0.244 | 1.646 | 7.6 | 4.29 | --- | 60 | 3.47 | 0.128 |
| 1009 | Mz 2 | 16 14 32.433 | -54 57 03.76 | --- | 1.763 | 23 | 5.50 | --- | 75 | 2 | 0.223 |
| 1010 | Mz 3 | 16 17 13.392 | -51 59 10.31 | 55.324 | 267.948 | 25.4 | 8.95 | --- | 649 | 1.03 | 0.127 |
| 1014 | NGC 1535 | 04 14 15.762 | -12 44 22.03 | 1.037 | 5.562 | 18.4 | 8.80 | 167.8 | 160 | 1.77 | 0.158 |
| 1015 | NGC 2022 | 05 42 06.229 | +09 05 10.75 | 0.371 | 3.789 | 19.4 | 6.04 | 92.3 | 91 | 2.02 | 0.190 |
| 1017 | NGC 2346 | 07 09 22.547 | -00 48 22.98 | 0.505 | 0.595 | 54.6 | 3.64 | 37.6 | 86 | 2.07 | 0.548 |
| 1019 | NGC 2392 | 07 29 10.768 | +20 54 42.44 | 0.476 | 4.368 | 44.8 | 6.48 | 279.9 | 251 | 1.44 | 0.313 |
| 1021 | NGC 2452 | 07 47 26.248 | -27 20 07.16 | 0.288 | 1.798 | 18.8 | 4.71 | 56 | 57 | 2.31 | 0.211 |
| 1024 | NGC 2792 | 09 12 26.596 | -42 25 39.90 | 0.399 | 3.618 | 13 | 5.13 | --- | 116 | 1.26 | 0.079 |
| 1026 | NGC 2867 | 09 21 25.337 | -58 18 40.69 | 1.312 | 7.277 | 16 | 4.98 | --- | 252 | 1.69 | 0.131 |
| 1028 | NGC 3132 | 10 07 01.771 | -40 26 11.11 | 0.883 | 3.437 | 45 | 8.18 | --- | 230 | 1.5 | 0.327 |
| 1029 | NGC 3195 | 10 09 20.910 | -80 51 30.73 | --- | 0.933 | 40 | 5.07 | --- | 35 | 1.96 | 0.380 |
| 1030 | NGC 3211 | 10 17 50.549 | -62 40 14.57 | 0.286 | 2.113 | 16 | 4.52 | --- | 228 | 1.7 | 0.132 |
| 1031 | NGC 3242 | 10 24 46.138 | -18 38 32.26 | 3.211 | 19.833 | 37.2 | 8.87 | 757.7 | 860 | 0.94 | 0.170 |
| 1034 | NGC 3918 | 11 50 17.730 | -57 10 56.90 | 2.747 | 27.613 | 18.8 | 6.14 | --- | 859 | 1.17 | 0.107 |
| 1035 | NGC 40 | 00 13 01.023 | +72 31 19.07 | --- | 54.443 | 36.4 | 9.48 | 510.3 | 460 | 0.98 | 0.173 |
| 1037 | NGC 4361 | 12 24 30.754 | -18 47 05.51 | --- | 4.061 | 81 | 9.86 | 159.9 | 207 | 0.74 | 0.291 |
| 1039 | NGC 5307 | 13 51 03.273 | -51 12 20.62 | 0.260 | 2.349 | 12.6 | 4.41 | --- | 95 | 2.42 | 0.148 |
| 1040 | NGC 5315 | 13 53 56.972 | -66 30 50.96 | 3.477 | 34.483 | 6 | 5.93 | --- | 442 | 2.14 | 0.062 |
| 1042 | NGC 5882 | 15 16 49.941 | -45 38 58.60 | 2.331 | 17.353 | 14 | 5.78 | --- | 334 | 1.74 | 0.118 |
| 1048 | NGC 6210 | 16 44 29.521 | +23 47 59.55 | 1.429 | 13.331 | 16.2 | 5.64 | 297.8 | 311 | 1.55 | 0.122 |

| KN | NAME | R.A. (2000.0) H:M:S | DEC (2000.0) D:M:S | F(9μm) Jy | F(18μm) Jy | θ arcsec | MN_18 arcsec | F(1.4GHz) mJy | F(5GHz) 5GHz | DIST. Kpc | DIAM. pc |
|---|---|---|---|---|---|---|---|---|---|---|---|
| 1050 | NGC 6309 | 17 14 04.322 | -12 54 37.71 | --- | 6.375 | 17 | 5.63 | 131.5 | 151 | 2.35 | 0.194 |
| 1051 | NGC 6326 | 17 20 46.301 | -51 45 15.33 | 0.248 | 1.149 | 12 | 4.03 | --- | 70 | 2.71 | 0.158 |
| 1053 | NGC 6369 | 17 29 20.443 | -23 45 34.22 | 7.301 | 43.871 | 28 | 8.47 | 1825 | 2002 | 0.75 | 0.102 |
| 1054 | NGC 6439 | 17 48 19.797 | -16 28 44.21 | 0.370 | --- | 5 | --- | 51.5 | 55 | 4.64 | 0.112 |
| 1056 | NGC 6537 | 18 05 13.059 | -19 50 34.86 | 4.099 | 26.593 | 4.7 | 5.63 | 427.5 | 640 | 1.77 | 0.040 |
| 1057 | NGC 6543 | 17 58 33.409 | +66 37 58.79 | 5.539 | 57.323 | 18.8 | 7.39 | 771.5 | 850 | 1.12 | 0.102 |
| 1059 | NGC 6565 | 18 11 52.470 | -28 10 42.27 | --- | 1.057 | 9.2 | 3.88 | 47 | 37 | 4.36 | 0.194 |
| 1060 | NGC 6567 | 18 13 45.136 | -19 04 33.67 | 1.778 | 4.466 | 8.8 | 4.19 | 163 | 168 | 2.4 | 0.102 |
| 1061 | NGC 6572 | 18 12 06.404 | +06 51 12.17 | 9.597 | 93.442 | 10 | 7.06 | 557.5 | 1374 | 1.06 | 0.051 |
| 1062 | NGC 6578 | 18 16 16.517 | -20 27 02.67 | 1.357 | --- | 8.6 | --- | 158.2 | 170 | 2.31 | 0.096 |
| 1063 | NGC 6620 | 18 22 54.186 | -26 49 17.00 | 0.146 | 0.802 | 5 | 3.43 | 17.1 | 20.5 | 6.42 | 0.156 |
| 1064 | NGC 6629 | 18 25 42.449 | -23 12 10.59 | 0.799 | --- | 15 | --- | 257.8 | 275 | 1.84 | 0.134 |
| 1065 | NGC 6644 | 18 32 34.641 | -25 07 44.00 | --- | 5.547 | 2.8 | 5.22 | 64.3 | 97 | 4.09 | 0.056 |
| 1067 | NGC 6741 | 19 02 37.088 | -00 26 56.97 | 0.682 | 4.923 | 8 | 4.25 | 131.9 | 230 | 2.32 | 0.090 |
| 1068 | NGC 6751 | 19 05 55.560 | -05 59 32.92 | 1.398 | 11.455 | 21 | 5.54 | 55.1 | 63 | 2.18 | 0.222 |
| 1071 | NGC 6778 | 19 18 24.939 | -01 35 47.41 | 0.463 | 1.644 | 15.8 | 4.39 | 64.6 | 55 | 2.54 | 0.195 |
| 1073 | NGC 6790 | 19 22 56.965 | +01 30 46.45 | 6.636 | 21.477 | 1.8 | 5.22 | 52.6 | 290 | 4.1 | 0.036 |
| 1074 | NGC 6803 | 19 31 16.490 | +10 03 21.88 | 0.603 | 5.242 | 5 | 4.15 | 68.9 | 114 | 3.26 | 0.079 |
| 1076 | NGC 6807 | 19 34 33.543 | +05 41 02.58 | 0.220 | 2.850 | 0.8 | 3.94 | 8 | 29 | 11.18 | 0.043 |
| 1077 | NGC 6818 | 19 43 57.844 | -14 09 11.91 | 0.851 | 6.010 | 18.2 | 5.55 | 290.2 | 300 | 1.47 | 0.130 |
| 1078 | NGC 6826 | 19 44 48.161 | +50 31 30.33 | 2.539 | 24.006 | 25.4 | 8.75 | 414.4 | 385 | 1.2 | 0.148 |
| 1083 | NGC 6879 | 20 10 26.682 | +16 55 21.30 | 0.111 | 0.831 | 5 | 3.59 | 23.1 | 18 | 5.78 | 0.140 |
| 1084 | NGC 6881 | 20 10 52.464 | +37 24 41.18 | 0.972 | 9.369 | 2.6 | 4.50 | 70.4 | 120 | 4.48 | 0.056 |
| 1085 | NGC 6884 | 20 10 23.669 | +46 27 39.54 | 0.743 | 6.356 | 5.3 | 4.46 | 152.4 | 200 | 2.7 | 0.069 |
| 1086 | NGC 6886 | 20 12 42.813 | +19 59 22.65 | 0.543 | 4.595 | 5.5 | 4.24 | 77.6 | 108 | 2.94 | 0.078 |
| 1087 | NGC 6891 | 20 15 08.838 | +12 42 15.63 | 0.345 | 4.865 | 10.2 | 5.14 | 110.8 | 105 | 2.35 | 0.116 |
| 1089 | NGC 6905 | 20 22 22.940 | +20 06 16.80 | --- | 2.166 | 40.4 | 6.79 | 66.7 | 52 | 1.62 | 0.317 |
| 1091 | NGC 7009 | 21 04 10.877 | -11 21 48.25 | 5.275 | 31.698 | 28.2 | 7.24 | 682 | 750 | 1.09 | 0.149 |
| 1093 | NGC 7027 | 21 07 01.696 | +42 14 09.50 | 173.674 | 977.249 | 12.5 | 9.36 | 1362.4 | 5200 | 0.64 | 0.039 |
| 1098 | NGC 7354 | 22 40 19.940 | +61 17 08.10 | 2.748 | 15.223 | 20 | 6.85 | 581 | 579 | 1.19 | 0.115 |
| 1099 | NGC 7662 | 23 25 53.968 | +42 32 05.04 | 2.377 | 15.165 | 20 | 6.74 | 611.5 | 631 | 1.17 | 0.113 |
| 1100 | Na 1 | 17 12 51.900 | -03 16 00.13 | 0.083 | 0.569 | 8 | 3.37 | 23 | 22.5 | 4.2 | 0.163 |
| 1105 | PB 1 | 07 02 46.764 | -13 42 34.65 | --- | 0.639 | --- | 3.92 | 11.9 | 18 | --- | --- |
| 1106 | PB 10 | 19 28 14.391 | +12 19 36.17 | 0.322 | 2.045 | 8 | 3.94 | 50.4 | 50 | 3.53 | 0.137 |
| 1107 | PB 2 | 08 20 40.188 | -46 22 58.82 | 0.164 | 0.562 | 3 | 3.18 | --- | 40 | 5.75 | 0.084 |
| 1108 | PB 3 | 08 54 18.323 | -50 32 22.34 | 0.256 | 1.635 | 7 | 3.84 | --- | 70 | 3.4 | 0.115 |
| 1109 | PB 4 | 09 15 07.747 | -54 52 43.78 | 0.243 | 1.444 | 11.2 | 4.12 | --- | 71 | 2.81 | 0.153 |
| 1110 | PB 5 | 09 16 09.613 | -45 28 42.79 | 3.014 | 11.328 | 5 | 4.71 | --- | 107 | --- | --- |
| 1111 | PB 6 | 10 13 15.949 | -50 19 59.28 | 0.330 | 3.205 | 11 | 4.18 | --- | 30 | 3.55 | 0.189 |
| 1112 | PB 8 | 11 33 17.717 | -57 06 14.00 | 0.160 | 2.536 | 5 | 4.06 | --- | 27 | 5.15 | 0.125 |
| 1113 | PB 9 | 19 27 44.814 | +10 24 20.82 | 0.103 | 1.029 | 7 | 3.74 | 33.1 | 40 | 3.57 | 0.121 |
| 1114 | PC 11 | 16 37 42.697 | -55 42 26.49 | 1.164 | 5.760 | 5 | 4.03 | --- | 11 | --- | --- |
| 1115 | PC 12 | 16 43 53.781 | -18 57 11.89 | --- | 1.994 | 2.2 | 3.80 | 14.8 | 19 | 8.6 | 0.092 |
| 1117 | PC 14 | 17 06 14.767 | -52 30 00.40 | 0.136 | 0.809 | 7 | 3.64 | --- | 30 | 4.32 | 0.147 |

| KN | NAME | R.A. (2000.0) H:M:S | DEC (2000.0) D:M:S | F(9μm) Jy | F(18μm) Jy | θ arcsec | MN_18 arcsec | F(1.4GHz) mJy | F(5GHz) 5GHz | DIST. Kpc | DIAM. pc |
|---|---|---|---|---|---|---|---|---|---|---|---|
| 1118 | PC 17 | 17 35 41.677 | -46 59 48.54 | --- | 1.063 | 5 | 3.58 | --- | 14.7 | 6.12 | 0.148 |
| 1120 | PC 20 | 18 43 03.456 | -00 16 37.02 | 0.212 | 1.123 | --- | 3.59 | 27.3 | --- | --- | --- |
| 1122 | PC 22 | 19 42 03.514 | +13 50 37.33 | --- | 0.458 | --- | 3.73 | 9.2 | --- | --- | --- |
| 1123 | PC 23 | 19 51 52.770 | +32 59 17.48 | 0.053 | 0.691 | 2.2 | 3.27 | 10 | 21 | 8.06 | 0.086 |
| 1124 | PC 24 | 20 19 38.133 | +27 00 11.23 | 0.184 | 0.953 | 5 | 3.52 | 17.1 | 18 | 5.78 | 0.140 |
| 1127 | PM 1-188 | 17 54 21.100 | -15 55 51.78 | 6.838 | 4.779 | --- | 3.92 | --- | --- | --- | --- |
| 1128 | PM 1-276 | 19 02 17.857 | +10 17 34.49 | --- | 1.065 | --- | 4.78 | 15.3 | --- | --- | --- |
| 1129 | PM 1-295 | 19 19 18.760 | +17 11 48.09 | --- | 2.020 | --- | 5.32 | 13.9 | --- | --- | --- |
| 1130 | PM 1-310 | 19 38 52.135 | +25 05 32.63 | 0.400 | --- | --- | --- | --- | --- | --- | --- |
| 1132 | PM 1-89 | 15 19 08.724 | -53 09 50.05 | 0.584 | 6.983 | --- | 4.53 | --- | --- | --- | --- |
| 1136 | Pe 1-1 | 10 38 27.607 | -56 47 06.40 | 0.846 | 8.744 | 3 | 4.48 | --- | 125.3 | 4.16 | 0.061 |
| 1137 | Pe 1-11 | 18 01 42.767 | -33 15 26.30 | --- | 0.779 | 9 | 3.51 | 10.9 | 8 | 7.9 | 0.345 |
| 1141 | Pe 1-15 | 18 46 24.485 | -07 14 34.57 | 0.079 | 0.722 | 5 | 3.49 | 8.3 | 8 | 7.4 | 0.179 |
| 1142 | Pe 1-16 | 18 47 32.251 | -06 54 03.52 | --- | 0.555 | --- | 3.33 | 14.2 | --- | --- | --- |
| 1144 | Pe 1-18 | 18 48 46.454 | -05 56 07.71 | 0.486 | --- | 1.1 | --- | 6.6 | 42 | 8.44 | 0.045 |
| 1145 | Pe 1-19 | 18 49 44.643 | -07 01 35.03 | --- | 0.366 | 3.9 | 3.15 | 4.8 | 6 | 8.42 | 0.158 |
| 1146 | Pe 1-2 | 10 39 32.690 | -57 06 13.70 | 0.269 | 2.818 | --- | 4.10 | --- | --- | --- | --- |
| 1147 | Pe 1-20 | 18 57 17.351 | -05 59 51.79 | --- | 0.191 | 8.4 | 2.51 | 5.1 | 29.4 | 4.52 | 0.183 |
| 1148 | Pe 1-21 | 18 57 49.642 | -05 27 39.73 | --- | 0.396 | 8.6 | 2.81 | 6.3 | 30 | 3.95 | 0.165 |
| 1151 | Pe 1-7 | 16 30 25.852 | -46 02 51.10 | 2.801 | 23.646 | 5 | 5.27 | --- | 117 | --- | --- |
| 1152 | Pe 1-8 | 17 06 22.558 | -44 13 09.96 | --- | 2.470 | --- | 5.23 | --- | --- | --- | --- |
| 1158 | Pe 2-14 | 18 29 59.545 | -19 40 37.74 | --- | 0.527 | --- | 3.13 | 3.7 | --- | --- | --- |
| 1160 | Pe 2-4 | 09 30 48.402 | -53 09 59.33 | 0.110 | 0.663 | --- | 3.34 | --- | --- | --- | --- |
| 1162 | Pe 2-7 | 10 41 19.574 | -56 09 16.34 | 0.043 | 0.511 | --- | 3.18 | --- | --- | --- | --- |
| 1163 | Pe 2-8 | 15 23 42.860 | -57 09 25.02 | 6.342 | 39.225 | 1.6 | 5.99 | --- | 100 | 5.83 | 0.045 |
| 1167 | SaSt 1-1 | 08 31 42.877 | -27 45 31.70 | --- | 1.819 | --- | 3.77 | --- | 0.3 | --- | --- |
| 1168 | SaSt 2-12 | 17 03 02.870 | -53 55 54.16 | 0.175 | 1.161 | --- | 3.59 | --- | --- | --- | --- |
| 1169 | SaSt 2-3 | 07 48 03.683 | -14 07 40.40 | --- | 0.151 | --- | 2.00 | 4 | 1.4 | --- | --- |
| 1173 | Sa 1-5 | 17 11 27.372 | -47 25 01.58 | --- | 0.364 | --- | 2.40 | --- | --- | --- | --- |
| 1175 | Sa 1-8 | 18 50 44.312 | -13 31 02.18 | --- | 1.150 | 5.6 | 3.80 | 13.5 | 11 | 6.33 | 0.172 |
| 1178 | Sa 2-237 | 17 44 42.282 | -15 45 11.45 | 0.703 | 2.588 | --- | 4.00 | 5.4 | --- | --- | --- |
| 1179 | Sa 3-134 | 18 29 19.823 | -15 07 39.97 | 0.083 | 0.927 | --- | 3.54 | 2.7 | --- | --- | --- |
| 1185 | ShWi 2-5 | 18 03 53.660 | -29 51 21.89 | 0.468 | 0.800 | --- | 3.16 | --- | --- | --- | --- |
| 1191 | Sn 1 | 16 21 04.421 | -00 16 10.54 | --- | 0.426 | 3 | 2.95 | 10.4 | 7 | 9.42 | 0.137 |
| 1193 | Sp 3 | 18 07 15.793 | -51 01 10.41 | --- | 1.411 | 35.6 | 5.04 | --- | 61 | 1.75 | 0.302 |
| 1196 | StWr 4-10 | 16 02 13.042 | -41 33 35.94 | --- | 0.373 | --- | 2.85 | --- | --- | --- | --- |
| 1198 | Ste 2-1 | 10 11 57.662 | -52 38 17.09 | --- | 0.257 | --- | 2.91 | --- | --- | --- | --- |
| 1201 | SwSt 1 | 18 16 12.237 | -30 52 08.71 | 9.268 | 52.915 | 1.3 | 6.25 | 26.7 | 216 | 3.79 | 0.024 |
| 1203 | Tc 1 | 17 45 35.298 | -46 05 23.81 | 1.197 | 5.695 | 15 | 5.81 | --- | 801 | 1.39 | 0.101 |
| 1207 | Te 1580 | 17 43 39.442 | -25 36 42.51 | --- | 0.658 | --- | 3.35 | 11.3 | --- | --- | --- |
| 1208 | Te 2022 | 17 42 42.541 | -29 51 34.65 | --- | 46.919 | --- | 9.09 | --- | --- | --- | --- |
| 1209 | Te 2111 | 17 48 28.465 | -24 41 25.07 | --- | 0.627 | --- | 3.32 | 20.7 | --- | --- | --- |
| 1212 | Th 2-A | 13 22 33.854 | -63 21 00.96 | 0.457 | 1.866 | 23 | 5.33 | --- | 60 | 2.07 | 0.231 |
| 1214 | Th 3-10 | 17 24 40.846 | -30 51 59.47 | 0.153 | 1.635 | 2 | 3.50 | 21.6 | 29.5 | 7.11 | 0.069 |

Table 1 (cont.)

| KN | NAME | R.A. (2000.0) H:M:S | DEC (2000.0) D:M:S | F(9μm) Jy | F(18μm) Jy | θ arcsec | MN_18 arcsec | F(1.4GHz) mJy | F(5GHz) 5GHz | DIST. Kpc | DIAM. pc |
|---|---|---|---|---|---|---|---|---|---|---|---|
| 1216 | Th 3-12 | 17 25 06.093 | -29 45 16.87 | --- | 0.825 | 1.8 | 3.34 | 2.6 | 3.5 | --- | --- |
| 1217 | Th 3-13 | 17 25 19.341 | -30 40 41.79 | --- | 4.774 | 2 | 3.99 | 7.4 | 14.3 | --- | --- |
| 1218 | Th 3-14 | 17 25 44.060 | -26 57 47.69 | 0.122 | 2.248 | 1.4 | 3.80 | 5.8 | 4 | --- | --- |
| 1221 | Th 3-19 | 17 28 41.787 | -28 27 19.32 | --- | 0.967 | 2 | 3.40 | 9.9 | 4.2 | 14.64 | 0.142 |
| 1222 | Th 3-23 | 17 30 21.340 | -29 10 12.57 | --- | 0.847 | --- | 3.18 | 46.2 | --- | --- | --- |
| 1224 | Th 3-25 | 17 30 46.716 | -27 05 59.12 | --- | 0.841 | 2 | 3.40 | 14.9 | 18 | 8.77 | 0.085 |
| 1226 | Th 3-27 | 17 35 58.467 | -24 25 29.15 | --- | 1.729 | 3.2 | 3.69 | 14.4 | 13.5 | --- | --- |
| 1228 | Th 3-32 | 17 35 15.533 | -28 07 01.70 | --- | 1.842 | --- | 3.73 | 10.2 | --- | --- | --- |
| 1229 | Th 3-33 | 17 35 48.123 | -27 43 20.38 | --- | 2.064 | --- | 4.01 | 4.2 | --- | --- | --- |
| 1230 | Th 3-35 | 17 38 42.168 | -28 42 45.33 | 0.259 | 2.482 | --- | 3.80 | 10.7 | --- | --- | --- |
| 1232 | Th 3-55 | 17 30 58.838 | -31 01 05.73 | --- | 1.201 | --- | 3.65 | 9.8 | 6.1 | --- | --- |
| 1234 | Th 4-1 | 17 46 20.804 | -20 13 48.08 | 0.227 | --- | <10 | --- | --- | --- | --- | --- |
| 1235 | Th 4-10 | 17 57 06.599 | -18 06 43.43 | --- | 0.562 | --- | 2.94 | 4.3 | --- | --- | --- |
| 1238 | Th 4-3 | 17 48 37.390 | -22 16 48.79 | --- | 1.015 | --- | 3.53 | 2.7 | --- | --- | --- |
| 1239 | Th 4-6 | 17 50 57.220 | -18 46 48.13 | --- | 0.332 | --- | 2.81 | 5.9 | 6.1 | --- | --- |
| 1240 | Th 4-7 | 17 52 22.569 | -21 51 13.43 | --- | 0.276 | 7 | 2.89 | 8.8 | 18.4 | --- | --- |
| 1241 | Th 4-9 | 17 56 00.589 | -19 29 26.60 | 0.595 | 1.130 | --- | 3.41 | 3.1 | --- | --- | --- |
| 1244 | V-V 3-5 | 18 36 32.291 | -19 19 28.01 | 0.152 | 0.693 | --- | 3.67 | 11.6 | 10 | --- | --- |
| 1250 | VBe 3 | 15 52 59.209 | -56 24 27.19 | --- | 0.196 | --- | 2.44 | --- | --- | --- | --- |
| 1251 | VY 2-1 | 18 27 59.603 | -26 06 48.29 | --- | 2.003 | 3.7 | 3.81 | 32.4 | 37 | --- | --- |
| 1252 | Vd 1-1 | 16 42 33.429 | -38 54 31.83 | 0.149 | --- | <10 | --- | 6.8 | --- | --- | --- |
| 1253 | Vd 1-2 | 16 46 45.141 | -38 36 58.10 | 0.253 | 4.186 | --- | 3.98 | 2.7 | --- | --- | --- |
| 1255 | Vd 1-4 | 16 50 25.324 | -39 08 18.87 | 0.253 | 0.556 | --- | 3.09 | 4.2 | --- | --- | --- |
| 1256 | Vd 1-5 | 16 51 33.575 | -40 02 56.01 | --- | 0.139 | --- | 2.62 | 3.7 | --- | --- | --- |
| 1257 | Vd 1-6 | 16 54 27.327 | -38 44 11.23 | --- | 0.643 | --- | 3.21 | 11.2 | --- | --- | --- |
| 1259 | Ve 26 | 08 43 28.087 | -46 06 39.72 | 1.509 | 13.141 | --- | 4.86 | --- | --- | --- | --- |
| 1260 | Vo 1 | 06 59 26.405 | -79 38 47.20 | 17.523 | 47.264 | --- | 6.18 | --- | --- | --- | --- |
| 1261 | Vo 2 | 08 16 10.000 | -39 51 50.67 | 0.145 | 1.134 | --- | 3.69 | 60.9 | --- | --- | --- |
| 1263 | Vo 4 | 13 53 23.072 | -60 33 47.40 | --- | 0.849 | --- | 3.61 | --- | --- | --- | --- |
| 1264 | Vy 1-1 | 00 18 42.167 | +53 52 20.03 | 0.133 | 0.761 | 6 | 3.66 | 19.8 | 28.9 | 5.34 | 0.155 |
| 1265 | Vy 1-2 | 17 54 22.994 | +27 59 57.99 | 0.105 | 0.786 | 4.6 | 3.30 | 11.1 | --- | --- | --- |
| 1266 | Vy 1-4 | 18 54 01.899 | -06 26 19.81 | --- | 1.182 | 4.5 | 3.90 | 19.7 | 22 | 6.01 | 0.131 |
| 1267 | Vy 2-2 | 19 24 22.229 | +09 53 56.66 | 7.850 | 56.233 | 0.6 | 6.47 | 5.9 | 50 | 9.82 | 0.029 |
| 1268 | Vy 2-3 | 23 22 57.952 | +46 53 58.24 | --- | 0.380 | 4.6 | 3.29 | 6.5 | 3 | 9.95 | 0.222 |
| 1272 | WeSb 4 | 18 50 40.302 | -01 03 11.16 | --- | 0.566 | --- | 3.71 | 4.3 | --- | --- | --- |
| 1287 | WhMe 1 | 19 14 59.755 | +17 22 46.01 | 7.890 | 13.232 | --- | 4.79 | --- | --- | --- | --- |
| 1291 | Wray 16-128 | 13 24 21.922 | -57 31 19.29 | --- | 0.898 | --- | 4.08 | --- | --- | --- | --- |
| 1293 | Wray 16-199 | 16 00 22.002 | -48 15 35.29 | --- | 0.974 | --- | 3.49 | --- | --- | --- | --- |
| 1298 | Wray 16-286 | 17 33 00.666 | -36 43 52.54 | --- | 3.591 | --- | 3.87 | 10.7 | --- | --- | --- |
| 1303 | Wray 16-93 | 11 30 48.339 | -59 17 04.63 | --- | 0.248 | --- | 2.86 | --- | --- | --- | --- |
| 1306 | Wray 17-18 | 08 23 53.825 | -45 31 10.70 | --- | 0.311 | --- | 2.88 | --- | --- | --- | --- |
| 1311 | Y-C 2-5 | 08 10 41.628 | -20 31 32.16 | --- | 0.176 | --- | 2.55 | 7.1 | 4.5 | --- | --- |

# TABLE 2

## Galactic Planetary Nebulae Identified in the AKARI FIS Catalogue

| KN | NAME | R.A. (2000.0) H:M:S | DEC. (2000.0) D:M:S | F(65μm) Jy | F(90μm) Jy | F(140μm) Jy | F(160μm) Jy | QUAL. | DIAM. arcsec | F(1.4GHz) mJy | F(5GHz) mJy | DIST. kpc | DIAM. pc |
|---|---|---|---|---|---|---|---|---|---|---|---|---|---|
| 1  | 000.6+08.8 | 17 14 09.788 | -23 31 53.41 | 1.040  | 1.860  | 1.987  | 2.565  | 1311 | ---  | 10.7 | ---  | ---  | ---  |
| 3  | 008.1+08.2 | 17 33 35.332 | -17 41 16.23 | 1.046  | 1.678  | 0.757  | 2.497  | 1311 | ---  | 7.5  | ---  | ---  | ---  |
| 6  | 050.6+19.7 | 18 08 20.083 | +24 10 43.26 | 2.136  | 1.050  | 0.462  | ---    | 1311 | ---  | ---  | ---  | ---  | ---  |
| 7  | 099.3-01.9 | 22 04 12.301 | +53 04 01.36 | 10.186 | 8.225  | 2.369  | 1.474  | 3311 | ---  | ---  | ---  | ---  | ---  |
| 10 | 341.4-09.0 | 17 35 02.502 | -49 26 26.38 | 46.648 | 22.854 | 5.621  | 5.089  | 3331 | ---  | ---  | ---  | ---  | ---  |
| 11 | A 12 | 06 02 20.045 | +09 39 14.09 | 2.902  | 3.487  | 2.301  | 2.519  | 1311 | 37.0 | 38.3 | 36   | 2.04 | 0.37 |
| 18 | A 2  | 00 45 34.678 | +57 57 34.88 | ---    | 0.672  | 0.008  | 0.334  | 1311 | 30.9 | 7.6  | 2.3  | 4.68 | 0.70 |
| 21 | A 23 | 07 43 17.984 | -34 45 15.64 | ---    | 0.964  | 0.067  | 0.835  | 1311 | 54.0 | 4.2  | 6.8  | 2.74 | 0.72 |
| 23 | A 26 | 08 09 01.644 | -32 40 24.91 | 0.253  | 1.234  | 1.051  | 3.250  | 1311 | 40.0 | 22.5 | 2.4  | 4.11 | 0.80 |
| 27 | A 30 | 08 46 53.514 | +17 52 45.47 | 89.781 | 56.794 | 26.965 | 21.707 | 3333 | 4.47 | ---  | 2.3  | 2.55 | 0.06 |
| 35 | A 40 | 16 48 34.515 | -21 00 50.67 | 1.165  | 1.264  | 2.470  | 0.438  | 1311 | 34.0 | 4.5  | 2    | 10.2 | 1.68 |
| 41 | A 46 | 18 31 18.290 | +26 56 12.86 | 0.826  | 0.950  | 1.520  | 0.955  | 1321 | 63.4 | ---  | 5    | 2.74 | 0.84 |
| 44 | A 49 | 18 53 28.292 | -06 28 46.83 | 0.499  | 1.170  | 2.701  | ---    | 1311 | 45.0 | ---  | ---  | ---  | ---  |
| 47 | A 52 | 19 04 32.342 | +17 57 07.13 | 0.767  | 1.299  | ---    | ---    | 1311 | ---  | 3.4  | ---  | ---  | ---  |
| 48 | A 53 | 19 06 45.910 | +06 23 52.47 | 11.713 | 10.097 | 19.476 | ---    | 3331 | 31.0 | 33.6 | 85.7 | 1.68 | 0.25 |
| 50 | A 55 | 19 10 25.768 | -02 20 23.46 | 0.944  | 2.171  | 1.047  | 0.451  | 1311 | 54.0 | 10.8 | 6    | 2.98 | 0.78 |
| 52 | A 57 | 19 17 05.733 | +25 37 33.41 | 0.179  | 0.415  | 0.240  | ---    | 1311 | 36.9 | 2.2  | ---  | ---  | ---  |
| 53 | A 58 | 19 18 20.476 | +01 46 59.62 | 26.646 | 20.791 | 6.876  | 3.949  | 3331 | ---  | ---  | ---  | ---  | ---  |
| 54 | A 59 | 19 18 40.003 | +19 34 32.95 | 0.467  | 1.613  | 2.130  | 0.899  | 1311 | 86.7 | 13.7 | 18   | 1.69 | 0.71 |
| 56 | A 60 | 19 19 17.819 | -12 14 36.78 | 0.209  | 0.455  | 0.214  | ---    | 1311 | 74.0 | ---  | 11   | 2.07 | 0.74 |
| 59 | A 63 | 19 42 10.372 | +17 05 14.52 | 0.532  | 1.765  | 1.095  | 0.782  | 1311 | 40.0 | 4.5  | ---  | ---  | ---  |
| 63 | A 68 | 20 00 10.607 | +21 42 55.43 | 0.636  | 1.132  | ---    | 0.537  | 1311 | ---  | ---  | ---  | ---  | ---  |
| 67 | A 71 | 20 32 24.243 | +47 21 02.83 | ---    | 0.943  | 1.042  | 0.910  | 1311 | 158.0| ---  | 82.8 | 0.84 | 0.64 |
| 69 | A 73 | 20 56 27.032 | +57 26 03.05 | 0.452  | 1.309  | 1.011  | 0.272  | 1311 | 73.2 | 11   | 7.4  | 2.31 | 0.82 |
| 71 | A 75 | 21 26 23.540 | +62 53 31.83 | 0.977  | 2.389  | 1.605  | 2.082  | 1311 | 56.0 | 19.3 | 17   | 2.04 | 0.55 |
| 75 | A 8  | 05 06 38.378 | +39 08 10.77 | ---    | 0.371  | ---    | ---    | 1311 | 60.0 | 4.3  | 25.6 | 1.78 | 0.52 |
| 77 | A 82 | 23 45 47.814 | +57 03 59.22 | 0.249  | 0.755  | 2.635  | 1.160  | 1311 | 81.0 | 7.6  | 5.3  | 2.41 | 0.95 |
| 83 | Al 2-E | 17 30 14.398 | -27 30 19.41 | 1.227 | 2.023  | ---    | 0.022  | 1311 | ---  | 14.7 | ---  | ---  | ---  |
| 91 | Al 2-Q | 17 53 25.040 | -29 17 08.73 | 0.179 | 3.261  | 5.271  | 1.104  | 1311 | ---  | 2.3  | ---  | ---  | ---  |
| 94 | Ap 1-12 | 18 11 35.088 | -28 22 37.02 | 8.231 | 6.480 | 1.255 | 1.526 | 3311 | 12.0 | 6.6  | 18.7 | 5.34 | 0.31 |

| KN | NAME | R.A. (2000.0) | DEC. (2000.0) | F(65μm) | F(90μm) | F(140μm) | F(160μm) | QUAL. | DIAM. | F(1.4GHz) | F(5GHz) | DIST. | DIAM. |
|---|---|---|---|---|---|---|---|---|---|---|---|---|---|
| | | H:M:S | D:M:S | Jy | Jy | Jy | Jy | | arcsec | mJy | mJy | kpc | pc |
| 95 | Ap 2-1 | 18 58 10.459 | +01 36 57.15 | 873.272 | 609.290 | 793.279 | 958.062 | 3333 | 40.0 | 195 | 320.4 | 1.13 | 0.22 |
| 97 | BV 5-1 | 00 20 00.450 | +62 59 03.16 | 1.283 | 1.029 | 1.138 | 2.259 | 1311 | 41.6 | 10.6 | 17.7 | 4.05 | 0.82 |
| 99 | BV 5-3 | 01 53 03.432 | +56 24 17.11 | 1.010 | 0.622 | 0.576 | 1.718 | 1311 | --- | 2.7 | --- | --- | --- |
| 102 | Bl 2-1 | 22 20 16.638 | +58 14 16.59 | 2.659 | 2.965 | --- | 1.066 | 1311 | 1.6 | 21.6 | 54 | 6.94 | 0.05 |
| 105 | Bl 3-15 | 17 52 35.947 | -29 06 38.97 | 3.780 | 3.210 | 3.346 | 5.144 | 1311 | --- | --- | --- | --- | --- |
| 106 | Bl B | 17 36 59.836 | -29 40 08.93 | 5.691 | 5.450 | 6.249 | 2.798 | 1311 | --- | 8.9 | --- | --- | --- |
| 118 | CTS 1 | 18 06 59.785 | -08 55 32.80 | 4.706 | 4.048 | 0.471 | 1.589 | 3311 | --- | 24 | --- | --- | --- |
| 119 | Cn 1-1 | 15 51 15.936 | -48 44 58.67 | 13.214 | 10.516 | 4.745 | 0.806 | 3331 | --- | --- | 10 | --- | --- |
| 120 | Cn 1-3 | 17 26 12.392 | -44 11 24.52 | 3.458 | 3.029 | --- | --- | 3311 | --- | --- | --- | --- | --- |
| 122 | Cn 1-5 | 18 29 11.653 | -31 29 59.21 | 6.379 | 6.339 | 2.788 | --- | 3311 | 7.0 | 51 | 44 | 4.43 | 0.15 |
| 123 | Cn 2-1 | 17 54 32.978 | -34 22 20.84 | 2.975 | 2.954 | --- | 0.611 | 3311 | 2.2 | 25.2 | 49 | 6.17 | 0.07 |
| 124 | Cn 3-1 | 18 17 34.133 | +10 09 03.99 | 7.298 | 5.991 | 1.936 | 2.188 | 3311 | 4.5 | 59.5 | 75 | 4 | 0.09 |
| 127 | Dd 1 | 20 08 43.080 | +42 30 11.19 | 0.819 | 1.087 | --- | --- | 1311 | --- | --- | 6 | --- | --- |
| 133 | DuRe 1 | 12 45 51.211 | -64 09 38.08 | 8.398 | 6.469 | 11.267 | 6.274 | 1331 | --- | --- | --- | --- | --- |
| 138 | ESO 040-11 | 13 34 14.128 | -75 46 31.33 | 1.049 | 0.991 | 1.758 | --- | 1311 | --- | --- | --- | --- | --- |
| 150 | Fg 1 | 11 28 36.205 | -52 56 04.01 | 5.624 | 10.809 | 6.454 | 4.621 | 3331 | 16.0 | --- | 55 | 2.02 | 0.16 |
| 151 | Fg 2 | 17 39 19.847 | -44 09 37.14 | 3.506 | 2.488 | 0.622 | 0.477 | 1311 | --- | --- | --- | --- | --- |
| 152 | Fg 3 | 18 00 11.819 | -38 49 52.73 | 14.692 | 11.152 | 4.286 | 3.226 | 3331 | 2.0 | 42 | 107 | --- | --- |
| 156 | G001.6-05.9 | 18 13 15.842 | -30 25 58.49 | --- | 0.428 | 0.472 | 0.942 | 1311 | --- | --- | --- | --- | --- |
| 158 | G002.5+05.1 | 17 32 12.788 | -24 04 59.56 | 15.981 | 11.822 | 3.348 | 1.594 | 3311 | --- | --- | --- | --- | --- |
| 163 | G009.3+05.7 | 17 45 14.155 | -17 56 46.57 | 56.190 | 40.015 | 15.844 | 12.064 | 3333 | --- | --- | --- | --- | --- |
| 164 | G010.2+07.5 | 17 41 00.035 | -16 18 12.50 | 1.362 | 0.956 | --- | 1.539 | 1311 | --- | --- | --- | --- | --- |
| 165 | G011.1-07.9 | 18 40 19.914 | -22 54 29.23 | 0.841 | 1.638 | 0.297 | 0.189 | 1311 | --- | --- | --- | --- | --- |
| 168 | G013.1-13.2 | 19 04 43.548 | -23 26 08.82 | 21.108 | 13.815 | 4.630 | 2.404 | 3331 | --- | --- | --- | --- | --- |
| 171 | G014.4-06.1 | 18 39 40.056 | -19 14 12.04 | 1.014 | 1.202 | 1.200 | --- | 1311 | --- | --- | --- | --- | --- |
| 175 | G017.0+11.1 | 17 42 14.430 | -08 43 18.61 | 5.365 | 5.666 | 1.605 | 0.349 | 3311 | --- | 23.4 | --- | --- | --- |
| 179 | G019.2-04.4 | 18 42 24.770 | -14 15 11.81 | 2.374 | 2.789 | 0.666 | 1.030 | 1311 | --- | 7.3 | --- | --- | --- |
| 181 | G022.0-04.3 | 18 47 03.954 | -11 41 11.74 | 8.676 | 8.306 | 2.024 | 1.337 | 3311 | --- | 8.4 | --- | --- | --- |
| 185 | G034.5-11.7 | 19 36 17.531 | -03 53 25.26 | 3.814 | 2.615 | 3.426 | 0.975 | 3311 | --- | 8.8 | --- | --- | --- |
| 186 | G038.4+01.8 | 18 54 54.139 | +05 48 11.29 | --- | 2.998 | --- | --- | 1311 | --- | 20.5 | --- | --- | --- |
| 188 | G039.1-02.2 | 19 10 53.284 | +04 27 28.50 | 1.248 | 2.490 | --- | --- | 1311 | --- | 11 | --- | --- | --- |
| 191 | G044.1+01.5 | 19 06 32.146 | +10 43 23.93 | 1.413 | 3.034 | 4.193 | --- | 1311 | --- | 19.4 | --- | --- | --- |
| 193 | G047.2+01.7 | 19 11 35.828 | +13 31 11.58 | 5.784 | 4.889 | --- | --- | 3311 | --- | --- | --- | --- | --- |

| KN | NAME | R.A. (2000.0) H:M:S | DEC. (2000.0) D:M:S | F(65μm) Jy | F(90μm) Jy | F(140μm) Jy | F(160μm) Jy | QUAL. | DIAM. arcsec | F(1.4GHz) mJy | F(5GHz) mJy | DIST. kpc | DIAM. pc |
|---|---|---|---|---|---|---|---|---|---|---|---|---|---|
| 202 | G069.2+01.2 | 20 00 41.998 | +32 27 41.23 | 1.094 | 1.588 | --- | --- | 1311 | --- | 17 | --- | --- | --- |
| 204 | G076.6-05.7 | 20 48 16.627 | +34 27 24.36 | 8.115 | 7.002 | 2.067 | --- | 3311 | --- | --- | --- | --- | --- |
| 207 | G095.0-05.5 | 21 56 32.945 | +47 36 13.40 | 0.858 | 0.472 | --- | --- | 1311 | --- | --- | --- | --- | --- |
| 211 | G124.0+02.9 | 01 02 24.488 | +65 46 32.81 | 2.184 | 2.741 | 4.239 | 3.567 | 1311 | --- | 7.4 | --- | --- | --- |
| 221 | G222.8-04.2 | 06 54 13.406 | -10 45 38.27 | 1.878 | 1.168 | 0.744 | 0.631 | 1311 | --- | 2.7 | --- | --- | --- |
| 226 | G255.3-03.6 | 08 06 28.339 | -38 53 23.93 | 1.622 | 1.085 | 0.289 | 1.120 | 1311 | --- | 2.8 | --- | --- | --- |
| 228 | G260.1+00.2 | 08 37 24.599 | -40 38 07.46 | 4.072 | 4.911 | 2.082 | 0.131 | 3311 | --- | --- | --- | --- | --- |
| 229 | G270.1-02.9 | 08 59 02.971 | -50 23 39.68 | 1.001 | 1.265 | | --- | 1311 | --- | --- | --- | --- | --- |
| 231 | G277.1-01.5 | 09 37 51.861 | -54 27 08.59 | 1.247 | 1.106 | --- | --- | 1311 | --- | --- | --- | --- | --- |
| 232 | G279.1-00.4 | 09 53 27.058 | -54 52 39.60 | 1.616 | 2.185 | 2.079 | 0.076 | 1311 | --- | --- | --- | --- | --- |
| 234 | G282.6-00.4 | 10 13 19.673 | -56 55 32.24 | 3.692 | 3.317 | 6.231 | 1.671 | 1311 | --- | --- | --- | --- | --- |
| 235 | G285.1-02.7 | 10 19 32.466 | -60 13 29.41 | 77.860 | 58.749 | 36.452 | 16.570 | 3333 | --- | --- | --- | --- | --- |
| 236 | G290.7+00.2 | 11 10 25.122 | -60 15 36.85 | 10.823 | 13.328 | 28.365 | 19.063 | 3333 | --- | --- | --- | --- | --- |
| 241 | G308.5-03.5 | 13 46 25.712 | -65 46 24.26 | 1.892 | 1.935 | 0.503 | 0.551 | 1311 | --- | --- | --- | --- | --- |
| 246 | G316.2+00.8 | 14 38 19.980 | -59 11 46.12 | 17.965 | 12.604 | 7.104 | 2.124 | 3311 | --- | --- | --- | --- | --- |
| 248 | G326.4+01.0 | 15 19 43.867 | -48 59 54.65 | 0.844 | 2.367 | --- | 0.838 | 1311 | --- | --- | --- | --- | --- |
| 251 | G328.4-02.8 | 16 09 20.143 | -55 36 09.84 | 10.647 | 9.904 | 5.721 | 6.032 | 3311 | --- | --- | --- | --- | --- |
| 253 | G331.3-12.1 | 17 16 21.107 | -59 29 23.35 | 2.504 | 1.926 | 0.895 | 1.700 | 3311 | --- | --- | --- | --- | --- |
| 264 | G347.9-06.0 | 17 40 03.329 | -42 24 05.58 | 1.674 | 1.070 | --- | 0.135 | 1311 | --- | --- | --- | --- | --- |
| 267 | G351.5-06.5 | 17 52 09.386 | -39 32 14.52 | 1.246 | 2.017 | 0.941 | 0.884 | 1311 | --- | --- | --- | --- | --- |
| 270 | G352.6-04.9 | 17 47 52.679 | -37 48 03.10 | 0.685 | 0.733 | --- | 1.812 | 1311 | --- | 3.7 | --- | --- | --- |
| 275 | G353.6-03.6 | 17 44 35.434 | -36 14 01.67 | 0.631 | 1.004 | 1.947 | 1.427 | 1311 | --- | --- | --- | --- | --- |
| 294 | G356.8-03.0 | 17 50 10.734 | -33 14 17.97 | 4.956 | 2.522 | 0.540 | 4.026 | 1311 | --- | 4.2 | --- | --- | --- |
| 299 | G358.3-07.3 | 18 11 39.900 | -34 00 21.99 | --- | 0.701 | --- | 0.763 | 1311 | --- | --- | --- | --- | --- |
| 302 | G358.7+05.1 | 17 22 43.615 | -27 13 36.67 | 7.125 | 4.881 | 3.383 | 2.801 | 3311 | --- | --- | --- | --- | --- |
| 307 | G359.4-08.5 | 18 19 26.331 | -33 37 04.86 | 0.298 | 0.974 | --- | 0.672 | 1311 | --- | 4.5 | --- | --- | --- |
| 310 | G000.5+01.9 | 17 39 31.219 | -27 27 46.77 | 5.529 | 3.087 | 3.930 | 10.456 | 1311 | --- | 6.8 | --- | --- | --- |
| 323 | G003.1+04.1 | 17 37 20.167 | -24 03 27.84 | 1.600 | 1.494 | 0.414 | 0.727 | 1311 | --- | 5.1 | --- | --- | --- |
| 326 | G003.6+04.9 | 17 35 31.198 | -23 11 47.62 | 0.551 | 1.164 | 2.137 | --- | 1311 | --- | 4.6 | --- | --- | --- |
| 332 | G035.4+03.4 | 18 43 36.599 | +03 46 39.24 | 5.368 | 5.503 | 3.326 | 0.518 | 3311 | --- | 5.1 | --- | --- | --- |
| 359 | H 1-11 | 17 21 17.693 | -22 18 35.32 | 3.175 | 2.960 | 2.676 | --- | 3311 | 6.0 | 20.7 | 13 | 7.27 | 0.21 |
| 367 | H 1-19 | 17 30 02.546 | -27 59 17.54 | 5.167 | 4.576 | --- | 1.333 | 3311 | 1.4 | 11.6 | 26 | 8.19 | 0.06 |
| 368 | H 1-2 | 16 48 54.091 | -35 47 09.01 | 6.754 | 7.334 | 3.178 | 5.698 | 3331 | 1.0 | 14.3 | 62 | --- | --- |

| KN | NAME | R.A. (2000.0) H:M:S | DEC. (2000.0) D:M:S | F(65μm) Jy | F(90μm) Jy | F(140μm) Jy | F(160μm) Jy | QUAL. | DIAM. arcsec | F(1.4GHz) mJy | F(5GHz) mJy | DIST. kpc | DIAM. pc |
|---|---|---|---|---|---|---|---|---|---|---|---|---|---|
| 369 | H 1-20 | 17 30 43.799 | -28 04 06.68 | 3.730 | 3.436 | --- | --- | 3311 | 3.3 | 26.9 | 43.9 | 5.12 | 0.08 |
| 372 | H 1-23 | 17 32 46.897 | -30 00 15.07 | 5.346 | 2.646 | 1.913 | 0.270 | 1311 | 2.6 | 24.5 | 34.8 | 6.12 | 0.08 |
| 373 | H 1-24 | 17 33 37.565 | -21 46 24.78 | 2.864 | 4.191 | 1.340 | 0.725 | 1311 | 5.0 | 5.1 | 15 | 7.32 | 0.18 |
| 375 | H 1-27 | 17 40 17.926 | -22 19 17.64 | 4.632 | 4.237 | 2.276 | 1.683 | 3311 | 5.2 | --- | 18 | 10.9 | 0.27 |
| 376 | H 1-28 | 17 42 54.066 | -39 36 24.04 | 0.741 | 1.613 | 0.910 | 0.820 | 1311 | --- | 5.4 | --- | --- | --- |
| 378 | H 1-3 | 16 53 31.347 | -42 39 23.45 | 6.802 | 6.116 | 14.597 | --- | 1311 | 15.8 | --- | --- | 1.36 | 0.10 |
| 379 | H 1-30 | 17 45 06.782 | -38 08 49.48 | 0.047 | 2.454 | 2.628 | 1.400 | 1311 | <5 | 7 | 81.3 | 5.21 | --- |
| 380 | H 1-31 | 17 45 32.104 | -34 33 55.32 | 1.182 | 1.808 | 3.421 | 2.872 | 1311 | 0.7 | 6 | 16 | 11.8 | 0.04 |
| 385 | H 1-36 | 17 49 48.172 | -37 01 29.47 | 3.996 | 3.046 | 1.295 | --- | 1311 | 0.8 | 11.8 | 50 | --- | --- |
| 390 | H 1-40 | 17 55 36.049 | -30 33 32.08 | 8.689 | 7.947 | 3.121 | 2.002 | 3311 | 3.0 | 8.1 | 31 | 7.84 | 0.11 |
| 391 | H 1-41 | 17 57 19.145 | -34 09 49.12 | 1.409 | 2.460 | --- | 1.307 | 1311 | 9.6 | 16.6 | 12 | 6.6 | 0.31 |
| 392 | H 1-42 | 17 57 25.169 | -33 35 42.94 | 1.873 | 2.163 | 1.065 | 1.600 | 1311 | 5.8 | 34.5 | 40 | 4.82 | 0.14 |
| 394 | H 1-44 | 17 58 10.641 | -31 42 56.08 | 3.517 | 3.355 | --- | --- | 3311 | <5 | 8 | --- | --- | --- |
| 396 | H 1-46 | 17 59 02.490 | -32 21 43.42 | 2.175 | 1.824 | 1.577 | --- | 3311 | 1.2 | 18.3 | 43 | 7.11 | 0.04 |
| 399 | H 1-50 | 18 03 53.461 | -32 41 42.15 | 2.527 | 1.911 | 0.001 | --- | 3311 | 1.4 | 19.6 | 31 | 7.68 | 0.05 |
| 404 | H 1-55 | 18 07 14.542 | -29 41 24.51 | 2.607 | 2.827 | 4.162 | 1.519 | 1311 | 10.0 | 2.5 | 5.3 | --- | --- |
| 405 | H 1-56 | 18 07 53.880 | -29 44 34.26 | 1.409 | 1.586 | 1.400 | --- | 1311 | 3.0 | 8.3 | 7.8 | 10.98 | 0.16 |
| 408 | H 1-59 | 18 11 29.263 | -27 46 15.69 | 1.006 | 1.115 | 0.501 | --- | 1311 | 6.0 | 3.4 | 4.2 | 11.12 | 0.32 |
| 412 | H 1-62 | 18 13 17.951 | -32 19 42.74 | 4.164 | 3.569 | --- | 2.715 | 3311 | --- | 11.9 | --- | --- | --- |
| 413 | H 1-63 | 18 16 19.336 | -30 07 35.98 | 1.399 | 0.981 | --- | 1.129 | 1311 | --- | 2.5 | 9 | --- | --- |
| 416 | H 1-66 | 18 24 57.539 | -25 41 55.76 | 1.680 | 2.448 | 2.442 | 4.279 | 1313 | <10 | 10.4 | 6 | 9.5 | --- |
| 417 | H 1-67 | 18 25 04.976 | -22 34 52.64 | 1.245 | 1.916 | 2.312 | --- | 1311 | 6.0 | 12.2 | 11.9 | 7.51 | 0.22 |
| 418 | H 1-7 | 17 10 27.389 | -41 52 49.42 | --- | 42.370 | 33.760 | 10.955 | 0131 | --- | --- | --- | --- | --- |
| 429 | H 2-17 | 17 40 07.428 | -24 25 42.57 | 3.915 | 2.947 | 1.795 | --- | 3311 | 4.0 | 8.2 | 10 | 8.88 | 0.17 |
| 437 | H 2-27 | 17 51 50.576 | -33 47 35.59 | 2.026 | 4.543 | 5.283 | 3.193 | 1311 | --- | 7.9 | --- | --- | --- |
| 438 | H 2-29 | 17 53 16.794 | -32 40 38.56 | 0.841 | 1.959 | 1.309 | 0.054 | 1311 | --- | 2.6 | --- | --- | --- |
| 442 | H 2-33 | 17 58 12.535 | -31 08 05.99 | 1.633 | 2.283 | --- | --- | 1311 | 8.0 | 8.1 | 5 | 9.71 | 0.38 |
| 443 | H 2-35 | 18 00 18.259 | -34 27 39.29 | 0.161 | 0.981 | 2.435 | 0.421 | 1311 | --- | --- | --- | --- | --- |
| 444 | H 2-36 | 18 04 07.936 | -31 39 13.70 | 0.939 | 1.627 | 1.844 | 1.400 | 1311 | --- | 3.4 | --- | --- | --- |
| 449 | H 2-42 | 18 12 23.255 | -26 32 54.02 | 1.024 | 1.992 | 0.430 | 3.923 | 1311 | --- | 3.1 | --- | --- | --- |
| 453 | H 2-46 | 18 18 37.444 | -31 54 45.42 | 0.751 | 0.772 | --- | 2.136 | 1311 | 25.0 | --- | --- | --- | --- |
| 454 | H 2-48 | 18 46 35.149 | -23 26 48.24 | 3.332 | 2.115 | 1.080 | --- | 3311 | 2.0 | 31.6 | 66 | 6.23 | 0.06 |
| 456 | H 2-8 | 17 24 45.766 | -21 33 35.80 | --- | 0.443 | 2.309 | --- | 1311 | --- | 4.8 | --- | --- | --- |

| KN | NAME | R.A. (2000.0) H:M:S | DEC. (2000.0) D:M:S | F(65μm) Jy | F(90μm) Jy | F(140μm) Jy | F(160μm) Jy | QUAL. | DIAM. arcsec | F(1.4GHz) mJy | F(5GHz) mJy | DIST. kpc | DIAM. pc |
|---|---|---|---|---|---|---|---|---|---|---|---|---|---|
| 457 | H 3-29 | 04 37 23.484 | +25 02 40.94 | 2.391 | 1.128 | --- | 3.097 | 1311 | 21.0 | 18.3 | 18 | 3.4 | 0.35 |
| 463 | HaTr 13 | 19 08 02.149 | +02 21 24.25 | 0.226 | 1.281 | 0.982 | 1.832 | 1311 | --- | 3.6 | --- | --- | --- |
| 464 | HaTr 14 | 19 09 13.670 | +07 05 44.86 | 1.974 | 1.872 | --- | --- | 1311 | --- | --- | --- | --- | --- |
| 465 | HaTr 2 | 15 30 18.542 | -61 01 38.79 | 0.808 | 1.883 | 1.534 | --- | 1311 | --- | --- | --- | --- | --- |
| 467 | HaTr 4 | 16 45 00.155 | -51 12 19.83 | 0.422 | 1.930 | 3.937 | 3.937 | 1331 | --- | --- | --- | --- | --- |
| 474 | Hb 12 | 23 26 14.814 | +58 10 54.65 | 28.818 | 19.707 | 6.499 | 3.075 | 3331 | 0.8 | --- | 45 | 10.46 | 0.04 |
| 475 | Hb 4 | 17 41 52.763 | -24 42 08.07 | 13.541 | 14.269 | 0.952 | 1.287 | 3311 | 7.3 | 157.6 | 170 | 2.68 | 0.09 |
| 477 | Hb 6 | 17 55 07.023 | -21 44 39.98 | 20.768 | 14.246 | 6.261 | 2.636 | 1311 | 6.0 | 190.5 | 243 | 2.45 | 0.07 |
| 479 | He 1-1 | 19 23 46.875 | +21 06 38.60 | 1.896 | 2.307 | --- | --- | 1311 | 8.0 | 16 | 14 | --- | --- |
| 480 | He 1-2 | 19 26 37.757 | +21 09 27.04 | 4.268 | 3.830 | 3.690 | 2.034 | 3311 | 4.7 | 14.6 | 15 | --- | --- |
| 481 | He 1-3 | 19 48 26.422 | +22 08 37.62 | 3.687 | 3.402 | --- | --- | 1311 | 8.0 | 53.5 | --- | --- | --- |
| 482 | He 1-4 | 19 59 18.014 | +31 54 39.14 | 2.196 | 3.218 | --- | --- | 3311 | --- | 16.1 | --- | --- | --- |
| 484 | He 1-6 | 20 17 21.424 | +25 21 46.78 | --- | 0.590 | 0.228 | --- | 1311 | 22.4 | 5 | 20.3 | 2.91 | 0.32 |
| 486 | He 2-102 | 13 58 13.866 | -58 54 31.78 | 3.526 | 5.212 | --- | 0.187 | 3311 | 9.0 | --- | 33 | 3.77 | 0.16 |
| 487 | He 2-104 | 14 11 52.077 | -51 26 24.18 | 5.304 | 5.412 | 2.607 | 1.568 | 3311 | 5.0 | --- | 15 | --- | --- |
| 488 | He 2-105 | 14 15 24.803 | -74 12 46.61 | 2.007 | 2.755 | 0.431 | 0.625 | 1311 | 31.0 | --- | 14 | 2.67 | 0.40 |
| 490 | He 2-108 | 14 18 08.889 | -52 10 39.76 | 11.056 | 11.255 | 2.980 | --- | 3311 | 11.0 | --- | 32 | 3.49 | 0.19 |
| 491 | He 2-109 | 14 20 48.916 | -55 27 59.21 | 0.507 | 0.868 | 1.644 | 1.676 | 1311 | 7.4 | --- | 15.8 | 6.41 | 0.23 |
| 492 | He 2-11 | 08 37 08.445 | -39 25 08.09 | 14.232 | 29.419 | 9.803 | 8.864 | 3333 | 65.0 | 343.6 | 444.6 | 0.77 | 0.24 |
| 494 | He 2-112 | 14 40 30.924 | -52 34 56.66 | 5.436 | 5.861 | 2.376 | 1.855 | 3311 | 14.6 | --- | 82 | 2.36 | 0.17 |
| 495 | He 2-113 | 14 59 53.523 | -54 18 07.20 | 146.924 | 75.968 | 29.685 | 20.859 | 3333 | --- | --- | 115 | --- | --- |
| 496 | He 2-114 | 15 04 08.800 | -60 53 18.78 | 2.378 | 3.831 | 2.893 | 13.696 | 1331 | 36.6 | --- | 11 | 2.8 | 0.50 |
| 497 | He 2-115 | 15 05 16.775 | -55 11 10.39 | 9.040 | 7.229 | 3.019 | 0.869 | 3311 | 3.0 | --- | 156 | 3.91 | 0.06 |
| 499 | He 2-117 | 15 05 59.193 | -55 59 16.53 | 30.411 | 20.756 | 7.710 | --- | 3331 | 5.0 | --- | 267 | 2.69 | 0.07 |
| 503 | He 2-123 | 15 22 19.360 | -54 08 12.77 | 13.454 | 14.537 | 6.106 | 5.808 | 3331 | 4.6 | --- | 110 | 3.59 | 0.08 |
| 504 | He 2-125 | 15 23 36.326 | -53 51 27.95 | 2.858 | 3.458 | 1.854 | 0.569 | 1311 | 3.0 | --- | 20.4 | 6.96 | 0.10 |
| 505 | He 2-128 | 15 25 07.841 | -51 19 42.29 | 1.343 | 1.317 | --- | 2.540 | 1311 | 5.0 | --- | 40 | --- | --- |
| 506 | He 2-129 | 15 25 32.677 | -52 50 37.96 | 1.332 | 1.958 | 0.083 | 1.087 | 1311 | 1.6 | --- | 35 | 7.85 | 0.06 |
| 507 | He 2-131 | 15 37 11.210 | -71 54 52.89 | 52.897 | 32.153 | 10.413 | 6.777 | 3331 | 6.0 | --- | 325 | 2.2 | 0.06 |
| 508 | He 2-132 | 15 38 01.182 | -58 44 42.07 | 5.143 | 6.911 | 3.164 | --- | 3311 | 17.8 | --- | 25 | 3.02 | 0.26 |
| 509 | He 2-133 | 15 41 58.788 | -56 36 25.72 | 19.455 | 12.659 | 2.857 | 3.351 | 3311 | <10 | --- | 210 | 3.18 | --- |
| 510 | He 2-136 | 15 52 10.666 | -62 30 46.94 | --- | 1.052 | --- | --- | 1311 | 10.0 | --- | 23 | --- | --- |
| 511 | He 2-138 | 15 56 01.694 | -66 09 09.23 | 37.174 | 27.055 | 8.317 | 5.173 | 3333 | 7.0 | --- | 76 | 3.61 | 0.12 |

| KN | NAME | R.A. (2000.0) H:M:S | DEC. (2000.0) D:M:S | F(65μm) Jy | F(90μm) Jy | F(140μm) Jy | F(160μm) Jy | QUAL. | DIAM. arcsec | F(1.4GHz) mJy | F(5GHz) mJy | DIST. kpc | DIAM. pc |
|---|---|---|---|---|---|---|---|---|---|---|---|---|---|
| 512 | He 2-140 | 15 58 08.063 | -55 41 50.13 | 11.771 | 8.420 | --- | 0.219 | 1311 | 2.6 | --- | 80 | 5.03 | 0.06 |
| 513 | He 2-141 | 15 59 08.762 | -58 23 53.15 | 2.406 | 2.797 | 0.987 | 1.007 | 3311 | 13.8 | --- | 51 | 2.77 | 0.19 |
| 514 | He 2-142 | 15 59 57.608 | -55 55 32.89 | 8.916 | 8.153 | 1.037 | 0.116 | 1311 | 3.6 | --- | 65 | 4.75 | 0.08 |
| 519 | He 2-149 | 16 14 24.266 | -54 47 38.82 | --- | 1.020 | --- | --- | 1311 | 3.0 | --- | 10 | 8.52 | 0.12 |
| 520 | He 2-15 | 08 53 30.701 | -40 03 42.08 | 8.855 | 13.475 | 7.766 | 8.121 | 3331 | 23.8 | 94.5 | 105 | 1.92 | 0.22 |
| 521 | He 2-151 | 16 15 42.269 | -59 54 00.96 | 4.698 | 3.035 | --- | --- | 3311 | 5.0 | --- | 10 | --- | --- |
| 522 | He 2-152 | 16 15 20.031 | -49 13 20.76 | 59.168 | 34.701 | 25.057 | 26.453 | 1333 | 11.0 | --- | 196 | 2.09 | 0.11 |
| 524 | He 2-155 | 16 19 23.101 | -42 15 35.98 | 3.768 | 6.096 | 1.291 | --- | 3311 | 14.6 | --- | 70 | 2.47 | 0.17 |
| 525 | He 2-157 | 16 22 14.264 | -53 40 54.09 | 2.612 | 2.420 | --- | 0.963 | 1311 | <5 | --- | 30 | 6.24 | --- |
| 526 | He 2-158 | 16 23 30.605 | -58 19 22.64 | 1.124 | 0.807 | --- | 0.634 | 1311 | 2.0 | --- | 2.7 | 14.71 | 0.14 |
| 528 | He 2-161 | 16 24 37.786 | -53 22 34.14 | 6.164 | 4.563 | 1.665 | --- | 1311 | 10.0 | --- | 32 | 3.64 | 0.18 |
| 530 | He 2-163 | 16 29 31.333 | -59 09 25.08 | 0.548 | 0.936 | --- | --- | 1311 | <25 | --- | 3.4 | 5.08 | --- |
| 531 | He 2-164 | 16 29 53.255 | -53 23 15.37 | 4.289 | 4.731 | 4.260 | 5.638 | 1311 | 16.0 | --- | 97 | 2.17 | 0.17 |
| 532 | He 2-165 | 16 30 00.057 | -54 09 28.00 | 1.498 | 2.651 | 2.271 | 5.820 | 1311 | 50.0 | --- | 15.9 | 2.2 | 0.53 |
| 534 | He 2-170 | 16 35 21.170 | -53 50 11.11 | 1.035 | 1.091 | --- | --- | 1311 | 5.0 | --- | 15 | --- | --- |
| 536 | He 2-175 | 16 39 28.112 | -36 34 16.41 | 8.230 | 8.479 | 3.012 | 1.863 | 3331 | 6.6 | 19.2 | 26.8 | 4.58 | 0.15 |
| 537 | He 2-18 | 09 08 40.050 | -53 19 13.91 | 0.712 | 1.440 | 0.206 | 0.118 | 1311 | 11.0 | --- | --- | --- | --- |
| 538 | He 2-182 | 16 54 35.167 | -64 14 28.43 | 0.723 | 0.970 | 0.501 | --- | 1311 | 10.0 | --- | 62 | --- | --- |
| 539 | He 2-185 | 17 01 17.254 | -70 06 03.35 | 0.310 | 0.532 | 0.323 | --- | 1311 | 10.0 | --- | 18 | --- | --- |
| 540 | He 2-186 | 16 59 36.064 | -51 42 06.46 | 1.573 | 1.378 | 0.342 | --- | 1311 | 3.0 | --- | 21 | 6.9 | 0.10 |
| 541 | He 2-187 | 17 01 36.976 | -50 22 57.49 | 0.606 | 2.288 | 1.562 | 1.152 | 1311 | 6.0 | --- | --- | --- | --- |
| 545 | He 2-25 | 09 18 01.308 | -54 39 29.01 | 1.808 | 1.355 | 1.190 | 0.091 | 1311 | 4.4 | --- | 827.9 | 2.07 | 0.04 |
| 546 | He 2-250 | 17 34 54.710 | -26 35 56.92 | 2.625 | 3.212 | --- | 0.965 | 3311 | 5.0 | 15.3 | 15 | 7.01 | 0.17 |
| 549 | He 2-28 | 09 22 06.825 | -54 09 38.60 | 1.481 | 1.190 | 0.504 | 0.840 | 1311 | 10.0 | --- | 20 | 4.15 | 0.20 |
| 551 | He 2-306 | 17 56 33.706 | -43 03 18.90 | 2.165 | 1.648 | 0.535 | --- | 1311 | --- | --- | --- | --- | --- |
| 553 | He 2-34 | 09 41 13.997 | -49 22 47.18 | 1.043 | 0.693 | 1.428 | --- | 1311 | <10 | --- | --- | --- | --- |
| 554 | He 2-35 | 09 41 37.504 | -49 57 58.71 | 1.402 | 1.463 | --- | --- | 1311 | 5.0 | --- | 20 | 5.61 | 0.14 |
| 555 | He 2-36 | 09 43 25.538 | -57 16 55.44 | 3.776 | 5.187 | 1.768 | 1.416 | 3311 | 22.0 | --- | 90 | 3.1 | 0.33 |
| 556 | He 2-37 | 09 47 24.987 | -48 58 11.96 | 0.111 | 1.566 | --- | --- | 1311 | 23.0 | --- | 22.6 | 2.8 | 0.31 |
| 557 | He 2-39 | 10 03 49.164 | -60 43 48.26 | 0.690 | 0.666 | --- | 1.180 | 1311 | --- | --- | --- | --- | --- |
| 561 | He 2-428 | 19 13 05.239 | +15 46 39.80 | 2.444 | 3.763 | 0.304 | --- | 1311 | 8.0 | 26.7 | 7 | 6.16 | 0.24 |
| 562 | He 2-429 | 19 13 38.422 | +14 59 19.21 | 5.302 | 5.489 | 1.651 | --- | 3311 | 4.2 | 58.5 | 64.9 | 4.33 | 0.09 |
| 563 | He 2-430 | 19 14 04.197 | +17 31 32.92 | 4.256 | 4.174 | 1.445 | 0.975 | 3311 | 1.7 | 18.2 | 40 | 7.36 | 0.06 |

| KN | NAME | R.A. (2000.0) H:M:S | DEC. (2000.0) D:M:S | F(65μm) Jy | F(90μm) Jy | F(140μm) Jy | F(160μm) Jy | QUAL. | DIAM. arcsec | F(1.4GHz) mJy | F(5GHz) mJy | DIST. kpc | DIAM. pc |
|---|---|---|---|---|---|---|---|---|---|---|---|---|---|
| 564 | He 2-432 | 19 23 24.821 | +21 08 00.50 | 1.213 | 1.211 | 0.805 | 1.393 | 1311 | 2.3 | 21.3 | 32 | 6.88 | 0.08 |
| 565 | He 2-434 | 19 33 49.420 | -74 32 58.66 | 2.155 | 0.770 | 0.869 | --- | 1311 | 10.0 | --- | --- | --- | --- |
| 567 | He 2-437 | 19 32 57.657 | +26 52 43.35 | 8.747 | 9.473 | 6.010 | 3.948 | 3331 | --- | 6.6 | --- | --- | --- |
| 568 | He 2-440 | 19 38 08.403 | +25 15 40.98 | 2.700 | 2.870 | 0.518 | 0.509 | 3311 | 2.2 | 26.4 | 43 | 6.45 | 0.07 |
| 570 | He 2-447 | 19 45 22.164 | +21 20 04.03 | 4.171 | 4.547 | 6.443 | 4.375 | 1131 | 1.2 | 22.9 | 60 | 7.63 | 0.04 |
| 571 | He 2-459 | 20 13 57.898 | +29 33 55.94 | 35.529 | 24.038 | 8.141 | 5.802 | 3331 | 1.3 | 13.9 | 64 | 7.24 | 0.05 |
| 572 | He 2-47 | 10 23 09.143 | -60 32 42.21 | 23.166 | 16.144 | 9.242 | 4.430 | 3331 | 5.0 | --- | 170 | --- | --- |
| 575 | He 2-50 | 10 34 19.002 | -53 41 03.75 | 1.386 | 1.491 | 1.105 | --- | 1311 | <25 | --- | 11.3 | 4.54 | --- |
| 576 | He 2-51 | 10 35 45.821 | -64 19 11.74 | 1.019 | 1.942 | 2.329 | 2.343 | 1311 | 9.0 | --- | 57 | 3.23 | 0.14 |
| 577 | He 2-55 | 10 48 43.168 | -56 03 10.21 | 0.961 | 1.151 | 0.498 | 0.539 | 1311 | <25 | --- | --- | --- | --- |
| 581 | He 2-64 | 11 27 24.259 | -57 17 58.99 | 0.779 | 0.495 | 2.263 | 0.248 | 1311 | --- | --- | --- | --- | --- |
| 582 | He 2-67 | 11 28 47.369 | -60 06 37.28 | 3.437 | 4.045 | 1.748 | 1.044 | 3311 | 5.0 | --- | 41 | --- | --- |
| 583 | He 2-68 | 11 31 45.427 | -65 58 13.67 | 1.823 | 1.576 | 0.778 | 0.445 | 1311 | 10.0 | --- | 34 | --- | --- |
| 584 | He 2-7 | 08 11 31.890 | -48 43 16.71 | 1.741 | 2.074 | 0.648 | --- | 1311 | 44.6 | --- | 47 | 2.91 | 0.63 |
| 586 | He 2-71 | 11 39 11.198 | -68 52 09.14 | 0.973 | 0.816 | 1.060 | 1.173 | 1311 | 5.0 | --- | 12 | --- | --- |
| 588 | He 2-73 | 11 48 38.191 | -65 08 37.33 | 4.998 | 4.438 | 2.214 | --- | 3311 | 4.0 | --- | 76 | 4.23 | 0.08 |
| 589 | He 2-76 | 12 08 25.433 | -64 12 09.32 | --- | 3.741 | 1.738 | 4.622 | 1311 | --- | --- | --- | --- | --- |
| 590 | He 2-78 | 12 09 10.196 | -58 42 37.44 | 0.664 | 0.842 | 1.869 | 1.298 | 1311 | --- | --- | --- | --- | --- |
| 591 | He 2-81 | 12 23 01.237 | -64 01 45.96 | 5.655 | 4.455 | --- | --- | 1311 | <25 | --- | --- | --- | --- |
| 593 | He 2-83 | 12 28 43.997 | -62 05 35.05 | 13.117 | 8.492 | 1.805 | 2.953 | 3311 | --- | --- | --- | --- | --- |
| 594 | He 2-84 | 12 28 46.822 | -63 44 37.22 | --- | 3.755 | 6.610 | 9.376 | 1311 | --- | --- | --- | --- | --- |
| 596 | He 2-86 | 12 30 30.426 | -64 52 05.58 | 18.480 | 14.073 | 3.369 | --- | 3311 | 3.6 | --- | 125 | 3.85 | 0.07 |
| 598 | He 2-9 | 08 28 27.988 | -39 23 40.27 | 5.928 | 4.829 | 0.147 | 1.568 | 3311 | 4.4 | 135.6 | 194.8 | 3.11 | 0.07 |
| 599 | He 2-90 | 13 09 36.252 | -61 19 35.96 | 12.683 | 9.680 | 7.373 | 11.110 | 1311 | 10.0 | --- | 25 | --- | --- |
| 600 | He 2-96 | 13 42 36.150 | -61 22 29.18 | 18.266 | 13.075 | 5.562 | 3.139 | 3311 | --- | --- | --- | --- | --- |
| 601 | He 2-97 | 13 45 22.394 | -71 28 56.13 | 1.942 | 1.843 | --- | --- | 1311 | 5.0 | --- | 30 | --- | --- |
| 602 | He 2-99 | 13 52 30.683 | -66 23 26.49 | 8.545 | 7.715 | 3.001 | 2.340 | 3311 | 17.0 | --- | 18 | 3.4 | 0.28 |
| 603 | He 3-1333 | 17 09 00.932 | -56 54 48.25 | 174.846 | 94.412 | 34.594 | 29.428 | 1333 | --- | --- | 26 | --- | --- |
| 606 | Hf 39 | 10 53 59.586 | -60 26 44.31 | 28.910 | 22.101 | 13.034 | 10.185 | 3331 | --- | --- | --- | --- | --- |
| 607 | Hf 48 | 11 03 55.979 | -60 36 04.57 | 2.802 | 2.693 | --- | 7.749 | 1311 | --- | --- | --- | --- | --- |
| 608 | Hu 1-1 | 00 28 15.614 | +55 57 54.71 | 2.450 | 2.793 | 3.616 | 3.719 | 1311 | 10.0 | 28 | 26 | 3.86 | 0.19 |
| 609 | Hu 1-2 | 21 33 08.349 | +39 38 09.57 | 2.478 | 2.558 | 2.059 | 1.414 | 3311 | 6.5 | 107 | 155 | 2.53 | 0.08 |
| 610 | Hu 2-1 | 18 49 47.562 | +20 50 39.46 | 4.192 | 3.208 | 2.134 | --- | 3311 | 1.8 | 42.6 | 110 | 4.52 | 0.04 |

| KN | NAME | R.A. (2000.0) H:M:S | DEC. (2000.0) D:M:S | F(65μm) Jy | F(90μm) Jy | F(140μm) Jy | F(160μm) Jy | QUAL. | DIAM. arcsec | F(1.4GHz) mJy | F(5GHz) mJy | DIST. kpc | DIAM. pc |
|---|---|---|---|---|---|---|---|---|---|---|---|---|---|
| 611 | IC 1295 | 18 54 37.206 | -08 49 39.08 | 0.714 | 2.886 | 2.997 | 0.009 | 1311 | 100.0 | 45.9 | 44 | 1.18 | 0.57 |
| 612 | IC 1297 | 19 17 23.459 | -39 36 46.40 | 3.558 | 4.067 | 0.836 | --- | 3311 | 7.0 | 59.4 | 68.9 | 3.75 | 0.13 |
| 613 | IC 1454 | 22 42 24.997 | +80 26 31.96 | 0.582 | 0.570 | --- | --- | 1311 | 34.0 | 3.3 | 1.3 | 5.29 | 0.87 |
| 614 | IC 1747 | 01 57 35.896 | +63 19 19.36 | 5.610 | 5.495 | 0.416 | 4.799 | 3311 | 13.0 | 85.8 | 124 | 2.23 | 0.14 |
| 615 | IC 2003 | 03 56 21.984 | +33 52 30.59 | 2.384 | 2.129 | --- | --- | 1311 | 10.0 | 54.8 | 30 | 3.89 | 0.19 |
| 616 | IC 2149 | 05 56 23.908 | +46 06 17.32 | 8.351 | 7.080 | 3.290 | 1.772 | 3331 | 8.4 | 177 | 280 | 2.03 | 0.08 |
| 617 | IC 2165 | 06 21 42.775 | -12 59 13.96 | 4.002 | 3.607 | --- | --- | 1311 | 9.0 | 180 | 188 | 2.47 | 0.11 |
| 618 | IC 2448 | 09 07 06.261 | -69 56 30.74 | 3.367 | 3.231 | 0.917 | 0.015 | 3311 | 10.0 | --- | 73 | 3.41 | 0.17 |
| 619 | IC 2501 | 09 38 47.213 | -60 05 30.92 | 10.743 | 9.401 | 2.776 | 0.924 | 3331 | 2.0 | --- | 261 | 2.09 | 0.02 |
| 620 | IC 2553 | 10 09 20.856 | -62 36 48.40 | 10.367 | 9.642 | 3.572 | 2.955 | 3311 | 9.0 | --- | 92 | 2.85 | 0.12 |
| 621 | IC 2621 | 11 00 20.111 | -65 14 57.77 | 16.660 | 13.745 | 6.657 | 2.474 | 3331 | 5.0 | --- | 195 | 2.94 | 0.07 |
| 622 | IC 289 | 03 10 19.273 | +61 19 00.91 | 7.280 | 8.816 | 1.371 | 0.208 | 3311 | 36.8 | 152 | 170 | 1.18 | 0.21 |
| 623 | IC 351 | 03 47 33.143 | +35 02 48.50 | 0.940 | 1.230 | 1.560 | 1.662 | 1311 | 7.0 | 31.9 | 27 | 4.45 | 0.15 |
| 624 | IC 3568 | 12 33 06.834 | +82 33 50.29 | 4.750 | 4.540 | 2.027 | --- | 3311 | 18.0 | 94.1 | 95 | 2.47 | 0.22 |
| 626 | IC 4191 | 13 08 47.343 | -67 38 37.67 | 14.074 | 12.272 | 3.336 | 0.740 | 3331 | 14.0 | --- | 172 | 1.95 | 0.13 |
| 627 | IC 4406 | 14 22 26.278 | -44 09 04.35 | 18.454 | 21.940 | 16.635 | 15.527 | 3333 | 20.0 | --- | 110 | 1.5 | 0.15 |
| 628 | IC 4593 | 16 11 44.545 | +12 04 17.06 | 13.840 | 11.520 | 3.349 | 1.929 | 3311 | 12.8 | 90.6 | 92 | 2.42 | 0.15 |
| 629 | IC 4634 | 17 01 33.572 | -21 49 32.77 | 8.034 | 6.439 | 2.370 | --- | 3311 | 5.5 | 116 | 100 | 3.47 | 0.09 |
| 630 | IC 4637 | 17 05 10.506 | -40 53 08.44 | 9.044 | 4.840 | 31.528 | 6.032 | 1331 | 18.6 | --- | 132.5 | 2.29 | 0.21 |
| 631 | IC 4642 | 17 11 45.025 | -55 24 01.47 | 5.784 | 5.526 | 2.109 | 1.953 | 3311 | 16.6 | --- | 60 | 2.52 | 0.20 |
| 633 | IC 4673 | 18 03 18.408 | -27 06 22.61 | 9.902 | 9.916 | 6.940 | 6.241 | 1311 | 16.0 | 53.4 | 62 | 3.12 | 0.24 |
| 634 | IC 4699 | 18 18 32.024 | -45 59 01.70 | 0.680 | 1.065 | --- | --- | 1311 | 5.0 | --- | 20 | 4.91 | 0.12 |
| 636 | IC 4846 | 19 16 28.220 | -09 02 36.57 | 2.221 | 1.882 | 0.336 | 0.547 | 1311 | 2.9 | 38.7 | 43 | 5.72 | 0.08 |
| 637 | IC 4997 | 20 20 08.741 | +16 43 53.71 | 6.372 | 4.577 | 1.158 | 2.358 | 3311 | 2.0 | 30.1 | 127 | 5.45 | 0.05 |
| 638 | IC 5117 | 21 32 31.027 | +44 35 48.53 | 13.535 | 11.489 | 3.659 | 0.543 | 3111 | 1.5 | 34.1 | 210 | 5.01 | 0.04 |
| 639 | IC 5217 | 22 23 55.725 | +50 58 00.43 | 3.748 | 3.111 | 1.901 | 0.711 | 3311 | 6.5 | 50.4 | 163 | 2.71 | 0.09 |
| 640 | IC 972 | 14 04 25.863 | -17 13 40.84 | 0.633 | 0.604 | 0.214 | 0.676 | 1311 | 47.0 | 3.9 | 6.8 | 2.87 | 0.65 |
| 644 | J 900 | 06 25 57.275 | +17 47 27.19 | 5.565 | 4.194 | 2.942 | 1.558 | 3311 | 6.0 | 107.8 | 100 | 3.29 | 0.10 |
| 651 | KFL 13 | 18 12 44.988 | -25 44 23.85 | 0.222 | 1.359 | --- | 1.112 | 1311 | --- | 4 | 3.5 | --- | --- |
| 652 | KFL 14 | 18 13 01.106 | -29 25 15.59 | 0.088 | 0.977 | 0.676 | --- | 1311 | --- | --- | 1.5 | --- | --- |
| 660 | KFL 9 | 18 07 19.381 | -31 42 56.95 | --- | 0.838 | --- | --- | 1311 | --- | 2.8 | 3.1 | --- | --- |
| 666 | K 1-14 | 17 42 36.757 | +21 27 02.06 | 0.454 | 0.708 | 0.813 | 0.763 | 1311 | 47.0 | --- | --- | --- | --- |
| 669 | K 1-17 | 19 03 37.362 | +19 21 22.60 | 0.610 | 1.071 | 0.621 | 2.994 | 1311 | 45.0 | 4.6 | --- | --- | --- |

| KN | NAME | R.A. (2000.0) H:M:S | DEC. (2000.0) D:M:S | F(65μm) Jy | F(90μm) Jy | F(140μm) Jy | F(160μm) Jy | QUAL. | DIAM. arcsec | F(1.4GHz) mJy | F(5GHz) mJy | DIST. kpc | DIAM. pc |
|---|---|---|---|---|---|---|---|---|---|---|---|---|---|
| 672 | K 1-21 | 08 04 14.193 | -34 16 07.32 | 0.272 | 0.796 | --- | --- | 1311 | 29.0 | 6.5 | 5 | 3.88 | 0.55 |
| 681 | K 2-16 | 16 44 49.050 | -28 04 04.56 | 22.836 | 17.837 | 4.856 | 1.122 | 3331 | --- | --- | --- | --- | --- |
| 684 | K 2-5 | 17 54 26.179 | -12 48 35.95 | 0.181 | 0.551 | 2.464 | 2.452 | 1311 | 24.6 | 4 | 3 | --- | --- |
| 686 | K 3-1 | 18 23 21.721 | +03 36 27.78 | 0.220 | 0.637 | 0.399 | 4.204 | 1311 | 5.0 | 5.4 | 5 | 8.31 | 0.20 |
| 687 | K 3-11 | 18 41 07.311 | -08 55 58.97 | 8.769 | 4.539 | 1.898 | 14.470 | 3111 | 3.0 | 14.5 | 17 | 7.33 | 0.11 |
| 693 | K 3-18 | 19 00 34.829 | -02 11 57.62 | 4.782 | 4.264 | --- | --- | 3311 | 3.5 | --- | 11 | 13.98 | 0.23 |
| 695 | K 3-2 | 18 25 00.576 | -01 30 52.64 | 2.934 | 1.956 | --- | --- | 3311 | 2.8 | 25.9 | 31 | 6.37 | 0.09 |
| 696 | K 3-20 | 19 02 10.154 | -01 48 45.31 | 1.358 | 1.085 | 0.791 | --- | 1311 | --- | 6.3 | --- | --- | --- |
| 697 | K 3-21 | 19 02 40.327 | +14 28 50.36 | 0.150 | 0.373 | --- | 0.189 | 1311 | --- | --- | --- | --- | --- |
| 699 | K 3-24 | 19 12 05.820 | +15 09 04.47 | 1.402 | 1.896 | 1.190 | 1.693 | 1311 | 6.2 | 8.5 | --- | --- | --- |
| 700 | K 3-26 | 19 14 39.178 | +00 13 36.29 | 1.422 | 1.442 | --- | --- | 1311 | --- | 6.4 | 0.5 | --- | --- |
| 702 | K 3-29 | 19 15 30.561 | +14 03 49.83 | 5.471 | 5.005 | --- | --- | 1311 | 1.0 | 13.9 | 65 | 8.07 | 0.04 |
| 703 | K 3-3 | 18 27 09.338 | +01 14 26.88 | 5.058 | 7.101 | 1.615 | 2.432 | 3311 | 10.0 | 35.9 | 34 | 3.43 | 0.17 |
| 704 | K 3-30 | 19 16 27.691 | +05 13 19.47 | 0.539 | 1.423 | 4.218 | 1.382 | 1311 | 3.3 | 12.1 | 23 | 6.46 | 0.10 |
| 705 | K 3-31 | 19 19 02.666 | +19 02 20.85 | 1.665 | 1.718 | --- | --- | 1311 | 1.5 | 16.8 | 39 | 7.83 | 0.06 |
| 708 | K 3-35 | 19 27 44.036 | +21 30 03.83 | 37.657 | 19.471 | 4.673 | 0.634 | 3311 | 1.7 | 14.5 | 60 | 6.56 | 0.05 |
| 709 | K 3-36 | 19 32 39.557 | +07 27 51.57 | 0.342 | 1.393 | 1.147 | 2.954 | 1311 | --- | 3.1 | 0.2 | --- | --- |
| 710 | K 3-37 | 19 33 46.749 | +24 32 27.09 | 0.797 | 0.821 | 0.198 | 1.414 | 1311 | 2.5 | 14.4 | 17 | 7.93 | 0.10 |
| 711 | K 3-38 | 19 35 18.354 | +17 13 00.71 | 3.674 | 4.207 | 4.376 | 5.177 | 3311 | 5.0 | 28.7 | 31 | 5.09 | 0.12 |
| 712 | K 3-39 | 19 35 54.469 | +24 54 48.20 | 3.351 | 2.565 | 0.402 | 1.241 | 3311 | 1.0 | 3.6 | 11 | 13.35 | 0.06 |
| 713 | K 3-4 | 18 31 00.227 | +02 25 27.35 | 3.694 | 2.055 | 4.661 | 0.696 | 1311 | 21.0 | 17 | 21 | --- | --- |
| 714 | K 3-40 | 19 36 21.828 | +23 39 47.93 | 3.452 | 3.892 | --- | --- | 3311 | 4.0 | 17.1 | 20 | 6.18 | 0.12 |
| 717 | K 3-43 | 19 40 25.908 | +18 49 14.20 | 0.482 | 0.690 | 0.122 | --- | 1311 | 3.0 | 4.8 | 0.4 | --- | --- |
| 720 | K 3-49 | 19 54 00.702 | +33 22 12.95 | 1.097 | 1.077 | --- | --- | 1311 | --- | 5.5 | 7 | 34.5 | --- |
| 721 | K 3-5 | 18 31 45.833 | +04 05 09.12 | 1.543 | 1.481 | 2.463 | 2.296 | 1311 | 10.0 | 6.7 | 3 | 7.11 | 0.34 |
| 723 | K 3-52 | 20 03 11.435 | +30 32 34.15 | 35.941 | 24.008 | 10.002 | 8.825 | 3331 | 0.7 | 16.9 | 65 | 9.42 | 0.03 |
| 724 | K 3-53 | 20 03 22.475 | +27 00 54.73 | 5.133 | 4.756 | 0.601 | --- | 3311 | 0.9 | 10 | 69 | 9.27 | 0.04 |
| 726 | K 3-55 | 20 06 56.210 | +32 16 33.60 | 5.551 | 7.453 | 7.527 | 6.486 | 1311 | 8.2 | 86.6 | 90 | 2.96 | 0.12 |
| 728 | K 3-57 | 20 12 47.679 | +34 20 32.52 | 2.474 | 3.289 | 1.827 | --- | 1311 | 6.3 | 46.8 | 60 | 3.72 | 0.11 |
| 729 | K 3-58 | 20 21 58.324 | +29 59 22.15 | 2.219 | 3.014 | 2.792 | 1.481 | 1311 | 4.8 | 5.4 | --- | --- | --- |
| 730 | K 3-6 | 18 33 17.490 | +00 11 47.19 | 5.499 | 4.908 | 4.120 | 2.520 | 3311 | 0.7 | 10.4 | 55 | 10.2 | 0.03 |
| 732 | K 3-61 | 21 30 00.710 | +54 27 27.45 | 1.973 | 1.628 | --- | 0.663 | 1311 | 6.0 | 16.6 | 14 | 5.73 | 0.17 |
| 733 | K 3-62 | 21 31 50.203 | +52 33 51.64 | 4.669 | 3.505 | --- | --- | 3311 | 2.5 | 59.5 | 115 | 4.62 | 0.06 |

| KN | NAME | R.A. (2000.0) H:M:S | DEC. (2000.0) D:M:S | F(65μm) Jy | F(90μm) Jy | F(140μm) Jy | F(160μm) Jy | QUAL. | DIAM. arcsec | F(1.4GHz) mJy | F(5GHz) mJy | DIST. kpc | DIAM. pc |
|---|---|---|---|---|---|---|---|---|---|---|---|---|---|
| 734 | K 3-63 | 21 39 11.976 | +55 46 03.94 | 0.395 | 1.098 | 1.335 | --- | 1311 | 7.0 | 8.5 | 29 | --- | --- |
| 735 | K 3-64 | 04 13 27.255 | +51 51 00.92 | --- | 0.392 | 0.082 | 0.596 | 1311 | --- | --- | --- | --- | --- |
| 737 | K 3-66 | 04 36 37.243 | +33 39 29.87 | 1.272 | 0.594 | 0.104 | 0.766 | 1311 | 2.1 | 15.4 | 18 | 8.42 | 0.09 |
| 738 | K 3-67 | 04 39 47.905 | +36 45 42.85 | 1.832 | 1.820 | --- | --- | 1311 | 2.2 | 33.3 | 42 | 6.49 | 0.07 |
| 740 | K 3-69 | 05 41 22.147 | +39 15 08.09 | 1.857 | 2.087 | 2.649 | --- | 3311 | 0.5 | 2.9 | 0.4 | 13.87 | 0.03 |
| 741 | K 3-7 | 18 34 13.601 | -02 27 36.41 | 4.976 | 4.560 | --- | --- | 1311 | 6.3 | 29.5 | 30 | 4.52 | 0.14 |
| 742 | K 3-70 | 05 58 45.350 | +25 18 43.86 | 1.236 | 1.558 | 2.457 | --- | 1311 | 1.6 | 4.4 | 6 | --- | --- |
| 746 | K 3-76 | 20 25 04.865 | +33 34 50.70 | 2.488 | 2.112 | --- | 3.736 | 1311 | 0.2 | 5.1 | 12 | --- | --- |
| 747 | K 3-78 | 20 45 22.710 | +50 22 39.92 | 1.991 | 2.624 | 0.090 | 1.142 | 1311 | 3.8 | 14.9 | 15 | 6.85 | 0.13 |
| 750 | K 3-82 | 21 30 51.636 | +50 00 06.96 | 2.461 | 3.525 | 2.515 | 1.426 | 3311 | 24.0 | 35.6 | 30 | 2.49 | 0.29 |
| 751 | K 3-83 | 21 35 43.877 | +50 54 16.94 | 1.270 | 1.835 | --- | --- | 1311 | 5.0 | 4.7 | 6.5 | 7.71 | 0.19 |
| 752 | K 3-84 | 21 38 49.006 | +46 00 27.81 | 0.705 | 0.608 | 0.535 | --- | 1311 | --- | 2.7 | --- | --- | --- |
| 754 | K 3-88 | 23 12 15.494 | +64 39 17.73 | 0.599 | 0.948 | --- | 0.494 | 1311 | --- | 7.1 | 0.3 | --- | --- |
| 755 | K 3-90 | 01 24 58.628 | +65 38 36.14 | 1.083 | 0.569 | --- | --- | 1311 | 8.5 | 12.4 | 13.9 | 4.6 | 0.19 |
| 757 | K 3-92 | 02 03 41.171 | +64 57 37.88 | 0.237 | 0.527 | --- | 0.724 | 1311 | 13.0 | --- | 2 | 7.11 | 0.45 |
| 759 | K 3-94 | 03 36 08.086 | +60 03 46.29 | 0.364 | 1.022 | 1.299 | 0.377 | 1311 | 10.0 | 6.9 | 5.5 | 5.98 | 0.29 |
| 761 | K 4-19 | 19 13 22.624 | +03 25 00.30 | 2.142 | 1.463 | 0.425 | --- | 3311 | --- | --- | --- | --- | --- |
| 764 | K 4-41 | 19 56 34.027 | +32 22 12.96 | 2.439 | 2.500 | 0.831 | 0.145 | 3311 | 3.0 | 11.5 | 15 | 7.59 | 0.11 |
| 765 | K 4-47 | 04 20 45.157 | +56 18 11.57 | 5.138 | 5.585 | 3.408 | 1.143 | 1331 | 7.8 | --- | 7.7 | 22.46 | 0.85 |
| 766 | K 4-48 | 06 39 55.879 | +11 06 30.69 | 3.627 | 3.673 | 0.833 | --- | 3311 | 2.2 | 11.8 | 14 | 8.86 | 0.09 |
| 768 | K 4-53 | 20 42 16.744 | +37 40 25.58 | 3.450 | 5.024 | 0.890 | 0.592 | 3311 | --- | 34.4 | --- | --- | --- |
| 769 | K 4-55 | 20 45 10.021 | +44 39 14.58 | 0.211 | 1.688 | --- | --- | 1311 | --- | 6.2 | --- | --- | --- |
| 775 | KjPn 6 | 22 49 02.175 | +67 01 38.89 | 0.518 | 0.873 | 1.165 | 1.764 | 1311 | --- | 8 | --- | --- | --- |
| 779 | LoTr 10 | 14 46 20.239 | -61 13 35.37 | 4.579 | 2.985 | 2.254 | 2.351 | 1311 | --- | --- | --- | --- | --- |
| 782 | LoTr 7 | 14 15 24.025 | -67 31 55.87 | 0.130 | 0.861 | 1.249 | 1.025 | 1311 | --- | --- | --- | --- | --- |
| 787 | Lo 11 | 16 03 22.211 | -36 00 53.73 | 0.205 | 1.758 | 1.073 | --- | 1311 | --- | 2.7 | 3.5 | --- | --- |
| 790 | Lo 16 | 17 35 41.798 | -40 11 26.20 | 3.656 | 10.988 | 4.495 | 10.058 | 3311 | --- | 89.9 | --- | --- | --- |
| 797 | Lo 9 | 15 42 13.340 | -47 40 46.33 | 0.121 | 0.630 | --- | 0.090 | 1311 | --- | --- | --- | --- | --- |
| 798 | MA 13 | 18 30 30.388 | -07 27 38.29 | 10.933 | 5.667 | --- | --- | 3311 | --- | 28.7 | --- | --- | --- |
| 799 | MA 2 | 18 15 13.391 | -06 57 12.20 | 1.006 | 1.304 | 0.076 | 0.792 | 1311 | --- | 7.4 | --- | --- | --- |
| 802 | M 1-1 | 01 37 19.430 | +50 28 11.50 | 0.993 | 0.510 | --- | 1.437 | 1311 | 4.5 | 14 | 8 | 7.27 | 0.16 |
| 803 | M 1-11 | 07 11 16.693 | -19 51 02.87 | 18.800 | 15.084 | 3.463 | 1.214 | 3311 | 2.2 | 25.9 | 113 | 4.9 | 0.05 |
| 804 | M 1-12 | 07 19 21.471 | -21 43 55.46 | 3.723 | 2.383 | 1.515 | --- | 3311 | 1.8 | 21.9 | 41 | 7.13 | 0.06 |

| KN | NAME | R.A. (2000.0) H:M:S | DEC. (2000.0) D:M:S | F(65μm) Jy | F(90μm) Jy | F(140μm) Jy | F(160μm) Jy | QUAL. | DIAM. arcsec | F(1.4GHz) mJy | F(5GHz) mJy | DIST. kpc | DIAM. pc |
|---|---|---|---|---|---|---|---|---|---|---|---|---|---|
| 805 | M 1-13 | 07 21 14.952 | -18 08 36.91 | 2.662 | 4.368 | 3.749 | --- | 1311 | 10.0 | 18.9 | 15 | 4.51 | 0.22 |
| 806 | M 1-14 | 07 27 56.510 | -20 13 22.89 | 2.101 | 1.844 | 1.666 | 0.876 | 3311 | 4.7 | 59.1 | 60 | 4.23 | 0.10 |
| 807 | M 1-16 | 07 37 18.955 | -09 38 49.67 | 8.731 | 7.022 | 7.089 | 6.625 | 3331 | 3.6 | 31.3 | 31 | 5.71 | 0.10 |
| 808 | M 1-17 | 07 40 22.206 | -11 32 29.81 | 3.930 | 3.522 | 2.095 | 3.194 | 3311 | 2.5 | 18.3 | 17 | 8.23 | 0.10 |
| 811 | M 1-2 | 01 58 49.675 | +52 53 48.57 | 1.513 | 1.075 | 0.416 | 0.094 | 1311 | 0.5 | --- | 10 | --- | --- |
| 812 | M 1-22 | 17 35 10.239 | -18 34 20.41 | 2.072 | 1.918 | --- | 2.497 | 1311 | 6.0 | 7.4 | 3.5 | 11.92 | 0.35 |
| 813 | M 1-23 | 17 37 22.003 | -18 46 41.87 | 2.195 | 3.146 | 2.712 | --- | 1311 | --- | 14.1 | --- | --- | --- |
| 814 | M 1-24 | 17 38 11.588 | -19 37 37.64 | 7.928 | 7.466 | 0.478 | 0.707 | 3311 | <10 | 43.5 | --- | --- | --- |
| 815 | M 1-25 | 17 38 30.307 | -22 08 38.88 | 6.244 | 6.016 | 0.689 | --- | 3311 | 3.2 | 39.9 | 57 | 4.93 | 0.08 |
| 819 | M 1-30 | 17 52 58.946 | -34 38 22.99 | 5.714 | 5.292 | --- | --- | 3311 | 3.5 | 23.4 | 31 | 5.9 | 0.10 |
| 822 | M 1-33 | 17 58 58.794 | -15 32 14.79 | 6.185 | 7.632 | --- | 3.584 | 3311 | 4.0 | 49 | 60 | 4.63 | 0.09 |
| 823 | M 1-34 | 18 01 22.193 | -33 17 43.08 | 2.006 | 3.779 | 1.363 | 7.131 | 1311 | 11.2 | 13.6 | 14.7 | 5.95 | 0.32 |
| 826 | M 1-38 | 18 06 05.765 | -28 40 29.28 | --- | 3.022 | --- | --- | 1311 | 3.5 | 14.9 | 24 | 6.58 | 0.11 |
| 828 | M 1-4 | 03 41 43.428 | +52 17 00.28 | 2.201 | 2.263 | --- | --- | 3311 | 6.0 | 77.8 | 90 | 3.39 | 0.10 |
| 829 | M 1-40 | 18 08 25.994 | -22 16 53.25 | 44.078 | 20.960 | --- | 7.625 | 3311 | 4.5 | 163.3 | 208 | 2.83 | 0.06 |
| 831 | M 1-42 | 18 11 05.028 | -28 58 59.33 | 5.007 | 7.279 | 8.013 | 4.503 | 3331 | 9.0 | 28.3 | 28.5 | 4.89 | 0.21 |
| 833 | M 1-44 | 18 16 17.365 | -27 04 32.47 | 5.194 | 5.412 | 6.768 | 2.805 | 1131 | 4.0 | 9.4 | 9 | 9.24 | 0.18 |
| 835 | M 1-46 | 18 27 56.339 | -15 32 54.43 | 19.466 | 12.055 | 3.570 | 4.934 | 3311 | 11.5 | 78.1 | 81 | 2.58 | 0.14 |
| 836 | M 1-47 | 18 29 11.154 | -21 46 53.43 | 1.007 | 1.099 | --- | --- | 1311 | 5.5 | 11.7 | 14 | 5.95 | 0.16 |
| 837 | M 1-48 | 18 29 29.991 | -19 05 44.58 | 1.291 | 2.036 | 2.169 | 1.771 | 1311 | 4.8 | 5 | 6 | 7.57 | 0.18 |
| 838 | M 1-5 | 05 46 50.000 | +24 22 02.32 | 2.046 | 1.733 | 0.522 | --- | 1311 | 2.3 | 41.1 | 71 | 5.49 | 0.06 |
| 839 | M 1-50 | 18 33 20.886 | -18 16 36.86 | 2.416 | 3.361 | 0.749 | 0.299 | 1311 | 5.6 | 35.4 | 50 | 4.12 | 0.11 |
| 842 | M 1-53 | 18 35 48.267 | -17 36 08.71 | 3.355 | 4.380 | 4.325 | 0.782 | 3331 | 6.0 | 19.6 | 8 | 3.93 | 0.11 |
| 843 | M 1-54 | 18 36 08.357 | -16 59 57.02 | 5.805 | 6.572 | 5.373 | 4.726 | 3331 | 13.0 | 35.5 | 53 | 3.09 | 0.19 |
| 845 | M 1-56 | 18 37 46.250 | -17 05 46.55 | 3.028 | 3.273 | 0.560 | 3.358 | 3311 | 1.5 | 14.6 | 22 | 9.48 | 0.07 |
| 847 | M 1-58 | 18 42 56.979 | -11 06 53.10 | 5.187 | 4.822 | --- | 0.323 | 3311 | 6.4 | 37.7 | 60 | 3.69 | 0.11 |
| 849 | M 1-6 | 06 35 45.140 | -00 05 37.37 | 3.831 | 2.361 | 3.460 | --- | 3111 | 3.0 | 54.4 | 86 | 4.7 | 0.07 |
| 850 | M 1-60 | 18 43 38.107 | -13 44 48.64 | 6.538 | 6.621 | 0.146 | 1.540 | 3311 | 2.5 | 35.1 | 48 | 5.81 | 0.07 |
| 851 | M 1-61 | 18 45 55.072 | -14 27 37.93 | 11.311 | 9.759 | --- | 1.401 | 3311 | 1.8 | 32.5 | 97 | 5.59 | 0.05 |
| 853 | M 1-63 | 18 51 30.935 | -13 10 36.81 | 1.535 | 1.661 | 1.588 | --- | 1311 | 4.2 | 8.4 | 10.2 | 7.32 | 0.15 |
| 854 | M 1-64 | 18 50 02.090 | +35 14 36.10 | 0.761 | 0.510 | 0.321 | 2.018 | 1311 | 17.0 | 4.3 | 2.4 | 6 | 0.49 |
| 855 | M 1-65 | 18 56 33.639 | +10 52 10.05 | 2.434 | 1.681 | 1.962 | 0.760 | 3311 | 4.0 | 21 | 23 | 5.94 | 0.12 |
| 856 | M 1-66 | 18 58 26.247 | -01 03 45.70 | 2.414 | 2.583 | --- | 14.087 | 1311 | 2.7 | 45.7 | 59 | 5.39 | 0.07 |

| KN | NAME | R.A. (2000.0) H:M:S | DEC. (2000.0) D:M:S | F(65μm) Jy | F(90μm) Jy | F(140μm) Jy | F(160μm) Jy | QUAL. | DIAM. arcsec | F(1.4GHz) mJy | F(5GHz) mJy | DIST. kpc | DIAM. pc |
|---|---|---|---|---|---|---|---|---|---|---|---|---|---|
| 857 | M 1-67 | 19 11 30.881 | +16 51 38.21 | 26.235 | 28.609 | 16.751 | 8.822 | 3333 | 120.4 | 217.6 | 250 | 0.69 | 0.40 |
| 858 | M 1-69 | 19 13 53.961 | +03 37 41.90 | 3.234 | 3.032 | --- | --- | 3311 | --- | 31.4 | --- | --- | --- |
| 859 | M 1-7 | 06 37 20.955 | +24 00 35.38 | 3.427 | 5.261 | 4.004 | 2.220 | 3331 | 11.0 | 18.5 | 13 | 4.5 | 0.24 |
| 860 | M 1-71 | 19 36 26.927 | +19 42 23.99 | 14.339 | 11.593 | 3.032 | 4.860 | 1311 | 3.0 | 83.8 | 204 | --- | --- |
| 861 | M 1-72 | 19 41 33.975 | +17 45 17.71 | 3.039 | 2.610 | --- | 0.131 | 3311 | 0.7 | 5.3 | 30 | 12.54 | 0.04 |
| 862 | M 1-73 | 19 41 09.323 | +14 56 59.37 | 6.896 | 7.010 | 2.241 | --- | 3311 | 6.0 | 47.7 | 43 | 4.17 | 0.12 |
| 863 | M 1-74 | 19 42 18.874 | +15 09 08.16 | 2.335 | 1.682 | --- | --- | 1311 | 1.0 | 7.9 | 29 | 10.15 | 0.05 |
| 865 | M 1-77 | 21 19 07.360 | +46 18 47.24 | 8.792 | 7.776 | 1.680 | 0.532 | 3311 | 8.0 | 29.4 | 25 | 4.29 | 0.17 |
| 866 | M 1-78 | 21 20 44.809 | +51 53 27.88 | 843.838 | 742.643 | 243.538 | 250.943 | 1133 | 6.0 | 363.5 | 1103.9 | 1.62 | 0.05 |
| 867 | M 1-79 | 21 37 01.500 | +48 56 02.62 | 1.384 | 3.435 | 2.012 | 1.020 | 1311 | 31.0 | 21.8 | 19 | 2.62 | 0.39 |
| 868 | M 1-8 | 06 53 33.795 | +03 08 26.96 | 2.242 | 3.569 | 2.483 | 0.837 | 3311 | 18.4 | 16.3 | 23 | 3.06 | 0.27 |
| 869 | M 1-80 | 22 56 19.806 | +57 09 20.68 | 0.845 | 0.859 | --- | --- | 1311 | 8.0 | 17.9 | 25 | 4.29 | 0.17 |
| 870 | M 1-9 | 07 05 19.204 | +02 46 59.16 | 1.085 | 0.680 | --- | 0.802 | 1311 | 2.3 | 22.4 | 27 | 7.22 | 0.08 |
| 871 | M 2-10 | 17 14 07.022 | -31 19 42.59 | 3.438 | 3.612 | --- | 0.790 | 3311 | 2.2 | 11.6 | 9.1 | 10.71 | 0.11 |
| 872 | M 2-11 | 17 20 33.252 | -29 00 38.66 | 0.775 | 1.457 | 1.927 | 1.417 | 1311 | 2.2 | 17.4 | 22 | 7.32 | 0.08 |
| 873 | M 2-12 | 17 24 01.451 | -25 59 23.16 | 3.449 | 2.817 | --- | 1.752 | 3311 | 4.4 | 10.3 | 12.9 | --- | --- |
| 875 | M 2-14 | 17 41 57.252 | -24 11 15.64 | 7.255 | 7.444 | 0.303 | --- | 3311 | 2.2 | 22.4 | 39.1 | --- | --- |
| 876 | M 2-15 | 17 46 54.449 | -16 17 24.80 | 2.663 | 3.981 | 0.105 | --- | 3311 | 7.4 | 18.7 | 12.5 | 5.37 | 0.19 |
| 877 | M 2-16 | 17 52 34.362 | -32 45 51.10 | 1.819 | 4.330 | 0.702 | --- | 1311 | 2.5 | 20.4 | 24.8 | 5.86 | 0.07 |
| 881 | M 2-2 | 04 13 15.045 | +56 56 58.08 | 3.482 | 3.280 | --- | 0.761 | 3311 | 6.5 | 52 | 54 | 3.66 | 0.12 |
| 886 | M 2-24 | 18 02 02.881 | -34 27 47.07 | 3.003 | 3.345 | 0.489 | 0.747 | 3311 | 6.8 | 3.2 | 3 | 11.79 | 0.39 |
| 889 | M 2-27 | 18 03 52.588 | -31 17 46.54 | 5.258 | 5.361 | 1.533 | --- | 3311 | 3.8 | 22.5 | 50 | 5.65 | 0.10 |
| 892 | M 2-30 | 18 12 34.414 | -27 58 11.59 | 1.596 | 2.289 | 3.750 | 2.127 | 1311 | 3.5 | 13.9 | 14 | 8.1 | 0.14 |
| 893 | M 2-31 | 18 13 16.026 | -25 30 04.97 | 3.561 | 3.598 | --- | 0.951 | 3311 | 4.0 | 39.6 | 51 | 4.91 | 0.10 |
| 895 | M 2-33 | 18 15 06.534 | -30 15 32.89 | 1.963 | 1.768 | 2.255 | 0.189 | 1331 | 5.0 | 11.9 | 12 | 8.52 | 0.21 |
| 896 | M 2-34 | 18 17 15.947 | -23 58 54.51 | 1.342 | 2.897 | --- | --- | 1311 | <25 | 4.8 | --- | --- | --- |
| 897 | M 2-35 | 18 17 37.195 | -31 56 46.86 | --- | 0.643 | --- | 0.424 | 1311 | --- | --- | --- | --- | --- |
| 898 | M 2-36 | 18 17 41.418 | -29 08 19.59 | 3.903 | 3.904 | 0.449 | 1.042 | 3311 | 6.8 | 22.6 | 13 | 6.99 | 0.23 |
| 899 | M 2-37 | 18 18 38.352 | -28 08 01.00 | 0.830 | 1.565 | 2.636 | --- | 1331 | --- | --- | --- | --- | --- |
| 901 | M 2-39 | 18 22 01.148 | -24 10 40.18 | 2.212 | 1.762 | 0.729 | --- | 3311 | 3.2 | 7.2 | 8 | 10.21 | 0.16 |
| 902 | M 2-4 | 17 01 06.231 | -34 49 38.58 | 3.778 | 3.978 | 0.977 | --- | 3311 | 2.0 | 24.5 | 32 | 7.15 | 0.07 |
| 903 | M 2-40 | 18 21 23.851 | -06 01 55.79 | 7.267 | 7.246 | 3.168 | --- | 1311 | 5.5 | 45.5 | 33 | 4.67 | 0.12 |
| 904 | M 2-41 | 18 22 34.379 | -30 43 29.66 | 1.289 | 1.000 | 0.666 | 2.018 | 1311 | --- | 3 | --- | --- | --- |

| KN | NAME | R.A. (2000.0) H:M:S | DEC. (2000.0) D:M:S | F(65μm) Jy | F(90μm) Jy | F(140μm) Jy | F(160μm) Jy | QUAL. | DIAM. arcsec | F(1.4GHz) mJy | F(5GHz) mJy | DIST. kpc | DIAM. pc |
|---|---|---|---|---|---|---|---|---|---|---|---|---|---|
| 905 | M 2-42 | 18 22 32.020 | -24 09 28.40 | 1.337 | 1.655 | 1.410 | --- | 1311 | 3.9 | 9.6 | 14 | 7.83 | 0.15 |
| 906 | M 2-43 | 18 26 40.048 | -02 42 57.63 | 13.475 | 9.881 | 4.454 | 0.769 | 3331 | 1.5 | 20.4 | 237 | 5 | 0.04 |
| 907 | M 2-44 | 18 37 36.908 | -03 05 55.96 | 5.521 | 5.483 | --- | 2.237 | 1311 | 8.0 | 48.3 | 54 | 3.45 | 0.13 |
| 909 | M 2-46 | 18 46 34.620 | -08 28 01.85 | 2.976 | 4.938 | 2.008 | 2.894 | 3311 | 4.4 | 13.1 | 12 | 6.48 | 0.14 |
| 910 | M 2-47 | 19 13 34.559 | +04 38 04.45 | 3.707 | 4.756 | 0.071 | 0.869 | 3311 | 6.0 | 38.3 | 45 | 4.12 | 0.12 |
| 913 | M 2-5 | 17 02 19.069 | -33 10 05.01 | 5.407 | 5.084 | 2.977 | --- | 3311 | 5.0 | 14.1 | 12 | 7.69 | 0.19 |
| 914 | M 2-50 | 21 57 41.814 | +51 41 39.01 | 0.367 | 0.487 | --- | 0.470 | 1311 | 4.6 | 8.2 | 6.5 | 8.07 | 0.18 |
| 915 | M 2-51 | 22 16 03.890 | +57 28 33.75 | 2.498 | 5.809 | 7.035 | 6.729 | 3331 | 39.2 | 27.2 | 40.6 | 1.88 | 0.36 |
| 917 | M 2-53 | 22 32 17.720 | +56 10 26.12 | 1.772 | 2.662 | 1.949 | 1.464 | 1311 | 18.0 | 14.6 | 11 | 3.64 | 0.32 |
| 918 | M 2-54 | 22 51 38.923 | +51 50 42.50 | 4.945 | 3.546 | 0.724 | 0.357 | 3311 | --- | 5.4 | 8 | 14.61 | --- |
| 919 | M 2-55 | 23 31 52.708 | +70 22 10.11 | 1.845 | 3.756 | 3.575 | 1.759 | 1331 | 41.0 | 25.4 | 19 | 2.31 | 0.46 |
| 921 | M 2-7 | 17 05 13.729 | -30 32 18.46 | 1.162 | 1.162 | 1.019 | --- | 1311 | <25 | 6 | 7 | 8.55 | --- |
| 924 | M 3-1 | 07 02 49.980 | -31 35 31.50 | 0.680 | 1.457 | 0.270 | 1.050 | 1311 | 11.2 | 25.7 | 24 | 3.78 | 0.21 |
| 925 | M 3-10 | 17 27 20.152 | -28 27 51.15 | 2.650 | 2.658 | --- | 2.107 | 1311 | 3.1 | 35.2 | 29 | 6.33 | 0.10 |
| 926 | M 3-12 | 17 36 22.634 | -21 31 12.22 | 2.522 | 2.640 | 0.961 | 0.396 | 3311 | 7.5 | 11.2 | 12.5 | 6.98 | 0.25 |
| 927 | M 3-13 | 17 41 36.594 | -22 13 02.46 | 1.684 | 1.432 | --- | --- | 1311 | <7 | --- | --- | --- | --- |
| 933 | M 3-20 | 17 59 19.318 | -28 13 48.05 | 1.786 | 1.353 | --- | --- | 1311 | 4.0 | 16.2 | 40 | 5.3 | 0.10 |
| 934 | M 3-21 | 18 02 32.324 | -36 39 12.24 | 2.846 | 2.546 | 1.570 | 0.155 | 3311 | <5 | 15.6 | 30 | 5.57 | --- |
| 935 | M 3-22 | 18 02 19.238 | -30 14 25.38 | 1.334 | 1.319 | 0.941 | 0.998 | 1311 | 6.4 | 10.1 | 8.7 | 8.45 | 0.26 |
| 937 | M 3-24 | 18 07 53.914 | -25 24 02.71 | 5.998 | 3.891 | 0.071 | --- | 3311 | <25 | 17.4 | --- | --- | --- |
| 938 | M 3-25 | 18 15 16.967 | -10 10 09.47 | 10.367 | 9.618 | 3.429 | 10.727 | 3311 | 1.5 | 25 | 76 | 6.89 | 0.05 |
| 940 | M 3-27 | 18 27 48.273 | +14 29 06.06 | 1.087 | 0.606 | 2.032 | --- | 1311 | 1.0 | --- | 53.2 | 8.55 | 0.04 |
| 942 | M 3-29 | 18 39 25.772 | -30 40 36.71 | 0.337 | 0.899 | 0.928 | 0.985 | 1311 | 8.2 | 7.6 | 18.7 | 5.86 | 0.23 |
| 943 | M 3-3 | 07 26 34.228 | -05 21 52.16 | 0.185 | 0.596 | 0.466 | --- | 1311 | 12.2 | --- | 5.6 | 5.46 | 0.32 |
| 944 | M 3-30 | 18 41 14.938 | -15 33 43.58 | 2.824 | 3.184 | --- | --- | 3311 | 17.2 | 8.5 | 7 | 3.97 | 0.33 |
| 946 | M 3-32 | 18 44 43.127 | -25 21 33.85 | 2.932 | 2.875 | 3.040 | 0.488 | 3311 | 7.5 | 12.2 | 12 | 7.06 | 0.26 |
| 947 | M 3-33 | 18 48 12.131 | -25 28 52.39 | 2.216 | 1.695 | --- | --- | 1311 | 6.0 | 9.8 | 7.5 | 8.94 | 0.26 |
| 948 | M 3-34 | 19 27 01.897 | -06 35 04.63 | 2.664 | 2.402 | 0.475 | 0.162 | 1311 | 8.0 | 29.6 | 29 | 4.12 | 0.16 |
| 949 | M 3-35 | 20 21 03.769 | +32 29 23.86 | 4.837 | 3.650 | --- | --- | 3311 | 1.5 | 29.5 | 140 | 5.62 | 0.04 |
| 950 | M 3-36 | 17 12 39.152 | -25 43 37.39 | 1.387 | 1.603 | 1.946 | 2.054 | 1311 | 3.2 | 4.9 | 3.5 | 13.92 | 0.22 |
| 952 | M 3-38 | 17 21 04.459 | -29 02 59.21 | 6.630 | 5.961 | 0.778 | 0.384 | 3311 | 1.8 | 7.6 | 21 | 8.29 | 0.07 |
| 953 | M 3-39 | 17 21 11.505 | -27 11 38.13 | 22.638 | 24.194 | 7.809 | 1.158 | 3331 | 19.0 | 279.4 | 280 | 1.7 | 0.16 |
| 954 | M 3-4 | 07 55 11.398 | -23 38 12.46 | 1.061 | 1.155 | 0.983 | 0.470 | 1311 | 13.5 | 6.3 | 2.4 | 6.58 | 0.43 |

| KN | NAME | R.A. (2000.0) H:M:S | DEC. (2000.0) D:M:S | F(65μm) Jy | F(90μm) Jy | F(140μm) Jy | F(160μm) Jy | QUAL. | DIAM. arcsec | F(1.4GHz) mJy | F(5GHz) mJy | DIST. kpc | DIAM. pc |
|---|---|---|---|---|---|---|---|---|---|---|---|---|---|
| 955 | M 3-40 | 17 22 28.272 | -27 08 42.40 | 5.041 | 4.717 | 3.036 | 2.794 | 3331 | 2.6 | 15.6 | 17 | 8.22 | 0.10 |
| 963 | M 3-48 | 17 59 56.823 | -31 54 27.46 | 0.104 | 0.419 | --- | --- | 1311 | --- | --- | --- | --- | --- |
| 965 | M 3-5 | 08 02 28.932 | -27 41 55.44 | 2.259 | 2.633 | 1.385 | --- | 1311 | 7.0 | 11 | 10 | 5.9 | 0.20 |
| 967 | M 3-51 | 18 04 56.233 | -32 54 01.23 | 0.509 | 0.945 | --- | 1.502 | 1311 | --- | 3.2 | --- | --- | --- |
| 969 | M 3-53 | 18 24 07.890 | -11 06 42.08 | 3.979 | 5.123 | 7.331 | --- | 1311 | --- | 19.5 | --- | --- | --- |
| 975 | M 3-9 | 17 25 43.364 | -26 11 55.47 | 7.307 | 9.022 | 3.599 | --- | 3331 | 16.0 | 37.3 | 35 | 3.88 | 0.30 |
| 977 | M 4-11 | 18 54 17.671 | -10 05 14.49 | 1.644 | 3.396 | 2.079 | 0.429 | 1311 | 21.2 | 9.4 | --- | --- | --- |
| 978 | M 4-14 | 19 21 00.718 | +07 36 52.37 | 3.094 | 3.937 | 5.787 | 2.183 | 3311 | 7.4 | 11.8 | 8 | 6.28 | 0.23 |
| 980 | M 4-18 | 04 25 50.831 | +60 07 12.72 | 3.410 | 1.863 | 0.355 | 1.565 | 3311 | 3.8 | 18.5 | 22 | 6.37 | 0.12 |
| 981 | M 4-2 | 07 28 53.808 | -35 45 13.92 | 0.899 | 0.632 | 2.254 | --- | 1311 | 6.0 | 24.2 | 19 | 5.26 | 0.15 |
| 982 | M 4-3 | 17 10 41.751 | -27 08 43.75 | 1.276 | 1.264 | 2.359 | 3.210 | 1311 | 1.6 | 12.2 | 28 | 8.13 | 0.06 |
| 987 | M 4-9 | 18 14 18.360 | -04 59 21.30 | 4.308 | 8.570 | 5.261 | 2.096 | 3311 | 44.2 | 43.8 | 42 | 1.72 | 0.37 |
| 990 | MaC 1-13 | 18 28 35.242 | -08 43 22.82 | 8.671 | 8.271 | 7.573 | 3.768 | 1311 | --- | 30.4 | --- | --- | --- |
| 992 | MaC 1-2 | 13 54 27.067 | -64 59 36.14 | 1.084 | 1.416 | --- | 0.051 | 1311 | --- | --- | --- | --- | --- |
| 1000 | MeWe 1-7 | 16 47 57.070 | -50 42 48.32 | 3.531 | 2.735 | 4.566 | 1.965 | 1311 | --- | --- | --- | --- | --- |
| 1001 | MeWe 1-8 | 16 48 40.194 | -51 09 20.26 | --- | 1.503 | 0.576 | 2.963 | 1311 | --- | --- | --- | --- | --- |
| 1004 | Me 2-1 | 15 22 19.274 | -23 37 31.34 | 0.842 | 1.197 | --- | 0.115 | 1311 | 7.0 | 33.6 | 30 | 4.32 | 0.15 |
| 1005 | Me 2-2 | 22 31 43.686 | +47 48 03.96 | 0.965 | 0.625 | 0.607 | 1.937 | 1311 | 1.2 | 16 | 40 | 8.56 | 0.05 |
| 1006 | MyCn 18 | 13 39 35.116 | -67 22 51.45 | 17.170 | 17.185 | 6.808 | 6.315 | 3331 | 12.6 | --- | 106 | 3.23 | 0.20 |
| 1007 | My 60 | 10 31 33.390 | -55 20 50.87 | 2.580 | 3.539 | 0.410 | --- | 1311 | 7.6 | --- | 60 | 3.47 | 0.13 |
| 1009 | Mz 2 | 16 14 32.433 | -54 57 03.76 | 6.386 | 7.386 | 2.800 | 0.136 | 3111 | 23.0 | --- | 75 | 2 | 0.22 |
| 1012 | NGC 1501 | 04 06 59.190 | +60 55 14.34 | 5.662 | 12.013 | 2.975 | 1.373 | 3331 | 51.8 | 207.7 | 210 | 1.03 | 0.26 |
| 1013 | NGC 1514 | 04 09 16.990 | +30 46 33.43 | 0.779 | 7.380 | 3.384 | 4.231 | 1331 | 100.4 | 182.7 | 288 | 1.01 | 0.49 |
| 1014 | NGC 1535 | 04 14 15.762 | -12 44 22.03 | 8.439 | 11.349 | 2.444 | 1.103 | 1311 | 18.4 | 167.8 | 160 | 1.77 | 0.16 |
| 1015 | NGC 2022 | 05 42 06.229 | +09 05 10.75 | 4.676 | 5.515 | 2.809 | --- | 3311 | 19.4 | 92.3 | 91 | 2.02 | 0.19 |
| 1017 | NGC 2346 | 07 09 22.547 | -00 48 22.98 | 5.761 | 9.003 | 10.585 | 7.978 | 3331 | 54.6 | 37.6 | 86 | 2.07 | 0.55 |
| 1018 | NGC 2371/2 | 07 25 34.720 | +29 29 25.63 | 5.715 | 6.803 | 4.715 | 1.400 | 3331 | 43.6 | 48.1 | --- | 2.07 | 0.44 |
| 1019 | NGC 2392 | 07 29 10.768 | +20 54 42.44 | 14.760 | 18.032 | 8.169 | 3.250 | 3331 | 44.8 | 279.9 | 251 | 1.44 | 0.31 |
| 1021 | NGC 2452 | 07 47 26.248 | -27 20 07.16 | 3.586 | 5.514 | 4.190 | --- | 3311 | 18.8 | 56 | 57 | 2.31 | 0.21 |
| 1023 | NGC 2610 | 08 33 23.320 | -16 08 57.67 | 1.063 | 1.386 | 0.912 | 0.768 | 1311 | 43.0 | 29.8 | 30 | 1.86 | 0.39 |
| 1024 | NGC 2792 | 09 12 26.596 | -42 25 39.90 | 5.685 | 6.392 | 1.133 | 0.652 | 3311 | 13.0 | --- | 116 | 1.26 | 0.08 |
| 1025 | NGC 2818 | 09 16 01.656 | -36 37 38.76 | 1.665 | 2.485 | 2.807 | 2.324 | 1311 | 40.0 | 46.6 | 33 | 1.79 | 0.35 |
| 1026 | NGC 2867 | 09 21 25.337 | -58 18 40.69 | 11.091 | 10.774 | 2.804 | 2.917 | 3311 | 16.0 | --- | 252 | 1.69 | 0.13 |

| KN | NAME | R.A. (2000.0) H:M:S | DEC. (2000.0) D:M:S | F(65μm) Jy | F(90μm) Jy | F(140μm) Jy | F(160mm) Jy | QUAL. | DIAM. arcsec | F(1.4GHz) mJy | F(5GHz) mJy | DIST. kpc | DIAM. pc |
|---|---|---|---|---|---|---|---|---|---|---|---|---|---|
| 1027 | NGC 2899 | 09 27 03.123 | -56 06 21.22 | 2.701 | 5.130 | 5.775 | 4.372 | 1331 | 90.0 | --- | 86 | 1.06 | 0.46 |
| 1028 | NGC 3132 | 10 07 01.771 | -40 26 11.11 | 26.899 | 38.620 | 23.599 | 18.285 | 3131 | 45.0 | --- | 230 | 1.5 | 0.33 |
| 1029 | NGC 3195 | 10 09 20.910 | -80 51 30.73 | 3.459 | 6.441 | 3.460 | 1.028 | 3331 | 40.0 | --- | 35 | 1.96 | 0.38 |
| 1030 | NGC 3211 | 10 17 50.549 | -62 40 14.57 | 4.017 | 4.506 | 2.280 | 0.502 | 3311 | 16.0 | --- | 228 | 1.7 | 0.13 |
| 1031 | NGC 3242 | 10 24 46.138 | -18 38 32.26 | 33.512 | 38.693 | 7.971 | 6.400 | 3333 | 37.2 | 757.7 | 860 | 0.94 | 0.17 |
| 1033 | NGC 3699 | 11 27 57.745 | -59 57 27.71 | 3.540 | 7.862 | 5.690 | 3.722 | 3311 | 44.8 | --- | 67 | 1.54 | 0.33 |
| 1034 | NGC 3918 | 11 50 17.730 | -57 10 56.90 | 34.408 | 28.325 | 7.270 | 3.945 | 3331 | 18.8 | --- | 859 | 1.17 | 0.11 |
| 1035 | NGC 40 | 00 13 01.023 | +72 31 19.07 | 48.192 | 36.405 | 11.212 | 9.101 | 3333 | 36.4 | 510.3 | 460 | 0.98 | 0.17 |
| 1036 | NGC 4071 | 12 04 14.815 | -67 18 35.57 | 1.429 | 3.284 | 4.025 | 1.239 | 1331 | 63.0 | --- | 26 | 1.73 | 0.53 |
| 1038 | NGC 5189 | 13 33 32.974 | -65 58 26.67 | 15.323 | 23.021 | 9.626 | 7.972 | 3331 | 140.0 | --- | 455 | 0.55 | 0.37 |
| 1039 | NGC 5307 | 13 51 03.273 | -51 12 20.62 | 3.896 | 4.726 | 0.144 | 1.748 | 3311 | 12.6 | --- | 95 | 2.42 | 0.15 |
| 1040 | NGC 5315 | 13 53 56.972 | -66 30 50.96 | 67.003 | 47.078 | 16.764 | 8.931 | 3331 | 6.0 | --- | 442 | 2.14 | 0.06 |
| 1041 | NGC 5873 | 15 12 51.052 | -38 07 33.65 | 1.246 | 0.795 | --- | --- | 1311 | 7.0 | 45.5 | 48 | 4.23 | 0.14 |
| 1042 | NGC 5882 | 15 16 49.941 | -45 38 58.60 | 38.540 | 29.700 | 9.091 | 5.312 | 3331 | 14.0 | --- | 334 | 1.74 | 0.12 |
| 1045 | NGC 6058 | 16 04 26.597 | +40 40 56.05 | 1.308 | 1.251 | 1.208 | 0.906 | 1311 | 26.4 | 7 | 10 | 3.34 | 0.43 |
| 1046 | NGC 6072 | 16 12 58.079 | -36 13 46.06 | 14.896 | 25.114 | 14.516 | 12.712 | 3311 | 70.0 | 138.5 | 164 | 0.98 | 0.33 |
| 1047 | NGC 6153 | 16 31 30.825 | -40 15 14.22 | 88.194 | 69.461 | 19.749 | 10.285 | 3331 | 24.6 | --- | 477 | 1.17 | 0.14 |
| 1048 | NGC 6210 | 16 44 29.521 | +23 47 59.55 | 28.626 | 21.385 | 7.113 | 6.288 | 3331 | 16.2 | 297.8 | 311 | 1.55 | 0.12 |
| 1049 | NGC 6302 | 17 13 44.211 | -37 06 15.94 | 670.180 | 304.595 | 188.327 | 201.675 | 3333 | 45.0 | 1906.8 | 3034 | 0.52 | 0.11 |
| 1050 | NGC 6309 | 17 14 04.322 | -12 54 37.71 | 15.699 | 16.964 | 5.961 | 2.593 | 3331 | 17.0 | 131.5 | 151 | 2.35 | 0.19 |
| 1052 | NGC 6337 | 17 22 15.658 | -38 29 03.47 | 19.391 | 14.945 | --- | --- | 1311 | 47.0 | 45.9 | 103 | 1.33 | 0.30 |
| 1053 | NGC 6369 | 17 29 20.443 | -23 45 34.22 | 69.688 | 61.394 | 18.329 | 11.351 | 3333 | 28.0 | 1825 | 2002 | 0.75 | 0.10 |
| 1055 | NGC 650-51 | 01 42 19.948 | +51 34 31.15 | 2.536 | 7.243 | 4.019 | 2.100 | 1311 | 138.4 | 138.9 | 125 | 1.56 | 1.05 |
| 1057 | NGC 6543 | 17 58 33.409 | +66 37 58.79 | 117.286 | 71.842 | 21.650 | 13.890 | 3333 | 18.8 | 771.5 | 850 | 1.12 | 0.10 |
| 1058 | NGC 6563 | 18 12 02.753 | -33 52 07.14 | 8.647 | 13.865 | 10.414 | 3.326 | 3331 | 43.0 | 61 | 69 | 2.36 | 0.49 |
| 1060 | NGC 6567 | 18 13 45.136 | -19 04 33.67 | 2.308 | 3.040 | 1.306 | --- | 1311 | 8.8 | 163 | 168 | 2.4 | 0.10 |
| 1061 | NGC 6572 | 18 12 06.404 | +06 51 12.17 | 78.382 | 45.439 | 13.059 | 9.839 | 3333 | 10.0 | 557.5 | 1374 | 1.06 | 0.05 |
| 1063 | NGC 6620 | 18 22 54.186 | -26 49 17.00 | 2.124 | 2.806 | 1.033 | --- | 1311 | 5.0 | 17.1 | 20.5 | 6.42 | 0.16 |
| 1064 | NGC 6629 | 18 25 42.449 | -23 12 10.59 | 25.167 | 19.057 | 5.565 | 3.093 | 3331 | 15.0 | 257.8 | 275 | 1.84 | 0.13 |
| 1065 | NGC 6644 | 18 32 34.641 | -25 07 44.00 | 2.548 | 2.359 | 1.031 | 1.555 | 1311 | 2.8 | 64.3 | 97 | 4.09 | 0.06 |
| 1066 | NGC 6720 | 18 53 35.079 | +33 01 45.03 | 31.015 | 40.615 | 20.511 | 20.619 | 3333 | 69.2 | 420.8 | 360 | 1.13 | 0.38 |
| 1067 | NGC 6741 | 19 02 37.088 | -00 26 56.97 | 10.430 | 11.810 | --- | 0.433 | 1311 | 8.0 | 131.9 | 230 | 2.32 | 0.09 |
| 1068 | NGC 6751 | 19 05 55.560 | -05 59 32.92 | 15.440 | 16.206 | 6.072 | 4.448 | 3331 | 21.0 | 55.1 | 63 | 2.18 | 0.22 |

| KN | NAME | R.A. (2000.0) H:M:S | DEC. (2000.0) D:M:S | F(65μm) Jy | F(90μm) Jy | F(140μm) Jy | F(160μm) Jy | QUAL. | DIAM. arcsec | F(1.4GHz) mJy | F(5GHz) mJy | DIST. kpc | DIAM. pc |
|---|---|---|---|---|---|---|---|---|---|---|---|---|---|
| 1069 | NGC 6765 | 19 11 06.465 | +30 32 42.52 | 0.685 | 1.270 | 1.117 | --- | 1311 | 38.0 | 10.1 | 17.7 | 2.41 | 0.44 |
| 1070 | NGC 6772 | 19 14 36.373 | -02 42 25.04 | 4.950 | 10.678 | 4.095 | 3.955 | 3331 | 90.0 | 77.4 | 73 | 1.11 | 0.48 |
| 1071 | NGC 6778 | 19 18 24.939 | -01 35 47.41 | 8.968 | 12.541 | 6.421 | 2.524 | 3311 | 15.8 | 64.6 | 55 | 2.54 | 0.19 |
| 1073 | NGC 6790 | 19 22 56.965 | +01 30 46.45 | 7.792 | 6.359 | 2.105 | 2.493 | 3331 | 1.8 | 52.6 | 290 | 4.1 | 0.04 |
| 1074 | NGC 6803 | 19 31 16.490 | +10 03 21.88 | 7.540 | 7.287 | 3.109 | 0.150 | 3311 | 5.0 | 68.9 | 114 | 3.26 | 0.08 |
| 1075 | NGC 6804 | 19 31 35.175 | +09 13 32.01 | 11.111 | 11.345 | 4.413 | 1.410 | 3331 | 31.4 | 112.7 | 135 | 1.32 | 0.20 |
| 1076 | NGC 6807 | 19 34 33.543 | +05 41 02.58 | 0.645 | 0.718 | 0.650 | 1.521 | 1311 | 0.8 | 8 | 29 | 11.18 | 0.04 |
| 1077 | NGC 6818 | 19 43 57.844 | -14 09 11.91 | 12.621 | 14.240 | 5.929 | 3.970 | 1311 | 18.2 | 290.2 | 300 | 1.47 | 0.13 |
| 1078 | NGC 6826 | 19 44 48.161 | +50 31 30.33 | 27.865 | 26.039 | 7.148 | 3.726 | 3331 | 25.4 | 414.4 | 385 | 1.2 | 0.15 |
| 1081 | NGC 6852 | 20 00 39.213 | +01 43 40.05 | 0.960 | 0.953 | 0.564 | 0.786 | 1311 | 28.0 | 10.4 | 20 | 2.66 | 0.36 |
| 1083 | NGC 6879 | 20 10 26.682 | +16 55 21.30 | 1.513 | 1.296 | 2.735 | --- | 1311 | 5.0 | 23.1 | 18 | 5.78 | 0.14 |
| 1085 | NGC 6884 | 20 10 23.669 | +46 27 39.54 | 11.033 | 10.902 | 2.693 | 1.791 | 3311 | 5.3 | 152.4 | 200 | 2.7 | 0.07 |
| 1086 | NGC 6886 | 20 12 42.813 | +19 59 22.65 | 8.122 | 7.470 | 3.494 | 1.503 | 3331 | 5.5 | 77.6 | 108 | 2.94 | 0.08 |
| 1087 | NGC 6891 | 20 15 08.838 | +12 42 15.63 | 7.305 | 8.289 | 3.144 | 2.253 | 3311 | 10.2 | 110.8 | 105 | 2.35 | 0.12 |
| 1088 | NGC 6894 | 20 16 23.965 | +30 33 53.17 | 2.634 | 8.688 | 3.956 | --- | 3311 | 44.0 | 66.1 | 61 | 1.44 | 0.31 |
| 1089 | NGC 6905 | 20 22 22.940 | +20 06 16.80 | 5.483 | 5.801 | 3.971 | 1.940 | 3331 | 40.4 | 66.7 | 52 | 1.62 | 0.32 |
| 1090 | NGC 7008 | 21 00 32.503 | +54 32 36.18 | 25.227 | 25.494 | 10.836 | 9.013 | 3333 | 90.0 | 263.3 | 217 | 1.31 | 0.57 |
| 1091 | NGC 7009 | 21 04 10.877 | -11 21 48.25 | 70.225 | 66.268 | 14.071 | 11.285 | 3333 | 28.2 | 682 | 750 | 1.09 | 0.15 |
| 1092 | NGC 7026 | 21 06 18.209 | +47 51 05.35 | 30.060 | 29.863 | 9.726 | 5.815 | 3331 | 15.0 | 260.5 | 260 | 1.91 | 0.14 |
| 1094 | NGC 7048 | 21 14 15.224 | +46 17 17.52 | 3.555 | 6.284 | 4.067 | 5.812 | 3311 | 65.0 | 45.6 | 37 | 1.5 | 0.47 |
| 1096 | NGC 7139 | 21 46 08.586 | +63 47 29.45 | 0.450 | 1.675 | 1.275 | 0.807 | 1311 | 76.0 | 29.8 | 28.9 | 1.56 | 0.57 |
| 1098 | NGC 7354 | 22 40 19.940 | +61 17 08.10 | 33.268 | 30.543 | 9.496 | 4.222 | 3331 | 20.0 | 581 | 579 | 1.19 | 0.12 |
| 1099 | NGC 7662 | 23 25 53.968 | +42 32 05.04 | 22.148 | 21.669 | 6.041 | 4.011 | 3331 | 20.0 | 611.5 | 631 | 1.17 | 0.11 |
| 1100 | Na 1 | 17 12 51.900 | -03 16 00.13 | 1.755 | 2.016 | 0.027 | --- | 1311 | 8.0 | 23 | 22.5 | 4.2 | 0.16 |
| 1104 | PBOZ 34 | 18 05 25.509 | -25 13 37.23 | 1.296 | 5.988 | 1.724 | --- | 1311 | --- | 20.8 | 52 | --- | --- |
| 1105 | PB 1 | 07 02 46.764 | -13 42 34.65 | 1.675 | 1.619 | 1.828 | --- | 1311 | --- | 11.9 | 18 | --- | --- |
| 1106 | PB 10 | 19 28 14.391 | +12 19 36.17 | 5.336 | 6.444 | 4.651 | --- | 3311 | 8.0 | 50.4 | 50 | 3.53 | 0.14 |
| 1107 | PB 2 | 08 20 40.188 | -46 22 58.82 | 0.882 | 0.718 | --- | --- | 1311 | 3.0 | --- | 40 | 5.75 | 0.08 |
| 1108 | PB 3 | 08 54 18.323 | -50 32 22.34 | 4.526 | 4.492 | 2.307 | 2.818 | 3311 | 7.0 | --- | 70 | 3.4 | 0.12 |
| 1109 | PB 4 | 09 15 07.747 | -54 52 43.78 | 4.735 | 6.393 | 2.352 | 0.322 | 3311 | 11.2 | --- | 71 | 2.81 | 0.15 |
| 1110 | PB 5 | 09 16 09.613 | -45 28 42.79 | 9.504 | 7.797 | 2.420 | 3.249 | 3331 | 5.0 | --- | 107 | --- | --- |
| 1111 | PB 6 | 10 13 15.949 | -50 19 59.28 | 3.869 | 3.036 | 1.924 | --- | 3311 | 11.0 | --- | 30 | 3.55 | 0.19 |
| 1112 | PB 8 | 11 33 17.717 | -57 06 14.00 | 5.599 | 5.831 | 1.735 | --- | 3311 | 5.0 | --- | 27 | 5.15 | 0.12 |

| KN | NAME | R.A. (2000.0) H:M:S | DEC. (2000.0) D:M:S | F(65μm) Jy | F(90μm) Jy | F(140μm) Jy | F(160μm) Jy | QUAL. | DIAM. arcsec | F(1.4GHz) mJy | F(5GHz) mJy | DIST. kpc | DIAM. pc |
|---|---|---|---|---|---|---|---|---|---|---|---|---|---|
| 1113 | PB 9 | 19 27 44.814 | +10 24 20.82 | 1.854 | 2.683 | 2.603 | --- | 1311 | 7.0 | 33.1 | 40 | 3.57 | 0.12 |
| 1114 | PC 11 | 16 37 42.697 | -55 42 26.49 | 0.536 | 1.411 | 0.574 | --- | 1311 | 5.0 | --- | 11 | --- | --- |
| 1116 | PC 13 | 16 50 17.108 | -30 19 55.50 | 0.441 | 1.149 | --- | --- | 1311 | --- | 6.6 | --- | --- | --- |
| 1117 | PC 14 | 17 06 14.767 | -52 30 00.40 | 1.694 | 2.376 | 2.814 | 0.399 | 1331 | 7.0 | --- | 30 | 4.32 | 0.15 |
| 1118 | PC 17 | 17 35 41.677 | -46 59 48.54 | 2.489 | 1.923 | 0.244 | 0.629 | 1311 | 5.0 | --- | 14.7 | 6.12 | 0.15 |
| 1119 | PC 19 | 18 24 44.524 | +02 29 28.10 | 1.632 | 0.877 | 0.617 | 0.512 | 1311 | 2.8 | 22.4 | 25.3 | 6.75 | 0.09 |
| 1120 | PC 20 | 18 43 03.456 | -00 16 37.02 | 8.889 | 7.243 | --- | 2.627 | 1311 | --- | 27.3 | --- | --- | --- |
| 1121 | PC 21 | 18 45 35.215 | -20 34 58.32 | 1.961 | 2.703 | 2.015 | 1.245 | 3311 | 13.4 | 11.2 | --- | --- | --- |
| 1122 | PC 22 | 19 42 03.514 | +13 50 37.33 | 1.509 | 1.767 | --- | --- | 1311 | --- | 9.2 | --- | --- | --- |
| 1124 | PC 24 | 20 19 38.133 | +27 00 11.23 | 2.484 | 2.735 | 1.285 | --- | 1311 | 5.0 | 17.1 | 18 | 5.78 | 0.14 |
| 1127 | PM 1-188 | 17 54 21.100 | -15 55 51.78 | 4.799 | 4.750 | 0.318 | --- | 3311 | --- | --- | --- | --- | --- |
| 1128 | PM 1-276 | 19 02 17.857 | +10 17 34.49 | 3.793 | 5.214 | 2.516 | 2.136 | 3311 | --- | 15.3 | --- | --- | --- |
| 1129 | PM 1-295 | 19 19 18.760 | +17 11 48.09 | 7.702 | 7.546 | 7.139 | 0.960 | 1311 | --- | 13.9 | --- | --- | --- |
| 1130 | PM 1-310 | 19 38 52.135 | +25 05 32.63 | 6.213 | 4.603 | 1.867 | 7.758 | 1311 | --- | --- | --- | --- | --- |
| 1132 | PM 1-89 | 15 19 08.724 | -53 09 50.05 | 10.027 | 9.316 | 2.292 | --- | 3311 | --- | --- | --- | --- | --- |
| 1136 | Pe 1-1 | 10 38 27.607 | -56 47 06.40 | 8.965 | 8.009 | 7.825 | 3.319 | 3331 | 3.0 | --- | 125.3 | 4.16 | 0.06 |
| 1137 | Pe 1-11 | 18 01 42.767 | -33 15 26.30 | 2.643 | 2.175 | 0.823 | 1.553 | 1311 | 9.0 | 10.9 | 8 | 7.9 | 0.34 |
| 1138 | Pe 1-12 | 18 17 42.379 | -28 17 13.80 | --- | 0.476 | 0.660 | 0.797 | 1311 | 12.0 | --- | 3.4 | 10.2 | 0.59 |
| 1139 | Pe 1-13 | 18 34 51.659 | -22 43 16.97 | 0.076 | 0.731 | --- | 1.028 | 1311 | 7.0 | 4.4 | 3 | 8.29 | 0.28 |
| 1141 | Pe 1-15 | 18 46 24.485 | -07 14 34.57 | 2.534 | 2.029 | --- | --- | 3311 | 5.0 | 8.3 | 8 | 7.4 | 0.18 |
| 1142 | Pe 1-16 | 18 47 32.251 | -06 54 03.52 | 2.434 | 2.455 | 7.163 | 1.089 | 3311 | --- | 14.2 | --- | --- | --- |
| 1147 | Pe 1-20 | 18 57 17.351 | -05 59 51.79 | 0.354 | 0.931 | 0.129 | 0.342 | 1311 | 8.4 | 5.1 | 29.4 | 4.52 | 0.18 |
| 1148 | Pe 1-21 | 18 57 49.642 | -05 27 39.73 | 1.160 | 1.480 | 0.549 | 3.774 | 1311 | 8.6 | 6.3 | 30 | 3.95 | 0.16 |
| 1149 | Pe 1-3 | 10 44 31.116 | -61 39 38.47 | 1.051 | 2.368 | 3.299 | 3.222 | 1311 | 8.0 | --- | 24 | --- | --- |
| 1150 | Pe 1-6 | 16 23 54.309 | -46 42 15.28 | --- | 4.576 | --- | --- | 0311 | 7.2 | --- | 40 | 3.94 | 0.14 |
| 1151 | Pe 1-7 | 16 30 25.852 | -46 02 51.10 | 42.654 | 25.280 | 11.357 | 2.405 | 3311 | 5.0 | --- | 117 | --- | --- |
| 1152 | Pe 1-8 | 17 06 22.558 | -44 13 09.96 | 6.277 | 8.943 | 6.223 | 0.763 | 1311 | --- | --- | --- | --- | --- |
| 1160 | Pe 2-4 | 09 30 48.402 | -53 09 59.33 | 2.264 | 2.602 | 1.272 | 0.906 | 1311 | --- | --- | --- | --- | --- |
| 1162 | Pe 2-7 | 10 41 19.574 | -56 09 16.34 | 1.045 | 1.399 | 2.067 | 0.005 | 1311 | --- | --- | --- | --- | --- |
| 1167 | SaSt 1-1 | 08 31 42.877 | -27 45 31.70 | 1.739 | 1.157 | 1.544 | 1.723 | 1311 | --- | --- | 0.3 | --- | --- |
| 1170 | SaSt 3-166 | 18 29 11.268 | -17 27 12.90 | 1.586 | 2.059 | 0.665 | 4.097 | 1311 | --- | 3.2 | --- | --- | --- |
| 1171 | SaWe 2 | 17 27 00.197 | -27 40 35.11 | 0.562 | 2.440 | 1.572 | --- | 1311 | --- | 14.2 | 3 | --- | --- |
| 1175 | Sa 1-8 | 18 50 44.312 | -13 31 02.18 | 2.218 | 2.007 | --- | 0.590 | 3111 | 5.6 | 13.5 | 11 | 6.33 | 0.17 |

| KN | NAME | R.A. (2000.0) H:M:S | DEC. (2000.0) D:M:S | F(65μm) Jy | F(90μm) Jy | F(140μm) Jy | F(160μm) Jy | QUAL. | DIAM. arcsec | F(1.4GHz) mJy | F(5GHz) mJy | DIST. kpc | DIAM. pc |
|---|---|---|---|---|---|---|---|---|---|---|---|---|---|
| 1176 | Sa 2-21 | 08 08 44.278 | -19 14 03.13 | 0.069 | 0.763 | 1.683 | 0.197 | 1311 | 40.0 | 8.5 | 2.4 | 4.11 | 0.80 |
| 1177 | Sa 2-230 | 17 42 02.007 | -15 56 07.46 | 1.949 | 1.323 | --- | --- | 1311 | --- | 4.9 | 3.8 | --- | --- |
| 1178 | Sa 2-237 | 17 44 42.282 | -15 45 11.45 | 9.836 | 9.483 | --- | 5.196 | 1113 | --- | 5.4 | --- | --- | --- |
| 1182 | Sa 4-1 | 17 13 50.356 | +49 16 11.26 | 0.427 | 0.457 | 0.832 | 0.571 | 1311 | --- | --- | --- | --- | --- |
| 1187 | Sh 1-89 | 21 14 07.628 | +47 46 22.17 | 1.096 | 2.718 | 0.395 | 2.673 | 1311 | 120.0 | 13.5 | 42.5 | 1.88 | 1.09 |
| 1191 | Sn 1 | 16 21 04.421 | -00 16 10.54 | 0.277 | 0.596 | 1.471 | 0.928 | 1311 | 3.0 | 10.4 | 7 | 9.42 | 0.14 |
| 1192 | Sp 1 | 15 51 40.934 | -51 31 28.39 | 4.179 | 8.354 | 4.189 | 4.004 | 1311 | 72.0 | --- | 75 | 1.21 | 0.42 |
| 1193 | Sp 3 | 18 07 15.793 | -51 01 10.41 | 8.052 | 12.906 | 6.587 | 7.967 | 3333 | 35.6 | --- | 61 | 1.75 | 0.30 |
| 1196 | StWr 4-10 | 16 02 13.042 | -41 33 35.94 | 0.041 | 0.394 | --- | 0.523 | 1311 | --- | --- | --- | --- | --- |
| 1197 | St 3-1 | 07 06 50.888 | -03 05 09.50 | 0.753 | 0.932 | 0.385 | --- | 1311 | --- | 7.5 | --- | --- | --- |
| 1198 | Ste 2-1 | 10 11 57.662 | -52 38 17.09 | --- | 0.499 | --- | --- | 1311 | --- | --- | --- | --- | --- |
| 1201 | SwSt 1 | 18 16 12.237 | -30 52 08.71 | 17.294 | 13.034 | --- | 0.315 | 3311 | 1.3 | 26.7 | 216 | 3.79 | 0.02 |
| 1203 | Tc 1 | 17 45 35.298 | -46 05 23.81 | 10.648 | 6.850 | 0.575 | 0.533 | 3311 | 15.0 | --- | 801 | 1.39 | 0.10 |
| 1210 | Te 2337 | 17 48 45.272 | -26 43 28.80 | 10.800 | 8.500 | 19.643 | --- | 3311 | --- | 15.2 | --- | --- | --- |
| 1212 | Th 2-A | 13 22 33.854 | -63 21 00.96 | 4.078 | 5.185 | --- | 3.694 | 1311 | 23.0 | --- | 60 | 2.07 | 0.23 |
| 1216 | Th 3-12 | 17 25 06.093 | -29 45 16.87 | 3.613 | 4.512 | --- | --- | 3311 | 1.8 | 2.6 | 3.5 | --- | --- |
| 1217 | Th 3-13 | 17 25 19.341 | -30 40 41.79 | 8.348 | 2.537 | --- | --- | 3111 | 2.0 | 7.4 | 14.3 | --- | --- |
| 1218 | Th 3-14 | 17 25 44.060 | -26 57 47.69 | 2.922 | 2.598 | 1.308 | --- | 3311 | 1.4 | 5.8 | 4 | --- | --- |
| 1224 | Th 3-25 | 17 30 46.716 | -27 05 59.12 | 0.043 | 1.042 | 0.528 | 2.846 | 1311 | 2.0 | 14.9 | 18 | 8.77 | 0.09 |
| 1225 | Th 3-26 | 17 31 09.297 | -28 14 50.42 | 2.632 | 2.337 | --- | --- | 1311 | 6.6 | 10.4 | 10 | 7.86 | 0.25 |
| 1226 | Th 3-27 | 17 35 58.467 | -24 25 29.15 | 4.357 | 5.395 | 2.905 | --- | 3311 | 3.2 | 14.4 | 13.5 | --- | --- |
| 1228 | Th 3-32 | 17 35 15.533 | -28 07 01.70 | 3.982 | 3.019 | --- | 1.552 | 1311 | --- | 10.2 | --- | --- | --- |
| 1235 | Th 4-10 | 17 57 06.599 | -18 06 43.43 | 1.819 | 2.087 | --- | --- | 1311 | --- | 4.3 | --- | --- | --- |
| 1236 | Th 4-11 | 18 00 08.820 | -17 40 43.33 | 1.832 | 2.515 | 2.449 | --- | 1311 | <5 | --- | 0.2 | --- | --- |
| 1238 | Th 4-3 | 17 48 37.390 | -22 16 48.79 | 1.110 | 1.217 | 0.433 | --- | 1311 | --- | 2.7 | --- | --- | --- |
| 1244 | V-V 3-5 | 18 36 32.291 | -19 19 28.01 | 1.906 | 3.073 | 0.720 | 2.982 | 3311 | --- | 11.6 | 10 | --- | --- |
| 1246 | VBRC 1 | 08 30 54.197 | -38 18 06.98 | 1.648 | 3.486 | 1.496 | 3.464 | 1311 | --- | 37.9 | --- | --- | --- |
| 1248 | VBRC 6 | 14 41 35.992 | -56 15 13.74 | 0.778 | 1.629 | 2.020 | 1.714 | 1311 | --- | --- | --- | --- | --- |
| 1251 | VY 2-1 | 18 27 59.603 | -26 06 48.29 | 4.039 | 4.763 | 2.109 | --- | 3311 | 3.7 | 32.4 | 37 | --- | --- |
| 1259 | Ve 26 | 08 43 28.087 | -46 06 39.72 | 16.331 | 10.899 | 6.139 | 0.586 | 3331 | --- | --- | --- | --- | --- |
| 1260 | Vo 1 | 06 59 26.405 | -79 38 47.20 | 33.470 | 21.452 | 5.513 | 4.417 | 3331 | --- | --- | --- | --- | --- |
| 1261 | Vo 2 | 08 16 10.000 | -39 51 50.67 | 1.385 | 1.676 | --- | 0.006 | 1311 | --- | 60.9 | --- | --- | --- |
| 1262 | Vo 3 | 08 42 16.561 | -40 44 10.57 | 38.376 | 49.309 | 57.133 | 57.844 | 3333 | --- | --- | --- | --- | --- |

| KN | NAME | R.A. (2000.0) H:M:S | DEC. (2000.0) D:M:S | F(65μm) Jy | F(90μm) Jy | F(140μm) Jy | F(160μm) Jy | QUAL. | DIAM. arcsec | F(1.4GHz) mJy | F(5GHz) mJy | DIST. kpc | DIAM. pc |
|---|---|---|---|---|---|---|---|---|---|---|---|---|---|
| 1264 | Vy 1-1 | 00 18 42.167 | +53 52 20.03 | 1.077 | 1.394 | --- | 0.028 | 1311 | 6.0 | 19.8 | 28.9 | 5.34 | 0.16 |
| 1265 | Vy 1-2 | 17 54 22.994 | +27 59 57.99 | 1.704 | 1.184 | 0.697 | --- | 1311 | 4.6 | 11.1 | --- | --- | --- |
| 1266 | Vy 1-4 | 18 54 01.899 | -06 26 19.81 | 1.821 | 2.476 | --- | 1.253 | 1311 | 4.5 | 19.7 | 22 | 6.01 | 0.13 |
| 1267 | Vy 2-2 | 19 24 22.229 | +09 53 56.66 | 31.892 | 23.912 | 4.713 | 4.645 | 3111 | 0.6 | 5.9 | 50 | 9.82 | 0.03 |
| 1268 | Vy 2-3 | 23 22 57.952 | +46 53 58.24 | 0.655 | 0.785 | 1.564 | --- | 1311 | 4.6 | 6.5 | 3 | 9.95 | 0.22 |
| 1271 | WeSb 3 | 18 06 00.766 | +00 22 38.57 | 0.261 | 0.936 | 2.347 | --- | 1311 | --- | 13.1 | --- | --- | --- |
| 1274 | We 1-1 | 00 38 54.178 | +66 23 48.63 | 0.362 | 1.020 | 1.167 | 0.138 | 1311 | 19.0 | 5.2 | 1.5 | 6.57 | 0.61 |
| 1281 | We 1-7 | 18 44 06.233 | -12 12 57.15 | --- | 1.680 | 3.263 | 1.739 | 1311 | --- | 5.3 | --- | --- | --- |
| 1282 | We 1-9 | 20 09 04.958 | +26 26 54.13 | 0.990 | 2.131 | 1.816 | 2.896 | 1311 | --- | 9.9 | --- | --- | --- |
| 1287 | WhMe 1 | 19 14 59.755 | +17 22 46.01 | 6.393 | 5.811 | 1.662 | 3.378 | 3311 | --- | --- | --- | --- | --- |
| 1288 | Wray 16-120 | 12 45 54.930 | -60 20 17.45 | 1.296 | 1.393 | 0.901 | 1.815 | 1311 | --- | --- | --- | --- | --- |
| 1289 | Wray 16-121 | 12 48 30.564 | -63 50 01.79 | 0.016 | 3.465 | 3.108 | 3.314 | 1311 | --- | --- | --- | --- | --- |
| 1290 | Wray 16-122 | 13 00 41.215 | -56 53 40.19 | 0.147 | 0.809 | --- | 2.081 | 1311 | --- | --- | --- | --- | --- |
| 1291 | Wray 16-128 | 13 24 21.922 | -57 31 19.29 | 1.399 | 1.924 | 2.794 | 0.051 | 1311 | --- | --- | --- | --- | --- |
| 1292 | Wray 16-189 | 15 51 19.819 | -48 26 06.93 | 0.808 | 1.757 | 0.916 | 1.617 | 1311 | --- | --- | --- | --- | --- |
| 1293 | Wray 16-199 | 16 00 22.002 | -48 15 35.29 | 2.151 | 2.989 | --- | --- | 1311 | --- | --- | --- | --- | --- |
| 1296 | Wray 16-266 | 17 22 36.947 | -52 46 34.45 | 0.491 | 0.823 | 0.737 | 2.130 | 1311 | --- | --- | --- | --- | --- |
| 1301 | Wray 16-411 | 18 26 41.781 | -40 29 52.52 | 1.512 | 0.884 | 0.772 | 1.506 | 1311 | --- | --- | --- | --- | --- |
| 1303 | Wray 16-93 | 11 30 48.339 | -59 17 04.63 | 0.487 | 0.940 | 0.995 | --- | 1311 | --- | --- | --- | --- | --- |
| 1306 | Wray 17-18 | 08 23 53.825 | -45 31 10.70 | 0.690 | 1.462 | 2.112 | 0.000 | 1311 | --- | --- | --- | --- | --- |
| 1308 | Wray 17-31 | 09 31 20.485 | -56 17 39.39 | 0.343 | 0.881 | 0.225 | --- | 1311 | --- | --- | --- | --- | --- |
| 1309 | Wray 17-40 | 10 06 59.570 | -64 21 49.99 | 0.157 | 1.650 | 0.365 | 1.062 | 1311 | --- | --- | --- | --- | --- |
| 1310 | Wray 17-59 | 13 19 29.925 | -66 09 07.25 | 0.381 | 1.348 | --- | 0.099 | 1311 | --- | --- | --- | --- | --- |
| 1311 | Y-C 2-5 | 08 10 41.628 | -20 31 32.16 | 0.287 | 0.541 | --- | --- | 1311 | --- | 7.1 | 4.5 | --- | --- |

# Figure Captions

**Figure 1**

The variations in the numbers of PNe with size of the search radius $\theta_S$. Note how the numbers of IRC detections achieve a maximum within ~ 2 arcsec of the nominal PNe positions, whilst the number of FIS detections increases more slowly, and peaks at radii close to 15-20 arcsec. This difference in detection rates reflects the difference in the positional accuracies of the surveys. Finally, the flattening of detection rates towards larger search radii confirms the low levels of contamination by background sources.

**Figure 2**

Comparison between source diameters determined from radio and optical observations, and those determined through AKARI 18 $\mu$m observations (upper panel). We also indicate the diameter at which AKARI flags the sources as being extended ("Extended 18 $\mu$m Flag Limit"); the size of the virtual pixel scale; and the FWHM of the PSF (identified as the "Processed FWHM"). Although mean AKARI diameters are comparable to those for the optical/radio regime, it is clear that the range in $\theta$ is significantly smaller. This is illustrated in the lower panel, where we show distributions for the optical/radio and AKARI results.

**Figure 3**

Comparison of IRAS and AKARI photometric results, whence it is seen that there is a linear relation in all of the panels. However, there is also evidence for a systematic disparity between the trends. This may be partially attributable to differences in the wavelengths of observation, and the spectral gradients of the nebular SEDs.

**Figure 4**

An illustration of the variety of AKARI SEDs detected in the present analysis, where the sources are sorted with respect to the wavelength of peak emission. The sources in the top-most panel peak at 18 $\mu$m, whilst those in the middle and lower panels peak at 65 and 90 μm. There is therefore a cooling of the continua as one passes from the upper to lower PNe. We have also included histograms and mean values of the nebular diameters D; where estimates for this parameter are taken from optical/radio observations. It is apparent that diameters are smaller for the upper SEDs, and shift to higher values for the lower SEDs. This trend is consistent with previous observations, and our later investigation of grain temperatures $T_{GR}$.

**Figure 5**

Positions of PNe in the FIS/IRC colour-colour planes (filled circles), together with diagonal lines corresponding to colour-corrected grain emission loci. The temperatures of the grains are indicated by ticks and dashed lines, and represent increments of 10 K in $T_{GR}$. Similarly, the diagonal loci correspond to emissivity exponents $\gamma$ = 0, 1, and 2, as indicated in the panels. Finally, we have indicated the mean errors in the colours (error bars to the left-hand side), and the effects of not including colour corrections (the grey diagonal lines in the lower panel). Note that whilst a large fraction of PNe in the lower panel fall within the limits of the dust continuum loci, this is less the case in the upper panel, where most of the nebulae are positioned to the right. Some of these disparities are attributable to errors in the results, and to the contribution of line and band components.

**Figure 6**

The variation of grain temperatures as a function of nebular diameter, where we have evaluated $T_{GR}$ using results in the lower panel of Fig. 5 (upper panel). Notice how temperatures are larger for diameters < 0.05 pc, and fall within the range $T_{GR}$ ~ 85-120 K for larger shell diameters. Sources with angular sizes $\theta$ < 7 arcsec are indicated by filled circles, and those having $\theta \geq$ 7 arcsec are flagged using open circles. We also show temperatures calculated using 18 and 65 μm fluxes, and assuming an emissivity exponent $\gamma$ = 1 (lower panel).

**Figure 7**

The variation of the intrinsic 18 and 90 µm fluxes with nebular diameter. In both cases, there is a decrease of flux with increasing diameter, most clearly seen in the 18 µm trends. It is likely that most of this arises due to systematic cooling of the IR continua.

**Figure 8**

Variation of total IR luminosity with nebular diameter. It is apparent that luminosities decline with expansion of the PNe, and have maximum values comparable to the luminosities of the central stars. This higher luminosity for the more compact sources suggests appreciable levels of dust opacity. The dashed curve represents an eye-fitting to the results.

**Figure 9**

A comparison between observed 1.4 GHz and integrated FIR fluxes (upper panel), and the intrinsic (distance corrected) counterparts of these fluxes (lower panel). The diagonal lines correspond to the trends where Ly$\alpha$ heating dominates the results, and represent low and high values of the nebular density. It is clear that Ly$\alpha$ heating represents $\leq$ 0.5 of the energy budget of the IR emitting grains.

**Figure 10**

Variation of the Infrared Excess in Galactic PNe. Notice the evolution in the IRE as nebular dimensions increase.

**Figure 11**

The 1.4 and 5.0 GHz emission of the AKARI PNe, where we show a comparison between the fluxes (upper panel), and the evolution of the 5.0GHz/1.4GHz flux ratios (lower panel). The lower dashed line indicates the limit for optically thin sources, whilst the upper dashed line is for high opacity homogenous PNe. It is clear that most of the nebulae follow the locus for optically thin emission (upper panel), and that only a minority of sources are partially or totally optically thick. The evolution of the ratios (lower panel) indicates higher opacities where D < 0.07 pc.

**Figure 12**

The variation of infrared fluxes with respect to 1.4 GHz emission, where we illustrate the trends for four of the AKARI photometric channels. All of the results indicate linear trends. Note how the limiting AKARI and radio sensitivities restrict the numbers of observed sources, and appreciably modify the trends. Least-squares fits to the data are indicated using dashed lines, for which we indicate equations and correlation coefficients (r) within each of the panels.

**Figure 13**

The variation of intrinsic 18 and 90 $\mu$m emission with 1.4 GHz flux. There is still evidence for linearity in the 90 $\mu$m trends, although the 18 $\mu$m results appear to be random. We have indicated the equation and trend for the least-squares fit in the upper panel.

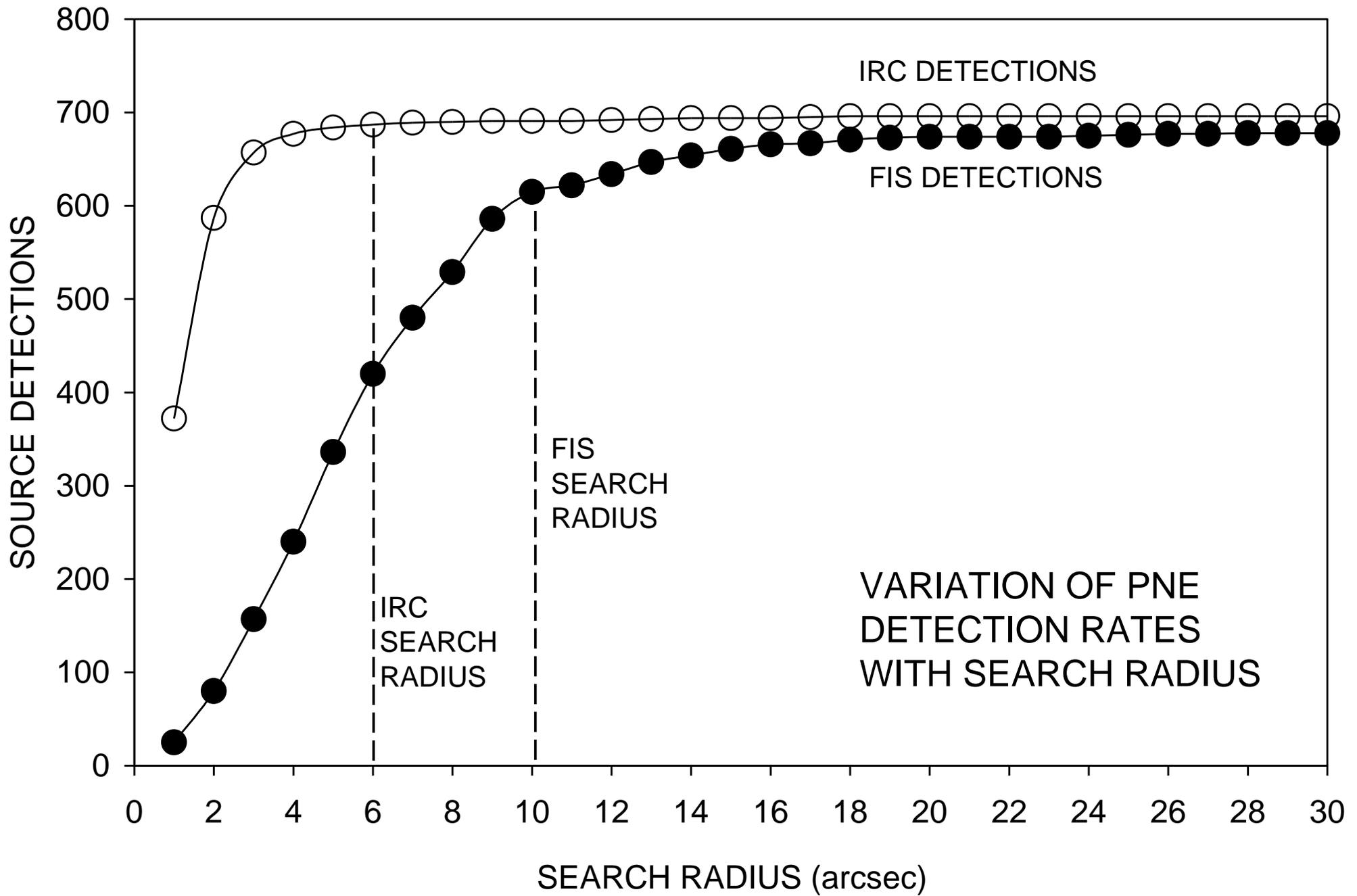

FIGURE 1

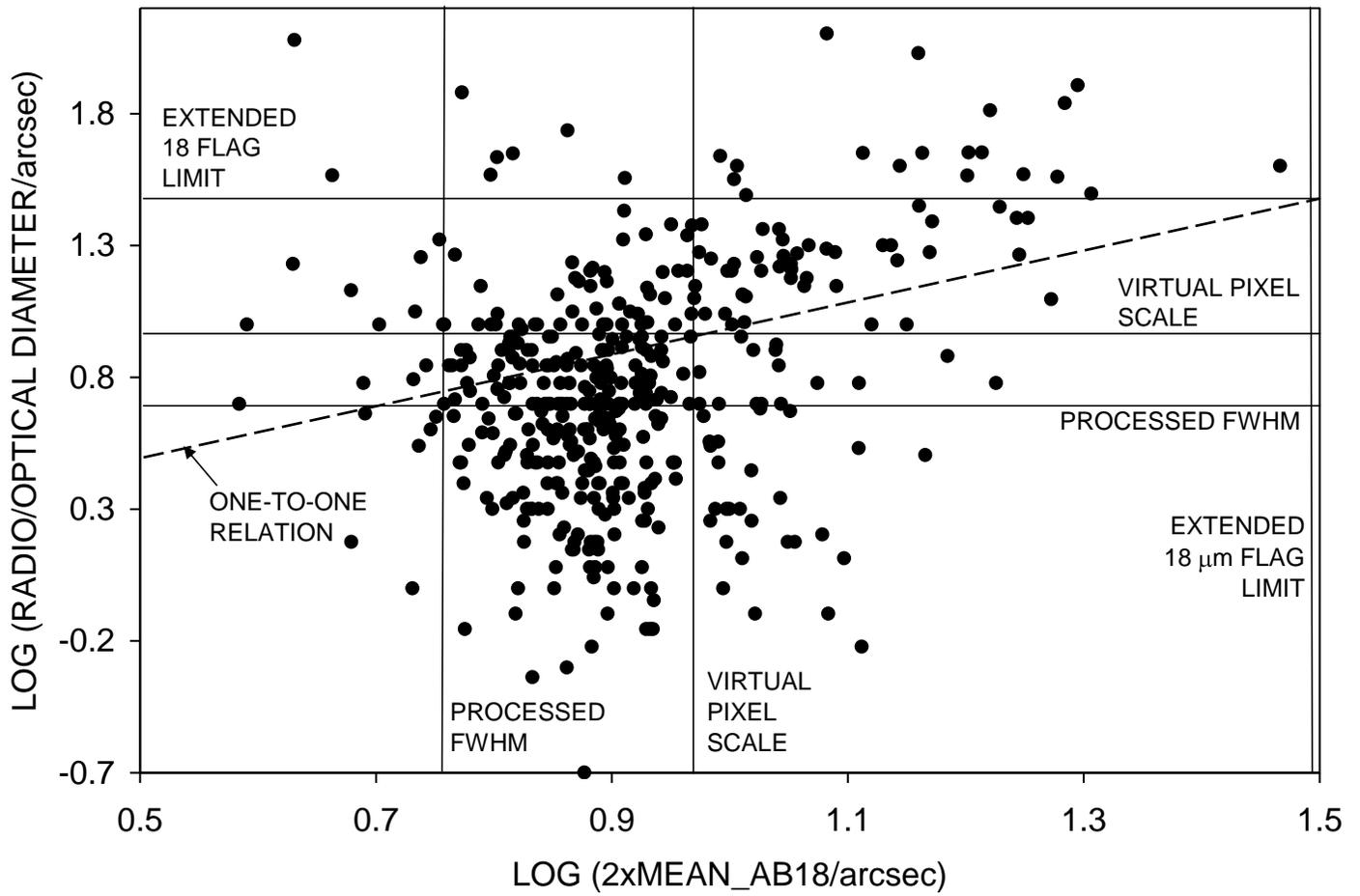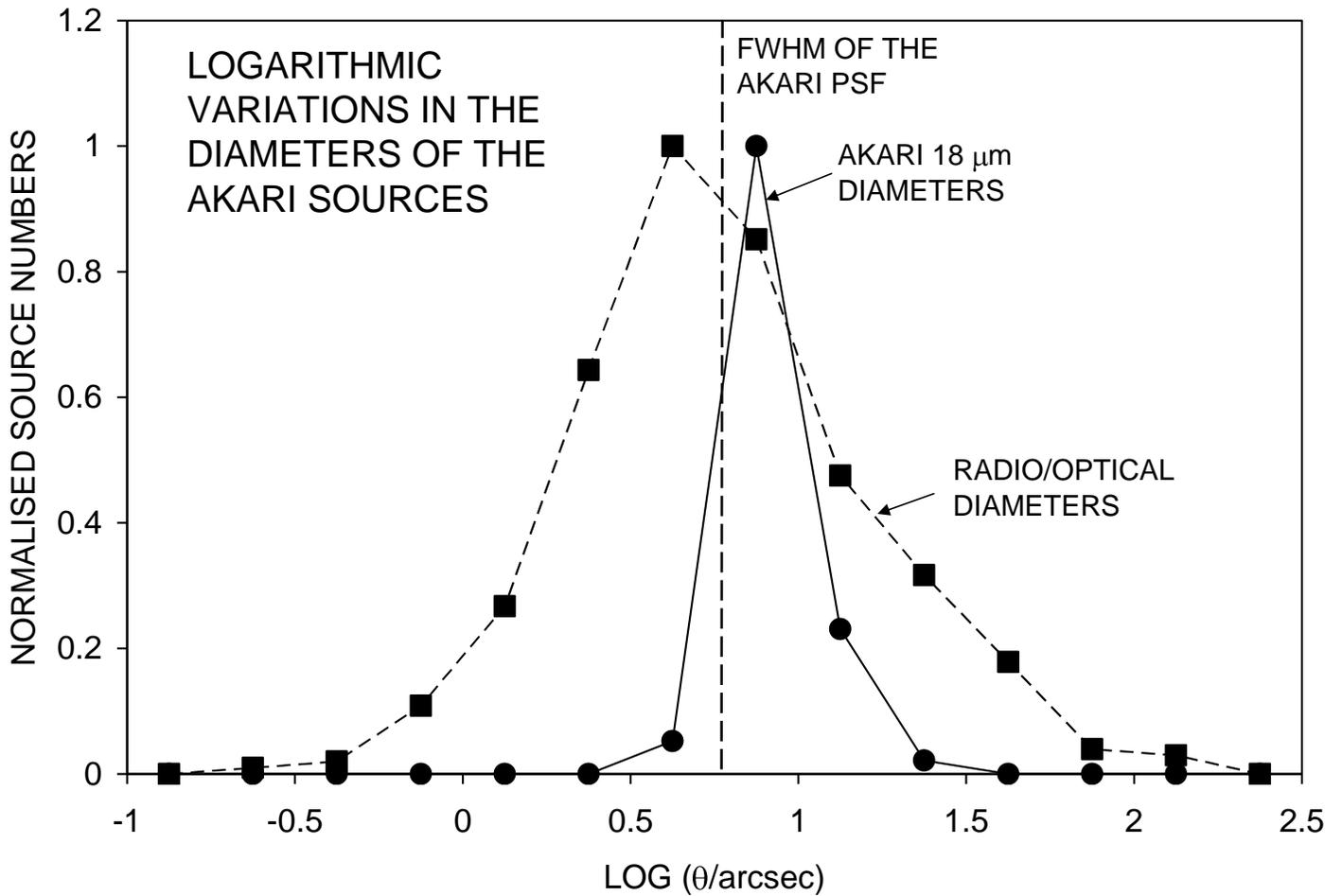

FIGURE 2

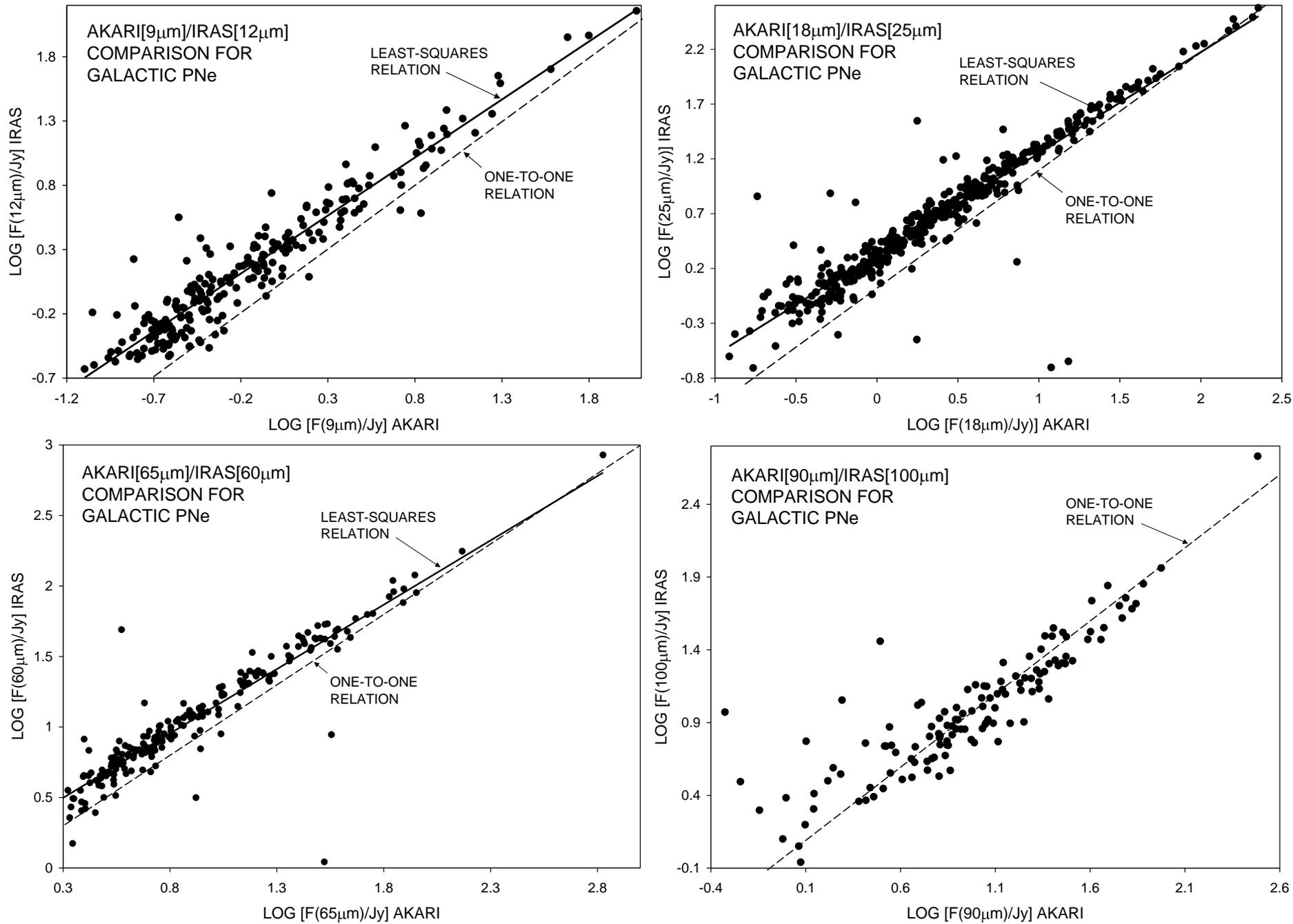

FIGURE 3

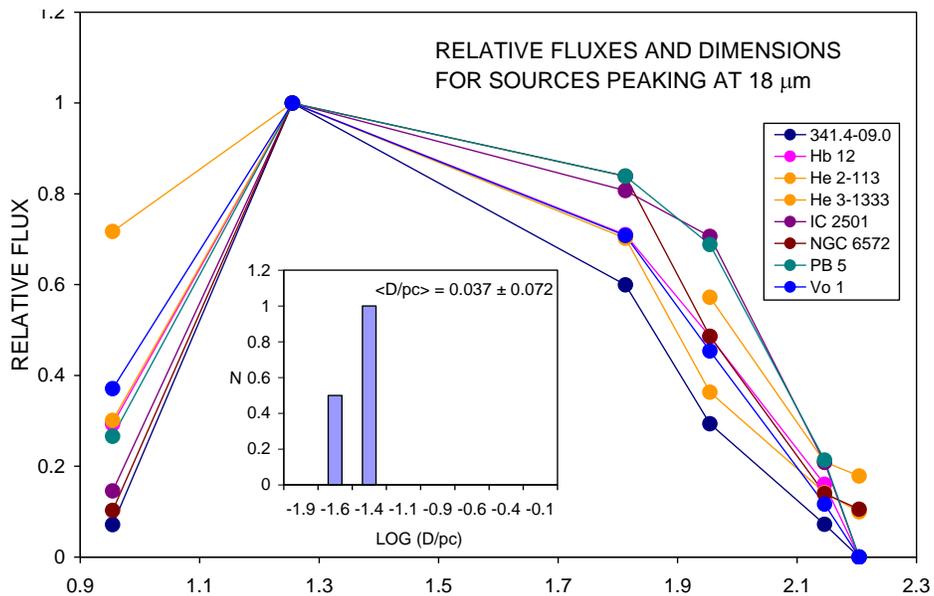
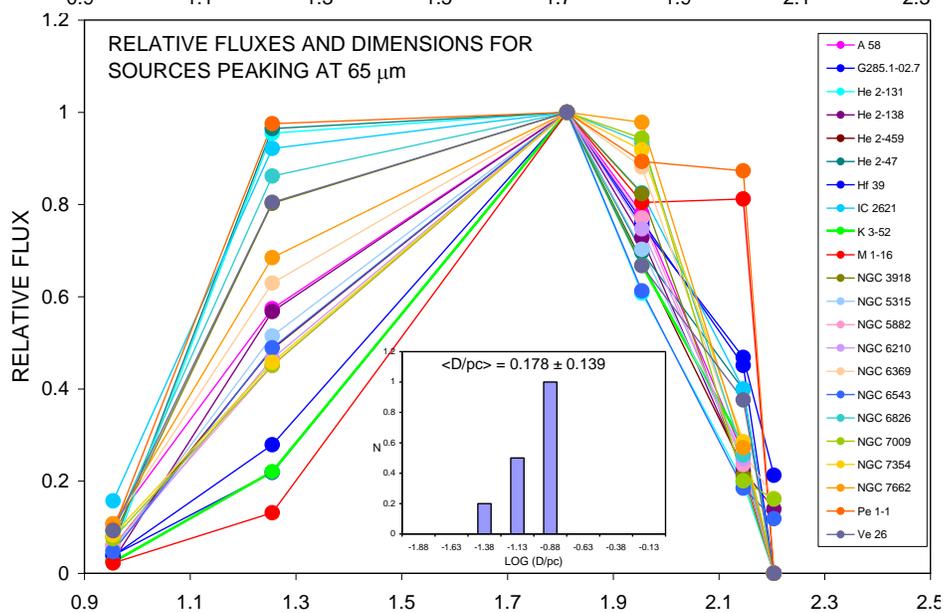
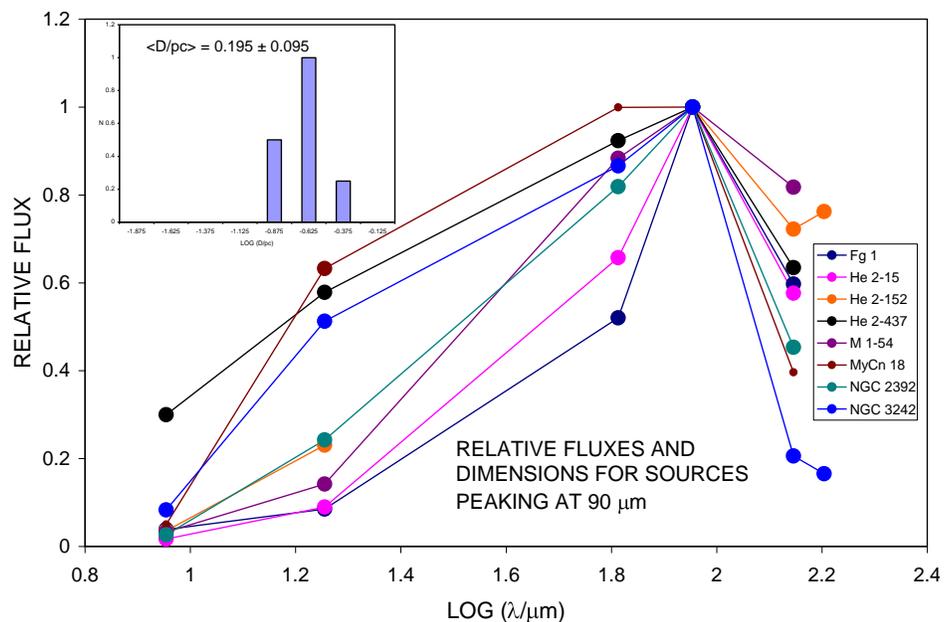

FIGURE 4

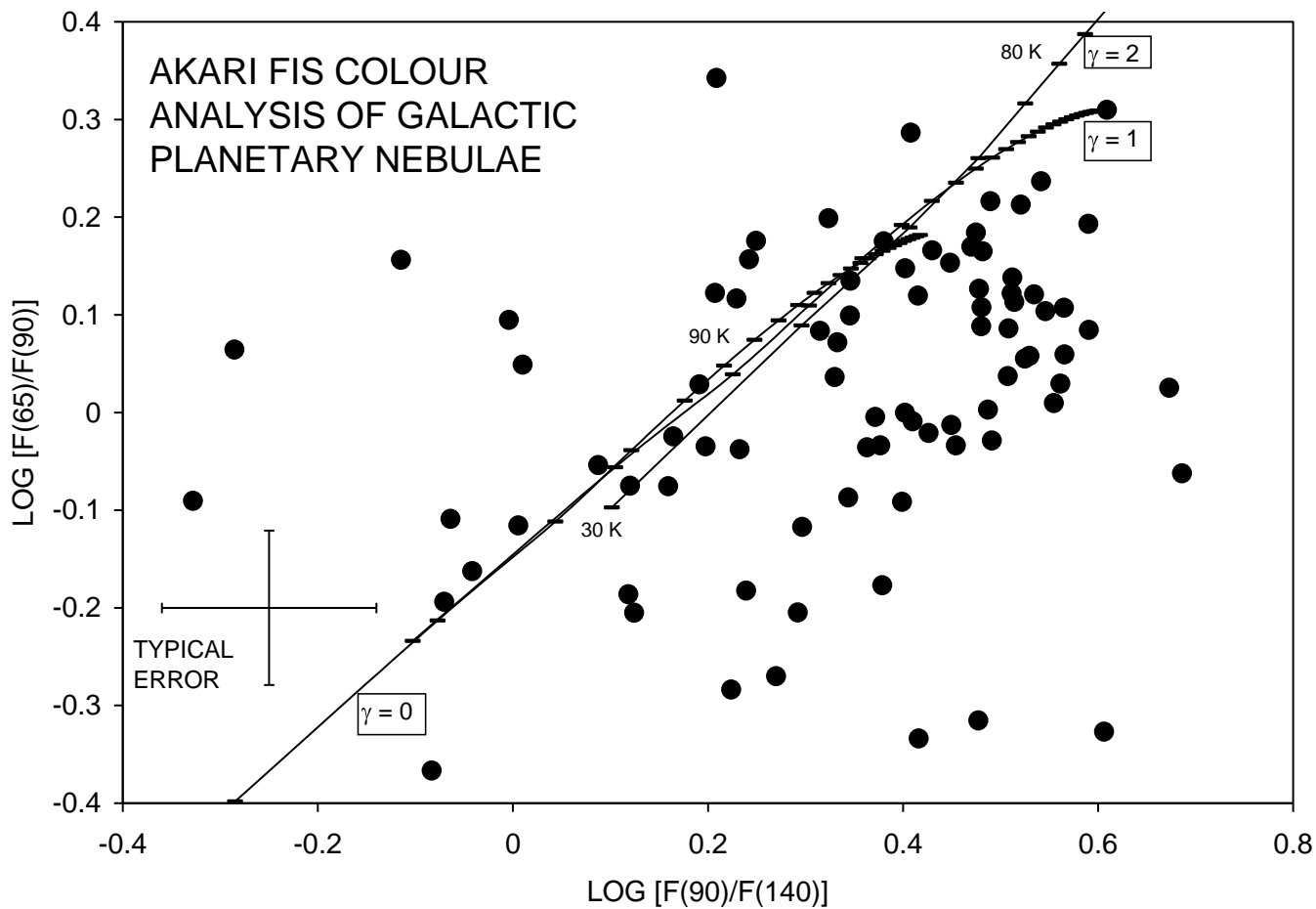
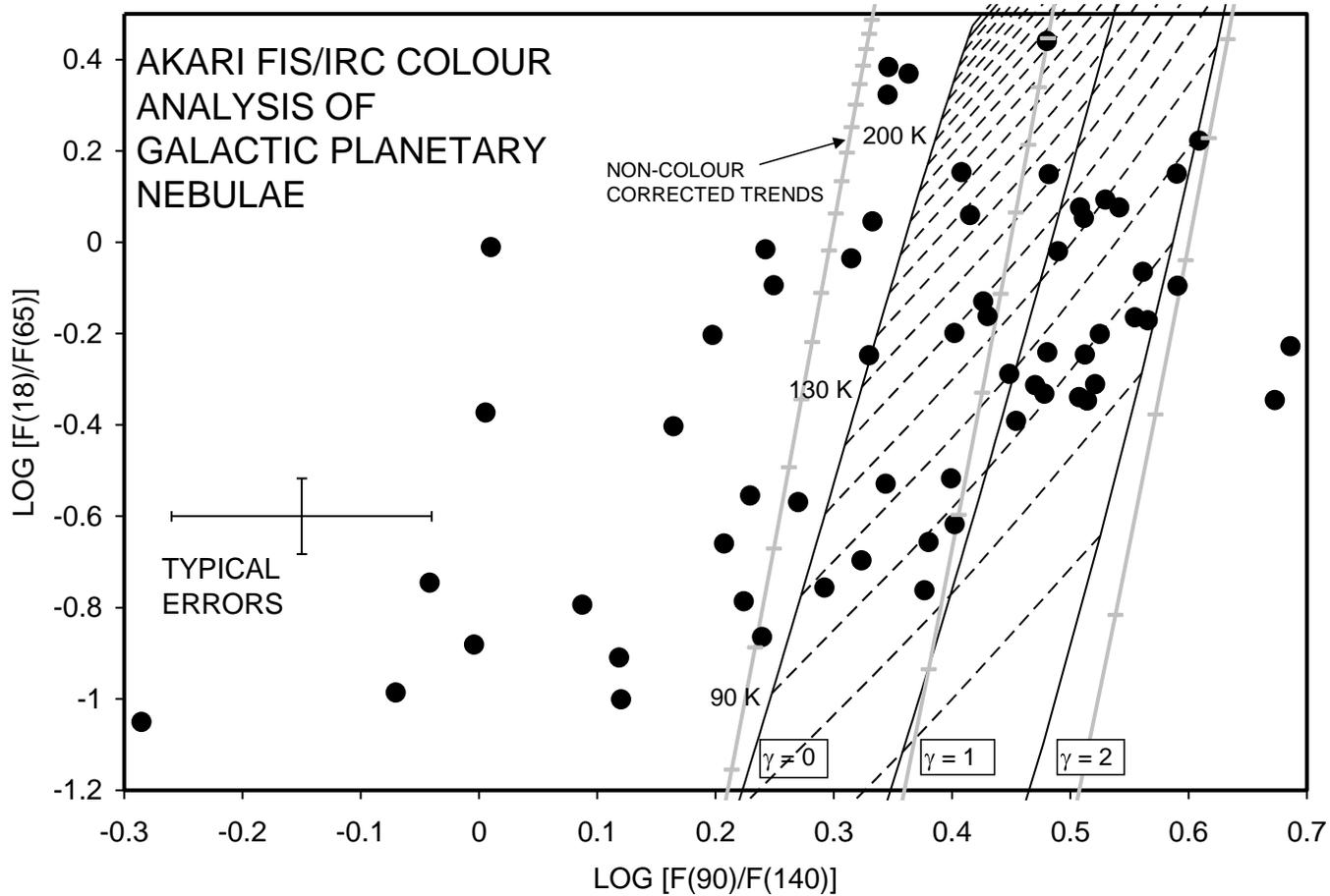

FIGURE 5

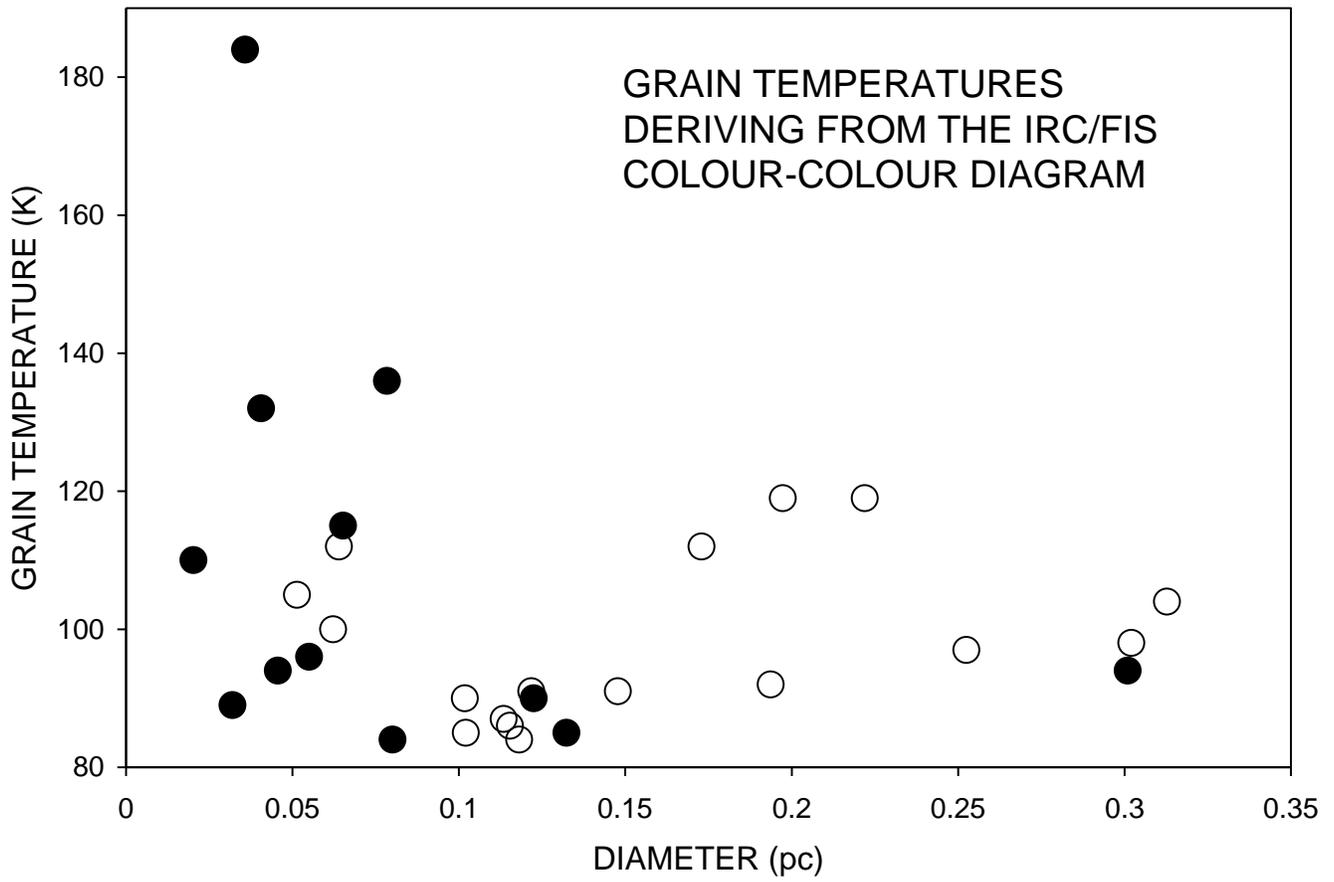
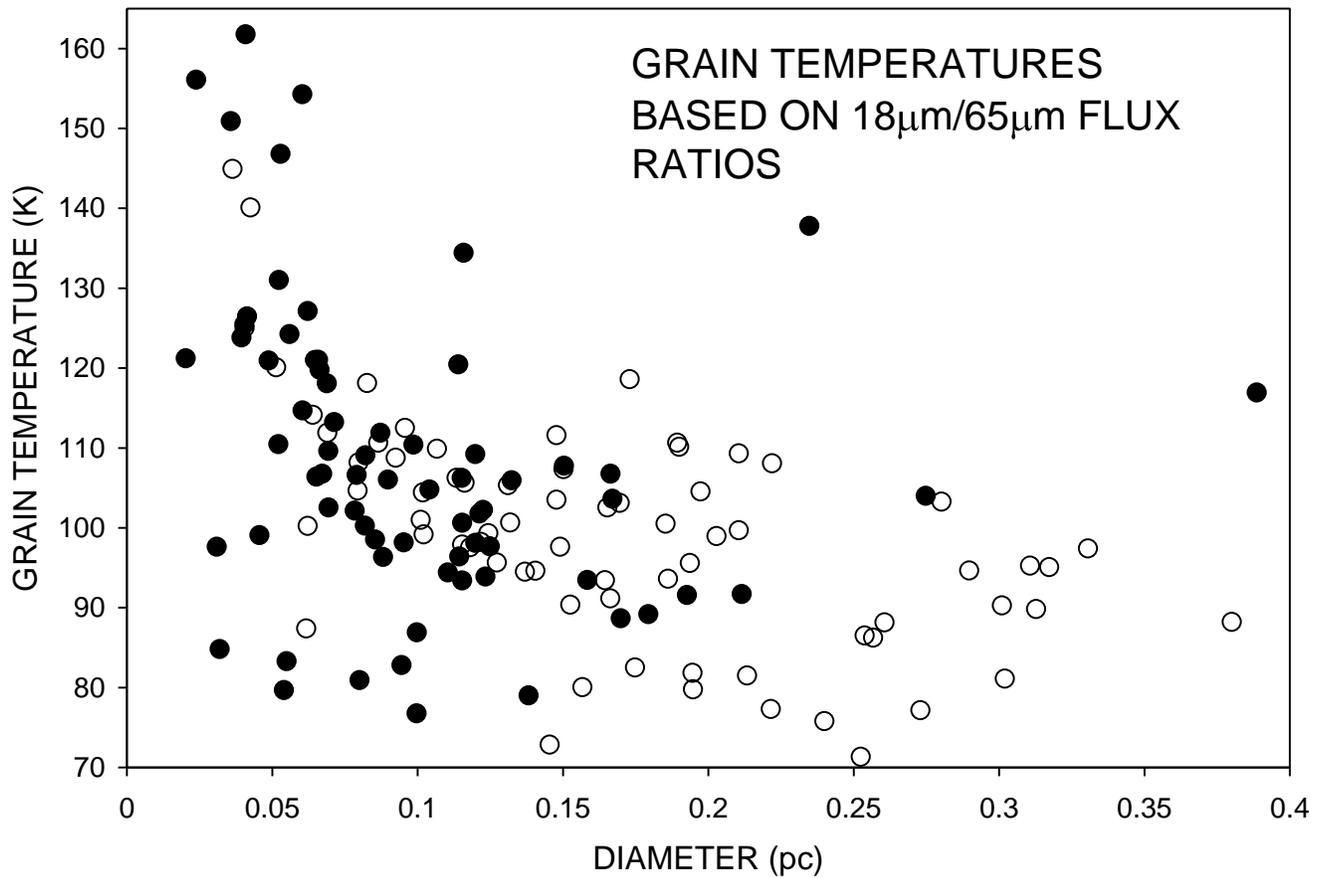

FIGURE 6

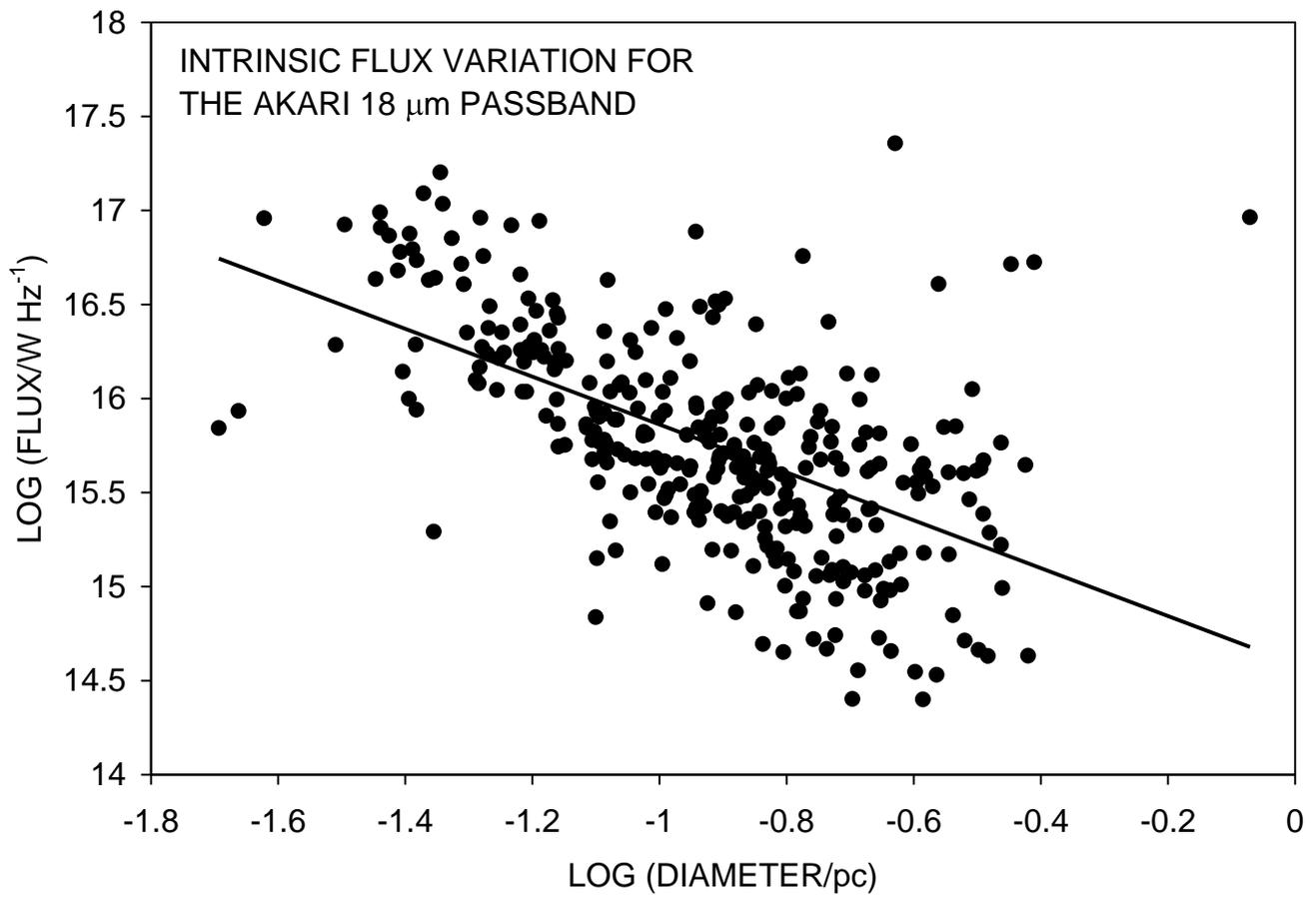
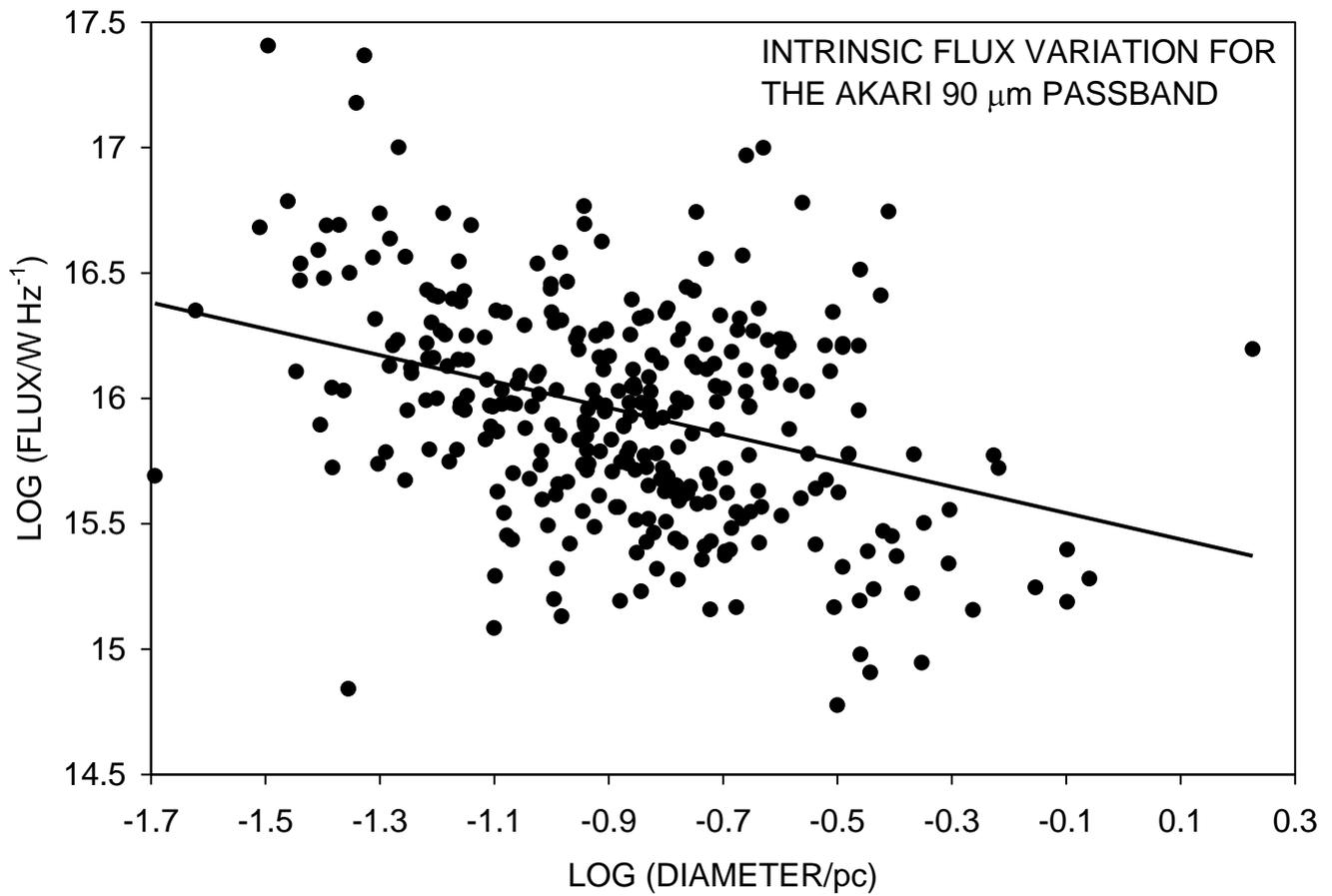

FIGURE 7

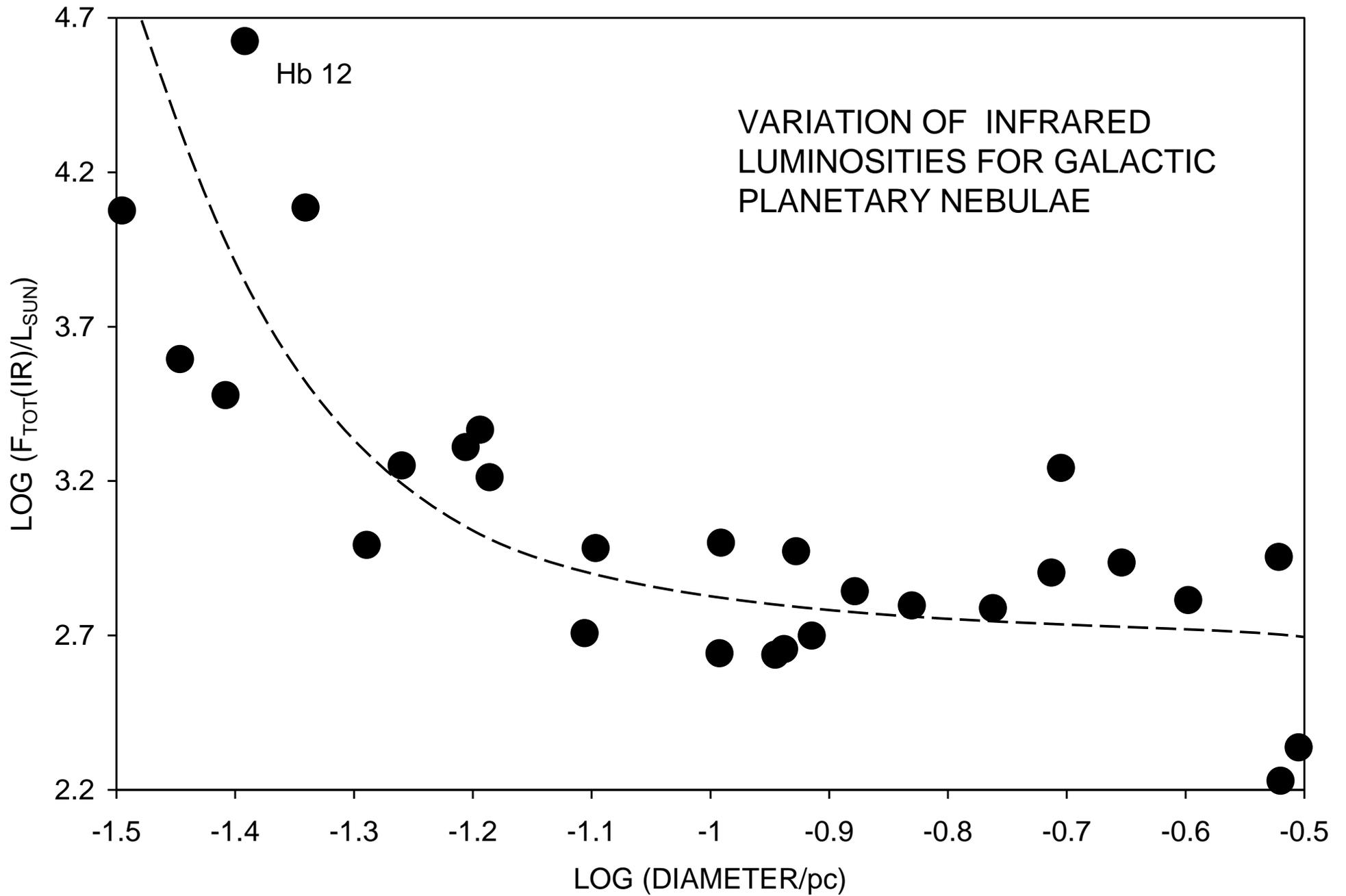

FIGURE 8

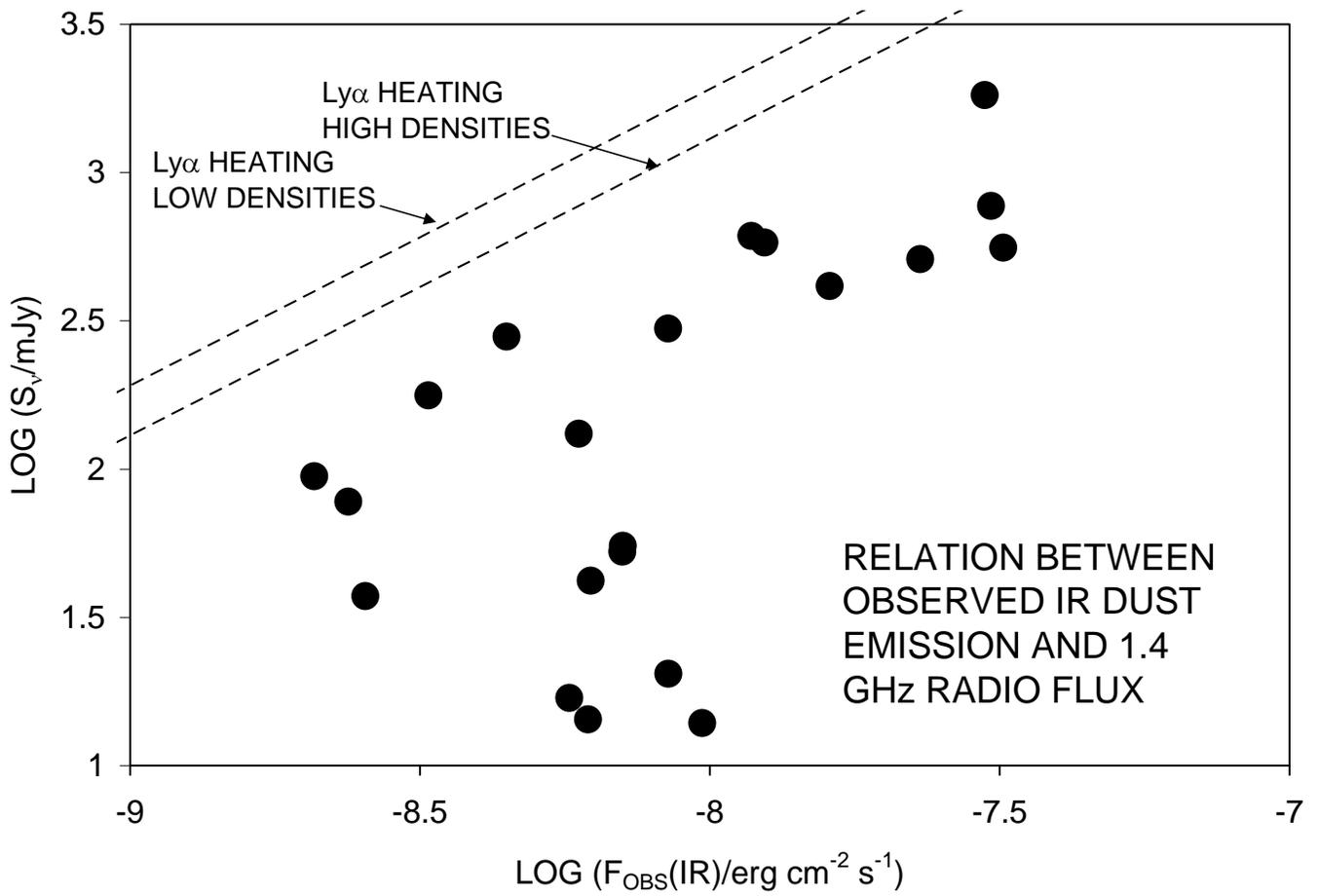
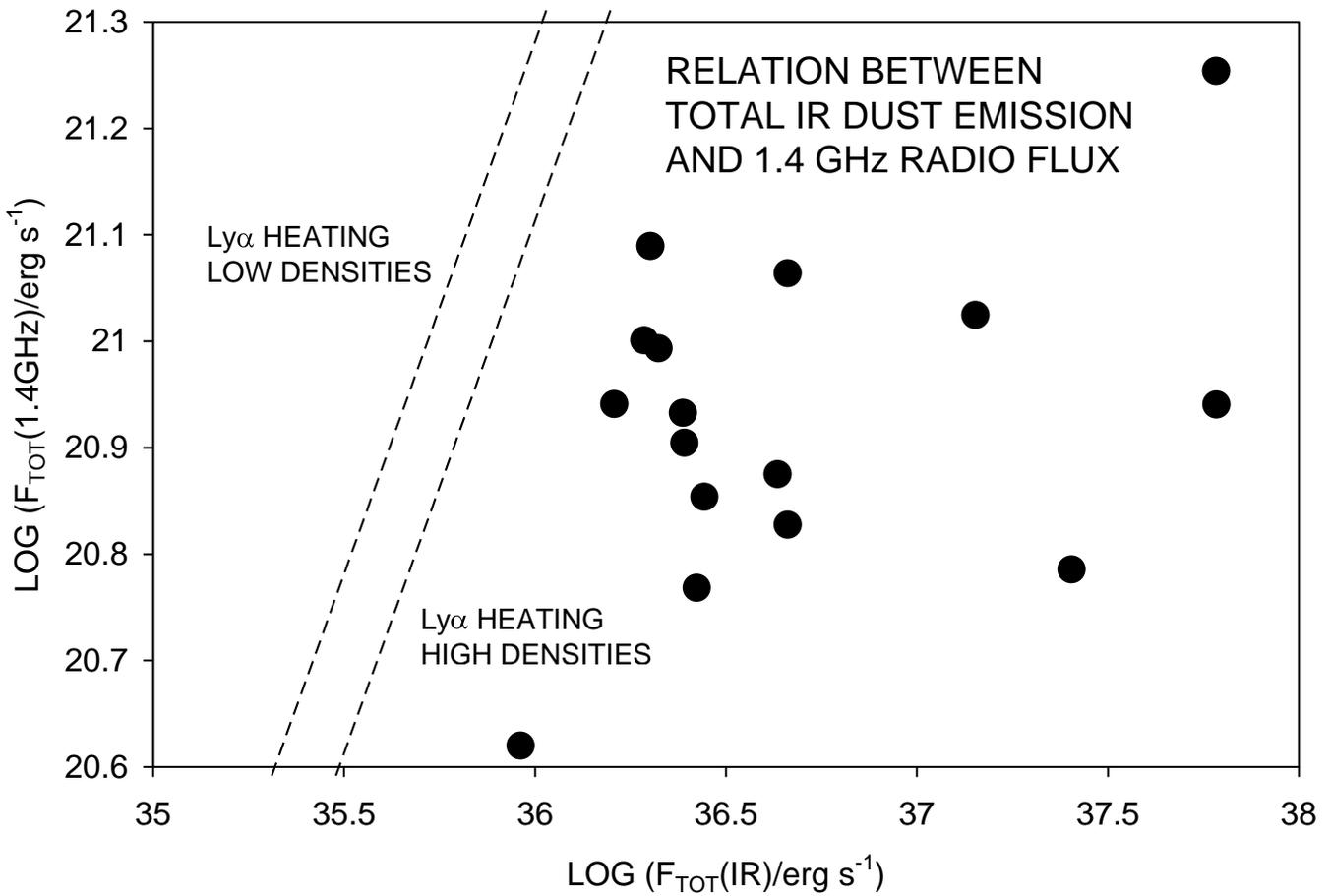

FIGURE 9

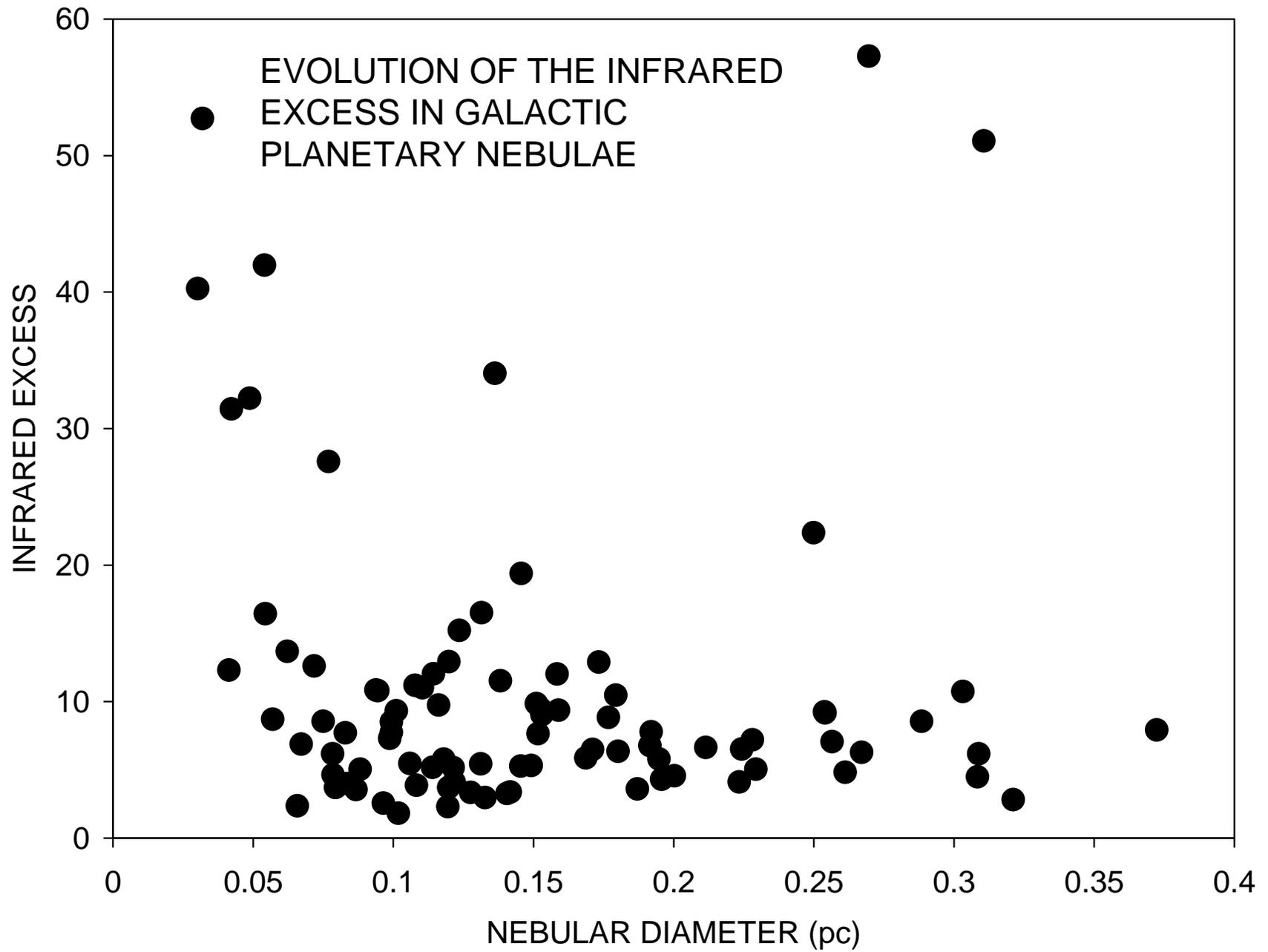

FIGURE 10

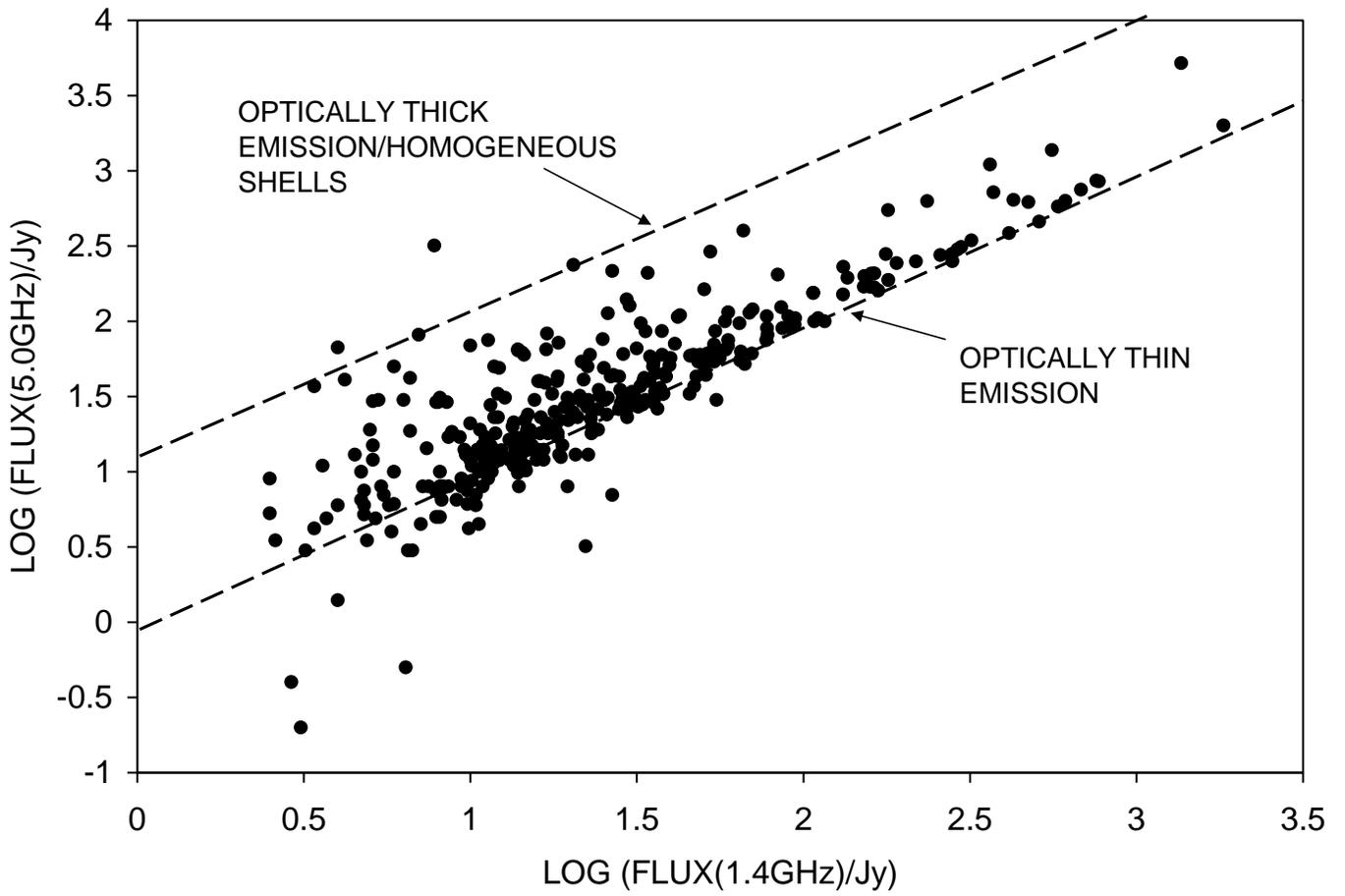
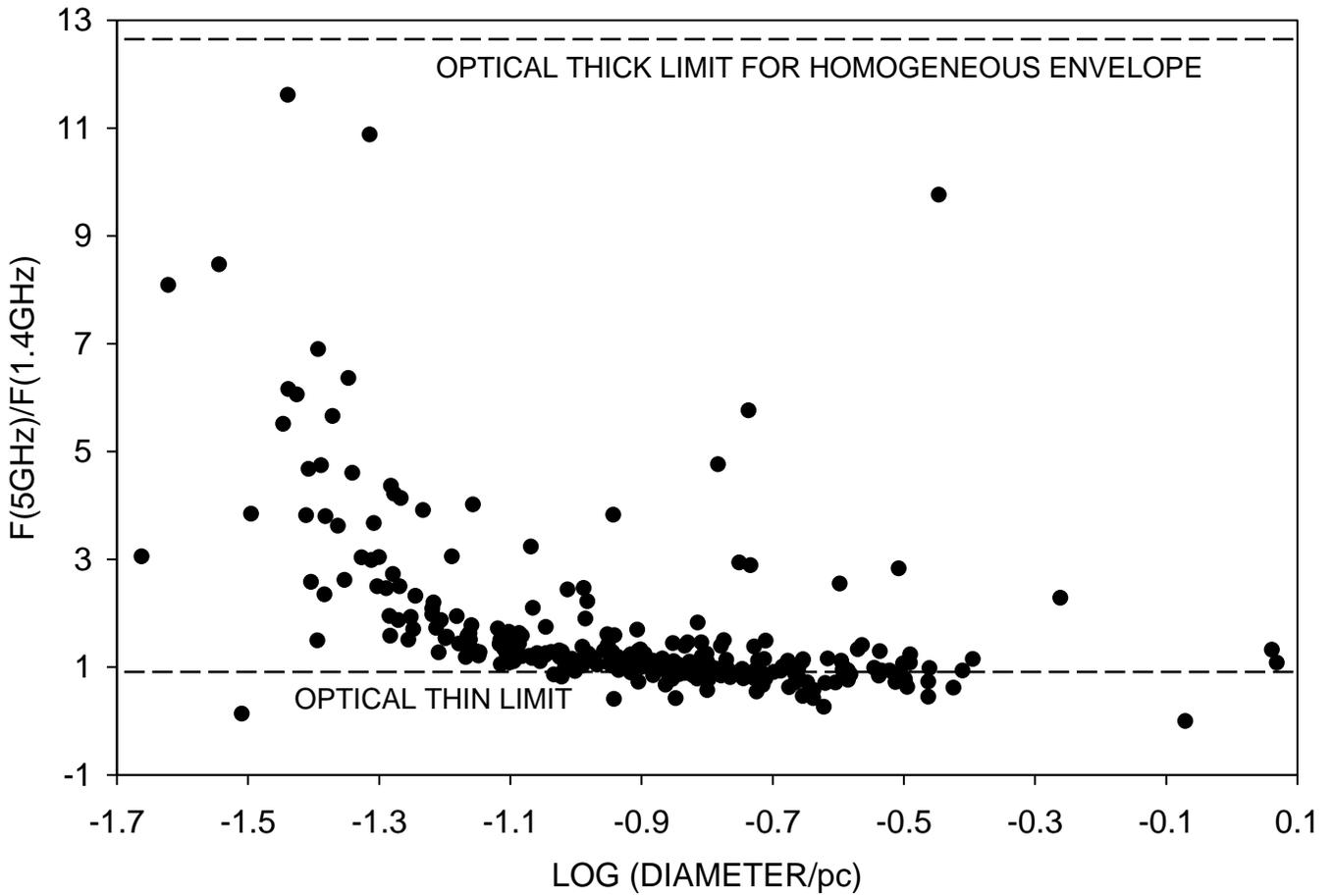

FIGURE 11

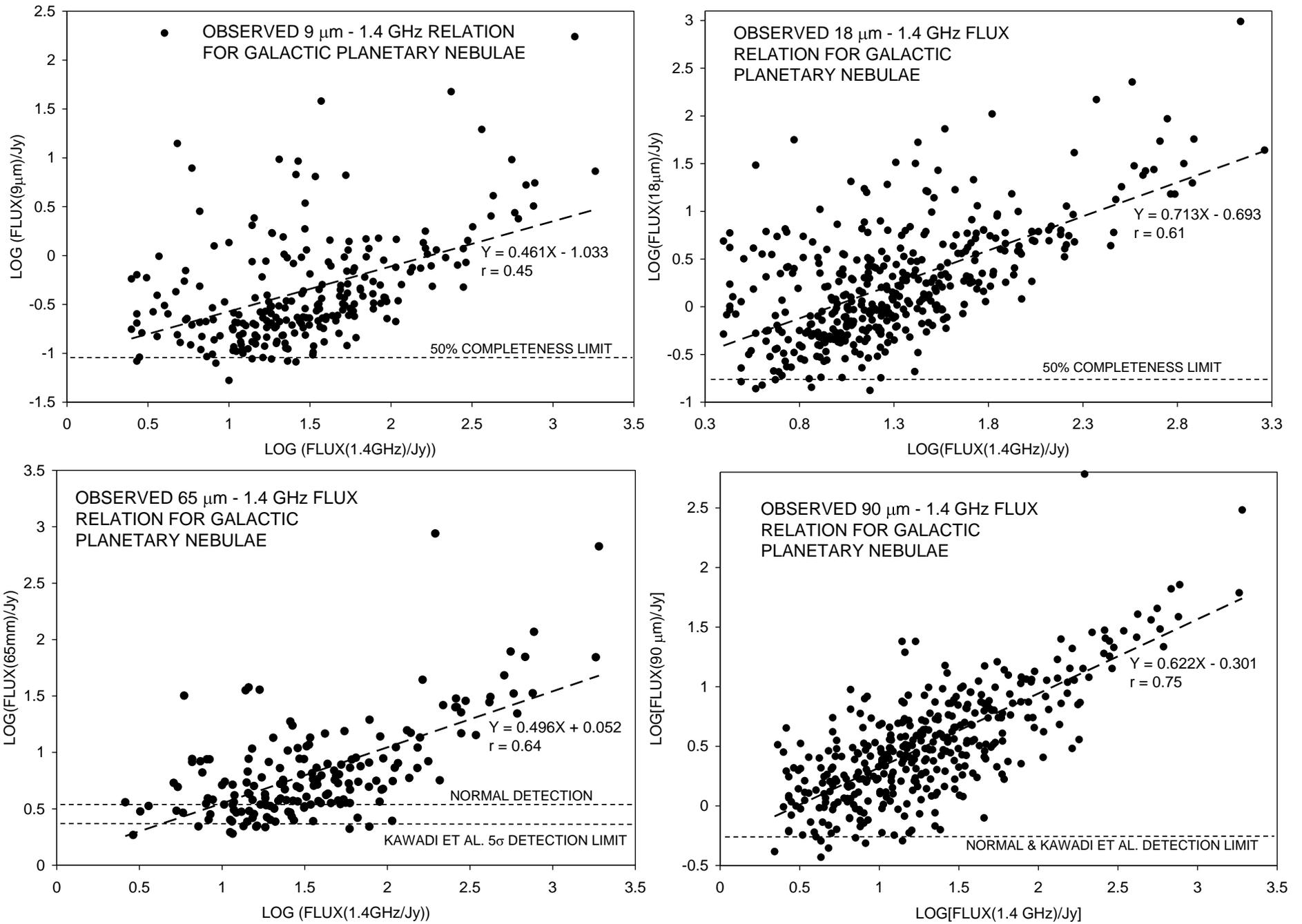

FIGURE 12

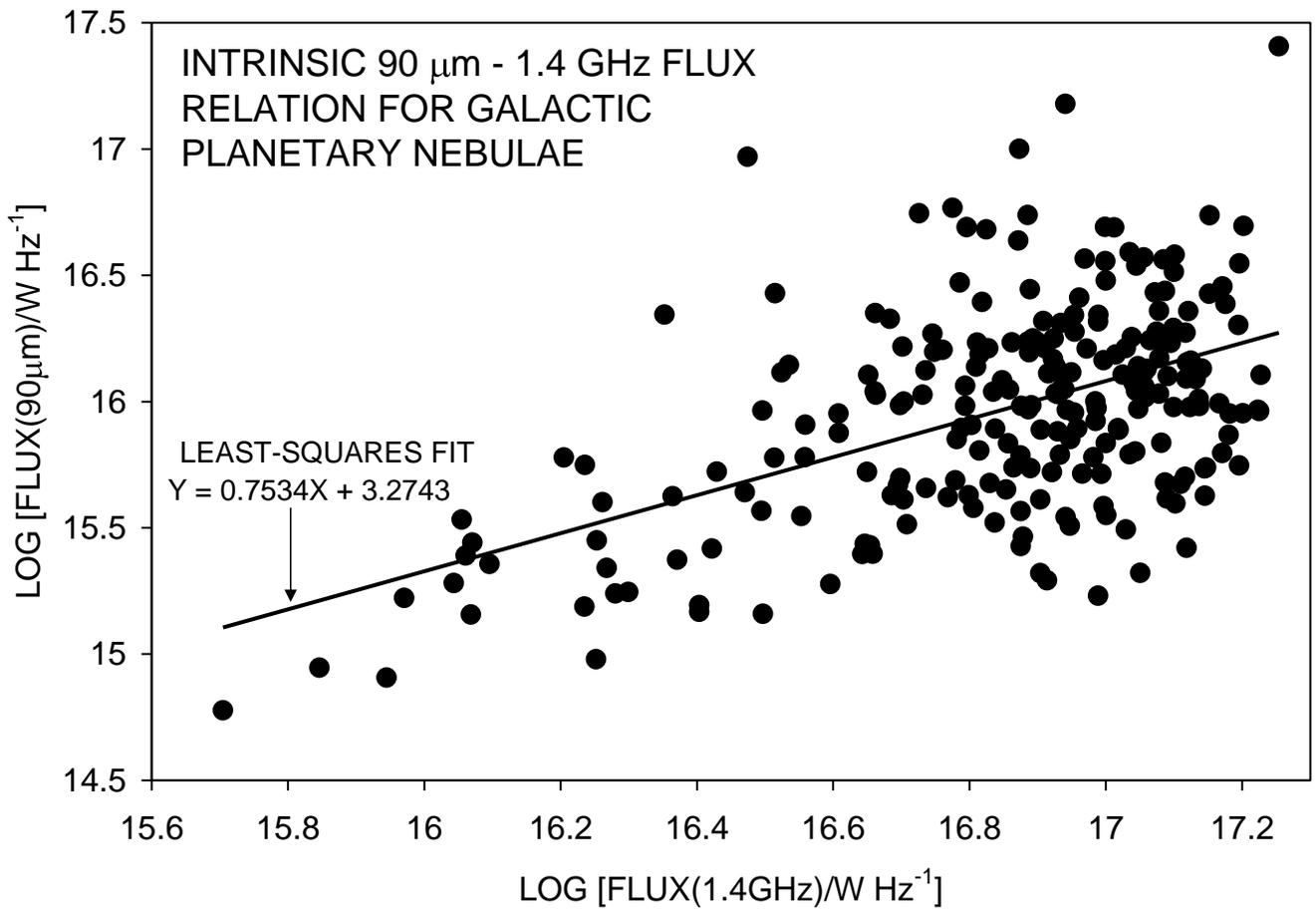

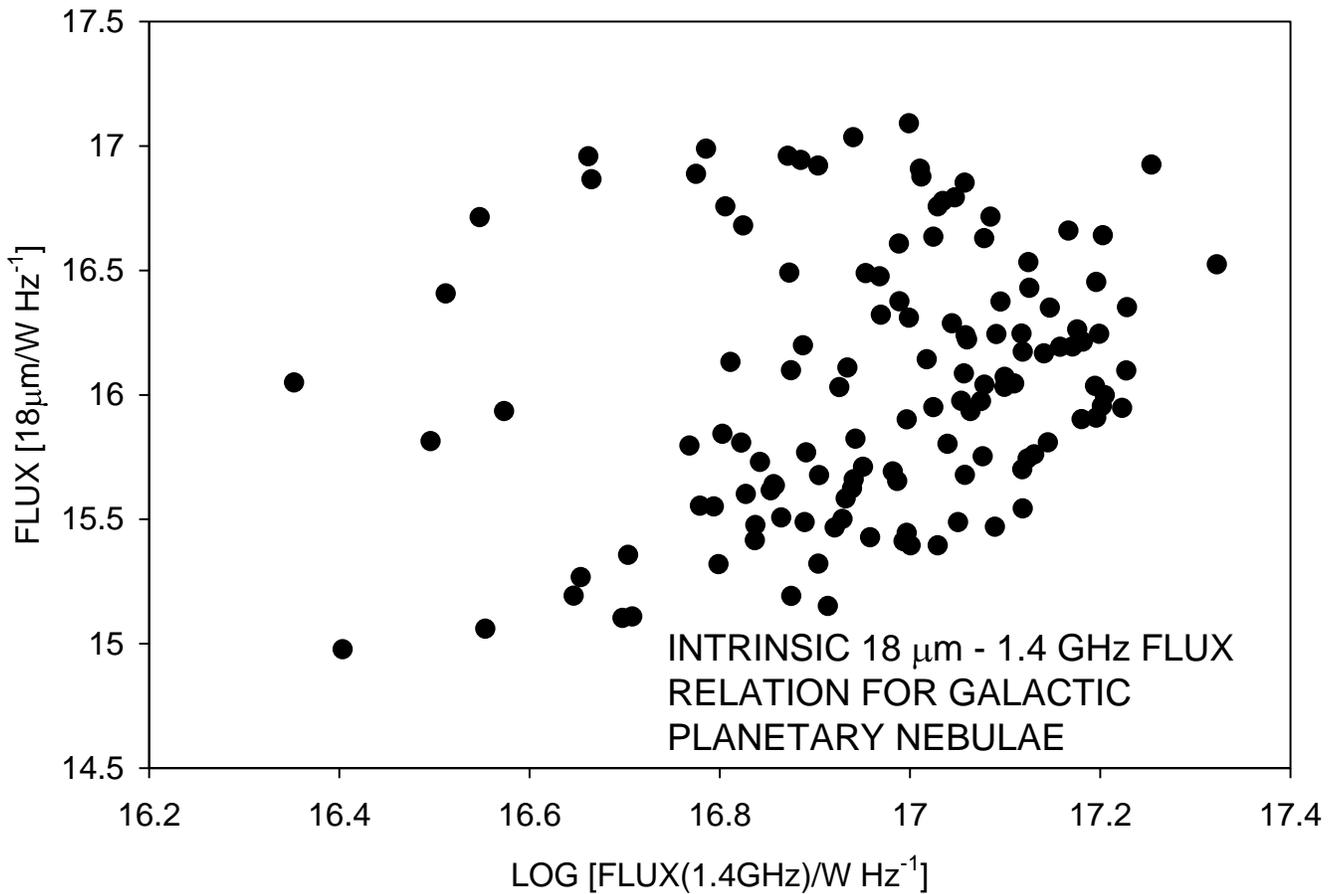

FIGURE 13